\documentclass[twocolumn]{aastex701}

\usepackage{longtable}
\usepackage{appendix}
\usepackage{amsmath}
\usepackage{amssymb}
\usepackage{graphicx}
\usepackage{mfirstuc}
\usepackage{rotating}
\usepackage{soul}

\graphicspath{{./}{Figures/}{Tables/}}

\newcommand{\ha}{H$\alpha$\ }
\newcommand{\hanospace}{H$\alpha$}
\newcommand\oiii{[\ion{O}{3}]\ }
\newcommand\nii{[\ion{N}{2}]\ }
\newcommand\sii{[\ion{S}{2}]\ }

\newcommand{\asym}{$\mathcal{A}$}
\newcommand{\gini}{$\mathcal{G}$}
\newcommand{\mtwenty}{$\mathcal{M}_{20}$}

\newcommand{\asymha}{$\mathcal{A}_{\text{H}\alpha}$}
\newcommand{\giniha}{$\mathcal{G}_{\text{H}\alpha}$}
\newcommand{\mtwentyha}{$\mathcal{M}_{20, \text{H}\alpha}$}

\begin{document}

\title{Galaxies Below the Fundamental Metallicity Relation Are Morphologically Disturbed: Merian \ha Morphologies and DESI Metallicities}

\author[0000-0002-9816-9300]{Abby Mintz}
\affiliation{Department of Astrophysical Sciences, Princeton University, 4 Ivy Lane, Princeton, NJ 08544, USA}
\email{abby.mintz@princeton.edu}

\author[0000-0002-5612-3427]{Jenny E. Greene}
\affiliation{Department of Astrophysical Sciences, Princeton University, 4 Ivy Lane, Princeton, NJ 08544, USA}
\email{}

\author[0000-0002-1841-2252]{Shany Danieli}
\affiliation{School of Physics and Astronomy, Tel Aviv University, Tel Aviv 69978, Israel}
\email{}

\author[0000-0002-0332-177X]{Erin Kado-Fong}
\affiliation{Kavli Institute for Particle Astrophysics \& Cosmology, Stanford University, Stanford, CA 94305, USA}
\email{}

\author[0000-0002-3677-3617]{Alexie Leauthaud}
\affiliation{Department of Astronomy and Astrophysics, University of California, Santa Cruz, 1156 High Street, Santa Cruz, CA 95064 USA}
\email{}

\author[0000-0001-9592-4190]{Jiaxuan Li}
\affiliation{Department of Astrophysical Sciences, Princeton University, 4 Ivy Lane, Princeton, NJ 08544, USA}
\email{}

\author[0000-0001-7729-6629]{Yifei Luo}
\affiliation{Lawrence Berkeley National Laboratory, 1 Cyclotron Road, Berkeley, CA 94720, USA}
\email{yifeiluo@lbl.gov}

\author[0000-0002-8040-6785]{Annika H. G. Peter}
\affiliation{Department of Physics, The Ohio State University, Columbus, OH 43210, USA}
\affiliation{Department of Astronomy, The Ohio State University, Columbus, OH 43210, USA}
\affiliation{Center for Cosmology and Astro-Particle Physics, The Ohio State University, Columbus, OH 43210, USA}
\email{}

\author[0000-0001-6442-5786]{Joy Bhattacharyya}
\affil{Department of Physics \& Astronomy, Amherst College, 6 East Drive, Amherst, MA 01002, USA}
\affil{Department of Astronomy, The Ohio State University, Columbus, OH 43210, USA}
\affil{Center for Cosmology and Astro-Particle Physics, The Ohio State University, Columbus, OH 43210, USA}
\email{}

\author{Mingyu Li}
\affiliation{Department of Astronomy, Tsinghua University, Beijing 100084, China}
\email{}

\author[0000-0001-7831-4892]{Akaxia Cruz}
\affiliation{Department of Physics, Princeton University, Princeton, NJ 08544, USA}

\affiliation{Department of Astrophysical Sciences, Princeton University, Princeton, NJ 08544, USA}

\affiliation{Center for Computational Astrophysics, Flatiron Institute, New York, NY 10010, USA}
\email{}

 \begin{abstract}
Empirical results have revealed a connection between a galaxy's metallicity, mass, and star formation rate, known as the fundamental metallicity relation (FMR), in which metallicity increases with stellar mass and decreases with SFR. Physical interpretations of the FMR suggest that actively star-forming galaxies are fueled by pristine gas, likely obtained via IGM accretion or minor mergers. Here we demonstrate the value of galaxy morphology as an additional axis in the FMR. Using medium-band imaging from the Merian Survey, we construct spatially resolved maps of stellar continuum and H$\alpha$ emission for a large sample of galaxies with DESI DR1 spectra, reaching stellar masses down to $10^8$~M$_\odot$. We measure non-parametric morphological parameters from the emission maps and derive gas-phase oxygen abundances from the spectra using strong-line calibrations. We fit a series of simple linear models to explore correlations among the metallicity, stellar mass, sSFR, and morphological parameters. We find that galaxies that are more metal-poor than predicted by the FMR appear particularly disturbed. At fixed stellar mass and sSFR, metallicity is inversely correlated with the asymmetry of both the continuum and H$\alpha$ morphology. The observed trends -- and the existence of particularly metal-poor, morphologically disturbed, highly star-forming galaxies -- are consistent with external burst triggering. Our results indicate that a galaxy's morphology is linked to its metallicity and star formation history, providing an additional observational probe of the physical drivers of elevated star-formation in low-mass galaxies.
\end{abstract}

\section{Introduction}  
Low-mass galaxies are expected to form stars in a fundamentally different mode than their massive counterparts. Rather than evolving smoothly along the star-forming main sequence, dwarf galaxies undergo repeated episodes of elevated star formation followed by quiescent lulls \citep{Lee2009, Weisz2012, McQuinn2010a, McQuinn2010b, Kauffmann2014, Emami2019}. Because of their shallow potential wells, these bursts can profoundly reshape low-mass galaxies, driving enriched outflows into the circumgalactic medium \citep{Mcquinn2015b, Collins2022, Piacitelli2025}, transforming central dark matter cusps into cores \citep{Pontzen2012, Teyssier2013, DiCintio2014, Read2016, Dutton2019, Sales2022}, and modulating subsequent star formation \citep{Hopkins2014}. Burstiness also complicates the interpretation of high-redshift observations, where measured properties of low-mass galaxies depend sensitively on the phase of the burst cycle at which they are observed \citep{Wang2024}. Understanding what triggers bursts of star formation in low-mass galaxies is therefore essential both for near-field cosmology and for interpreting the growing population of low-mass galaxies observed in the early universe.

Cosmological simulations generally predict that low-mass galaxies experience bursty star formation even in complete isolation: the shallow potential wells of dwarfs allow stellar feedback to drive large-scale outflows that temporarily suppress star formation, after which the expelled gas cools and returns, fueling the next episode \citep{Hopkins2014, El-Badry2016, Sparre2017, Cenci2024}. Notably, these simulations also find that bursts leave observable imprints on galaxy structure, with feedback-driven gas flows causing dwarfs to fluctuate in size and central density over the burst cycle \citep{El-Badry2016}. Recent work suggests that bursts can reshape galaxy morphology more broadly, modulating the degree of rotational support \citep{KadoFong2020, Benavides2025}. A galaxy's morphology -- and in particular the spatial distribution of its young and old stellar populations -- thus carries information about its recent star formation history and, potentially, about the mechanism that triggered its most recent burst. Different triggering channels should leave different signatures: bursts driven by mergers, interactions, or asymmetric gas accretion should disturb the existing stellar body in ways that internally driven cycles may not.

Observational evidence suggests that external processes do in fact play an important role in triggering bursts in low-mass galaxies. Studies of dwarf–dwarf pairs find systematically enhanced star formation relative to isolated controls \citep{Stierwalt2015, KadoFong2024c}, irregular HI morphologies are observed around starbursting dwarfs \citep{Lelli2014}, and chemical dilution at sites of intense star formation has been interpreted as evidence for accretion of relatively pristine gas \citep{SanchezAlmeida2015}. Establishing the relative importance of external and internal triggering across the low-mass population, however, remains difficult: directly detecting the culprits -- faint companions, gas-rich minor mergers, or cold inflows -- is often beyond current observational limits, and a burst's trigger cannot generally be identified on a galaxy-by-galaxy basis. Progress therefore requires statistical diagnostics, measurable for large samples, that respond differently to the different triggering channels.

Galaxy morphology provides one such diagnostic. External perturbations disturb the existing stellar body and redistribute the gas, while purely internal bursts should largely preserve the structure of the evolved stellar population. Beyond the morphology of the stellar continuum, the spatial distribution of recent star formation -- traced by \ha emission on $\sim$10 Myr timescales -- offers a complementary and more responsive probe. Individual case studies and statistical samples have established connections between morphology, stellar mass, and star formation activity \citep{Conselice2003, Fossati2013, Boselli2015, Nersesian2023, Yao2023}. Recent work from the Merian Survey \citep{Danieli2025} showed that low-mass galaxies with elevated specific star formation rates have asymmetric continuum emission and \ha emission that is clumpy, concentrated, and slightly offset from the galaxy center -- morphological signatures suggestive of destabilizing interactions or asymmetric accretion rather than purely internal cycling \citep{Mintz2024}.
 
If external processes are indeed responsible, they should leave a second, independent signature: in the gas-phase metallicity. Observations of star-forming galaxies reveal a tight relation among stellar mass, metallicity, and star formation rate -- the fundamental metallicity relation (FMR) -- in which metallicity increases with stellar mass and, at fixed mass, decreases with SFR \citep{Ellison2008, Mannucci2010, Andrews2013, Curti2020}. The prevailing physical interpretation is that elevated star formation is fueled by the arrival of relatively pristine gas, delivered by accretion from the intergalactic medium or by minor mergers with more metal-poor companions, which simultaneously boosts the SFR and dilutes the interstellar medium \citep{Dave2012, Lilly2013}. The FMR thus encodes, in a galaxy's chemical state, information about its recent fueling history.

This reasoning leads to a testable joint prediction. If bursts are externally triggered, the same event -- an interaction or an episode of pristine inflow -- should simultaneously disturb the galaxy's morphology and dilute its gas-phase metallicity. Galaxies that are more metal-poor than the FMR predicts for their mass and sSFR should then also be more morphologically disturbed. Purely internal triggering makes no such prediction: feedback-driven cycling recycles the galaxy's own enriched gas and should produce no correlation between metallicity residuals and morphological residuals at fixed mass and sSFR, particularly not for the evolved stellar population. The presence or absence of this correlation therefore provides a population-scale discriminant between triggering channels -- one that does not depend on detecting individual companions or inflows.

Here, we combine medium-band imaging from the Merian Survey \citep{Luo2024, Danieli2025} with spectroscopy from the first data release of the Dark Energy Spectroscopic Instrument \citep[DESI DR1,][]{DESIdr1} to search for this joint signature. Merian is an optical imaging survey conducted with two custom medium-band filters on the Dark Energy Camera, designed to capture the [O III] and \ha emission of star-forming galaxies at $0.06\lesssim z \lesssim 0.10$ over a planned $\sim750$ deg$^2$ of the Hyper Suprime-Cam Subaru Strategic Program footprint \citep[HSC SSP,][]{Aihara2018}. The medium-band photometry enables not only integrated \ha measurements for large samples of low-mass galaxies, but also spatially resolved maps of their \ha emission -- providing morphological information on the distribution of recent star formation that is inaccessible from broadband imaging alone. 

From these maps and the accompanying broadband imaging we measure non-parametric morphological statistics of both the recent star formation and the evolved stellar population. From the DESI spectra we derive gas-phase oxygen abundances using strong-line calibrations appropriate for the low-metallicity regime. The resulting sample of $\sim$2600 galaxies at $0.064 < z < 0.094$ with stellar masses down to $10^8\ M_\odot$ allows us to model morphology as a joint function of stellar mass, sSFR, and metallicity.

The paper is organized as follows. In Section~\ref{sec:data} we describe the Merian and DESI data and the construction of our sample. Section~\ref{sec:morphology} details the construction of \ha emission maps and the measurement of morphological parameters. Section~\ref{sec:metallicities} describes our metallicity estimates. In Section~\ref{sec:results} we present the dependence of continuum and \ha morphology on stellar mass, sSFR, and metallicity, and in Section~\ref{sec:discussion} we discuss the implications for burst-triggering mechanisms and summarize our conclusions. Throughout this paper, we assume a standard $\Lambda$CDM WMAP9 cosmology \citep{Hinshaw2013} with $H_0 = 69.32$ km s$^{-1}$ Mpc$^{-1}$, $\Omega_m=0.2865$, and $\Omega_\Lambda=0.7135$. All magnitudes are reported in the AB system \citep{Oke1983}.

\section{Data}\label{sec:data}
\subsection{Merian Photometric Catalog}

The Merian Survey is an optical medium-band imaging survey, which used two custom-built filters installed on the Dark Energy Camera (DECam) on the Blanco Telescope at the Cerro Tololo Inter-American Observatory. The two medium-band filters, N540 ($\lambda_c = 5400$ \AA, $\Delta\lambda=210$ \AA) and N708 ($\lambda_c = 7080$ \AA, $\Delta\lambda=275$ \AA), are designed to detect \oiii and \ha emission in galaxies in the redshift range $0.06\lesssim z\lesssim0.1$. The full survey footprint covers $\sim$ 750 deg$^2$ overlapping with HSC SSP wide, deep, and ultra-deep fields. For this work we use the first data release of the Merian Survey, which includes 234 deg$^2$ of the total footprint. Details on the filter design can be found in \citet{Luo2024} and the observing strategy and reduction pipeline are presented in \citet{Danieli2025}. In short, the Merian DECam data are processed jointly with the HSC \textit{grizy} broadband data using the Rubin Observatory LSST Science Pipelines. We include only sources from DR1 which have detections in all seven bands, resulting in a catalog with over 6.4 million objects. 

\subsection{DESI DR1}
To construct a sample of Merian DR1 galaxies with available spectroscopy, we crossmatch the Merian DR1 catalog with DESI DR1. DESI is a multi-object spectrograph on the 4-m Mayall Telescope at Kitt Peak National Observatory, with 5000 robotically positioned fibers of 1.5\arcsec\ diameter observing $\sim8 $ deg$^2$ field of view \citep{DESI2016, DESI2022}. The DESI spectra cover 3600--9800 \AA\ at a resolution of R $\sim$ 2000--5000. DESI DR1 includes all observations from the first 13 months of the main survey (May 2021 -- June 2022), together with a uniform reprocessing of the earlier Survey Validation data \citep{DESI2024}, and contains spectra and redshift measurements for over 18 million unique targets \citep{DESIdr1}.

At the redshifts of the Merian sample, the DESI targets are drawn primarily from the Bright Galaxy Survey \citep[BGS;][]{Hahn2023}: $\sim$66\% of our matched sources come from the magnitude-limited BGS Bright sample ($r < 19.5$) and $\sim$9\% from the fainter ($19.5 < r < 20.175$), color-selected BGS Faint sample, which is designed for high redshift efficiency and preferentially selects galaxies with strong emission lines. The remaining quarter of the sample entered through DESI's secondary target programs (13\%), Survey Validation observations (8\%), and the emission-line galaxy selection (2\%). The bright-time flux limit means that our spectroscopic sample is not mass-complete: galaxies near our lower mass limit of $10^8\,M_\odot$ enter the sample only if they are sufficiently luminous -- i.e., blue and actively star-forming. We return to the consequences of this selection in Section~\ref{subsec:caveats}.

We use a crossmatching radius of 1\arcsec\ and only include DESI sources that satisfy \texttt{ZCAT\_PRIMARY == True}, \texttt{SPECTYPE == GALAXY}, \texttt{ZWARN == 0}, and \texttt{DELTACHI2 >= 40}. These selections remove sources with poor redshift measurements or unreliable photometry. We find DESI matches for $\sim6.1$\% of the $\sim6.4$ million sources in the Merian DR1 catalog. Of these, 10,738 sources have spectroscopic redshifts in the Merian redshift range of [0.064, 0.094].

\subsubsection{Emission Line Measurements}\label{subsubsec:emlinemes}

We take emission line measurements from the publicly available DESI DR1 FastSpecFit Spectral Synthesis and Emission-Line value added catalog\footnote{\url{https://data.desi.lbl.gov/doc/releases/dr1/vac/FastSpecFit/}}. We use the latest version, which is version 3.0 at the time of writing. The catalog is based on fits using the \texttt{FastSpecFit} code \citep{fastspecfit}, which was developed specifically for DESI to model the stellar continuum and emission lines by jointly fitting all available DESI spectrophotometry with physically motivated templates. The reported emission line measurements are corrected for galactic extinction and stellar absorption. We estimate attenuation due to interstellar dust using the Balmer decrement and assuming an intrinsic ratio of H$\alpha$/H$\beta=2.86$ under Case B recombination for $T_e=10^4$ K \citep{Hummer1987, osterbrock2006}. We correct all emission line fluxes using the \citet{Cardelli1989} attenuation law. To ensure reliable dust corrections, we require S/N $>$ 3 for both H$\alpha$ and H$\beta$ emission line fluxes. This cut removes 2953 of the Merian-DESI in-band sources, leaving 7785 remaining objects.

\subsection{Stellar Mass}

We estimate the stellar mass of the galaxies in our sample using HSC rest-frame photometry following the calibration from the SAGA survey \citep{Mao2024}. 
We remove all sources with $\log M_\star/M_\odot < 8$ for which the Merian catalog becomes incomplete (Luo et al., in prep.), leaving 6956 remaining galaxies in the Merian-DESI catalog. 

\begin{figure}[t]
    \centering
     \includegraphics[width=0.8\linewidth]{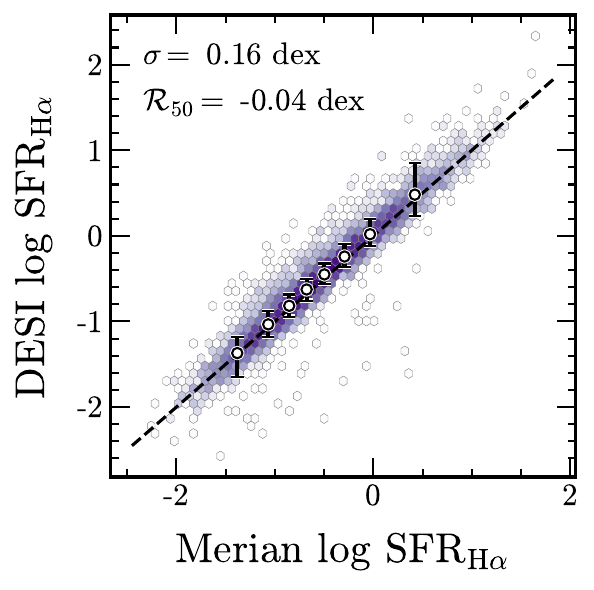}
    \caption{\hanospace-derived star formation rates measured from Merian medium-band photometry compared to those measured from DESI spectra. The black line shows the one-to-one line. The Merian photometric values are offset by a median of -0.04 dex from the DESI spectroscopic values with a 1~$\sigma$ scatter of 0.16 dex.}
    \label{fig:ha_cont_cat}
\end{figure}

\subsection{AGN}
We remove an additional 6\% of the sources in the sample, which are classified as AGN according to their position in the BPT diagram following the \citet{Kauffmann2003} criteria. About 1500 of the galaxies have S/N $<3$ in either \oiii 5007 or \nii 6584 and so cannot be reliably classified as AGN or star-forming galaxies using this approach. We do not exclude those sources from the catalog. The AGN cut leaves 6547 sources. 

\subsection{H$\alpha$ Emission From Photometry}\label{subsec:ha_measure}

We measure H$\alpha$ equivalent width, flux, and luminosity for the galaxies in the sample using the Merian photometry following a similar approach to \citet{Mintz2024}. We estimate the contribution of the stellar continuum through the N708 medium band by fitting a power law through the three neighboring broadbands (\textit{r}, \textit{i}, and \textit{z}) and integrating the fitted continuum through the medium-band transmission curve. The (uncorrected) H$\alpha$+\nii flux is then found as

    \begin{equation}
    {\rm F}_{{\rm H} \alpha+[\text{N II}], 0} =  \left(F_{\nu}^{(\text{N708})} - F_{\nu, \text{cont}}^{(\text{N708})}\right) \ \frac{\int \lambda R_{\text{N708}}(\lambda) \ \text{d} \lambda }{\lambda_{{\rm H} \alpha, {\rm obs}} \cdot R_{\text{N708}}(\lambda_{{\rm H} \alpha, {\rm obs}})},
    \end{equation}
where $F_{\nu}^{(\text{N708})}$ is the observed flux in the N708 medium-band, $F_{\nu,\text{cont}}^{(\text{N708})}$ is the continuum estimate as described above, $R_\text{N708}$ is the transmission curve of the N708 filter, $\lambda_{\text{H}\alpha, {\rm obs}}$ is the observed-frame wavelength of \hanospace. At a redshift of $z$, we then measure the rest-frame equivalent width as

\begin{equation}
    {\rm EW}_{{\rm H}\alpha+[\text{N II}]} = \frac{{\rm F}_{{\rm H} \alpha+[\text{N II}], 0}}{F_{\nu,\text{cont}}^{(\text{N708})} \cdot (1+z)}.
\end{equation}

We apply a number of corrections to obtain our final estimates of H$\alpha$ emission from the Merian photometry. We

\begin{itemize}
    \item correct for galactic extinction using Galactic $A_V$ measured by \citet{Schlafly2011} and assuming a \citep{Cardelli1989} extinction curve with $R_V=3.1$;
    \item correct for ISM attenuation using the Balmer decrement as described in Section~\ref{subsubsec:emlinemes};
    \item correct for line contamination in the medium band from \nii and \sii using an empirically derived mass- and redshift-dependent relation presented in \autoref{app:lines};
    \item correct for stellar absorption assuming a constant absorption EW of 2\,\AA\ \citep{GildePaz2003, Gavazzi2012, Hopkins2013};
    \item and apply an aperture correction.
\end{itemize}

Using the corrected H$\alpha$ flux F$_{{\rm H}\alpha}$, we compute the H$\alpha$ luminosity as 
\begin{equation}
    L_{{\rm H}\alpha} = F_{{\rm H}\alpha} \cdot 4\pi D_L^2
\end{equation}
where $D_L$ is the luminosity distance computed from the spectroscopic redshift. The star formation rates are calculated from $L_{{\rm H}\alpha}$ using the calibration from \citet{Kennicutt2012}. 

\begin{figure}
    \centering
    \includegraphics[width=1\linewidth]{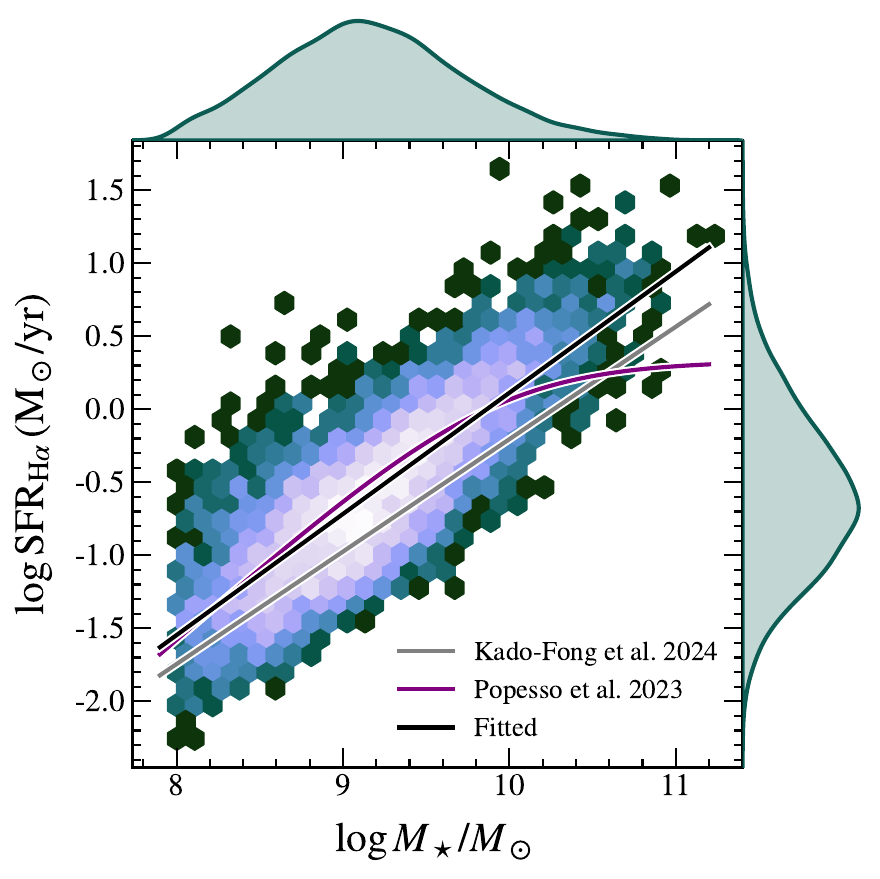}
    \caption{\hanospace-based SFRs derived from Merian medium-band photometry as a function of stellar mass for the galaxies in the Merian-DESI sample, with points colored by density. Marginal distributions for $\log$ SFR$_{\text{H}\alpha}$ and $\log M_\star/M_\odot$ are shown on the top and right of the figure.  Three SFMS fits are shown for context: in purple, the \citet{Popesso2023} SFMS at $z=0.08$; in grey, \citet{KadoFong2024b} SFMS at $0.07 <z <0.1$ that was calculated with a thorough incompleteness correction at low stellar mass; and in black, a simple linear fit to the Merian-DESI sample. }
    \label{fig:sfms}
\end{figure}

We compare our photometric measurements of \ha emission and SFR to the (aperture-corrected) spectroscopic values from \texttt{FastSpecFit} and find them to be in strong agreement, with minimal systematic deviations and a scatter of 0.14 dex in $\log$ EW, 0.18 dex in $\log F_{{\rm H}\alpha}$ and 0.16 dex in $\log$ SFR (see \autoref{fig:ha_cont_cat}). To ensure reliable H$\alpha$ measurements, we remove all sources with EW$_{{\rm H}\alpha}<10$\,\AA, requiring that both the Merian-photometry derived EW and the \texttt{FastSpecFit} EW meet this criterion. This cut removes just over 450 sources, primarily galaxies at high mass with low sSFR -- the median mass of galaxies removed in this step is $\log(M_\star/M_\odot) = 10.06$. This leaves a sample of 6092 galaxies. \autoref{fig:sfms} shows the distribution of these galaxies in the SFR -- $M_\star$ plane with literature fits to the SFMS plotted for comparison.

\section{H$\alpha$ maps and morphology} \label{sec:morphology}

\begin{figure*}[ht]
    \centering
    \includegraphics[width=1\linewidth]{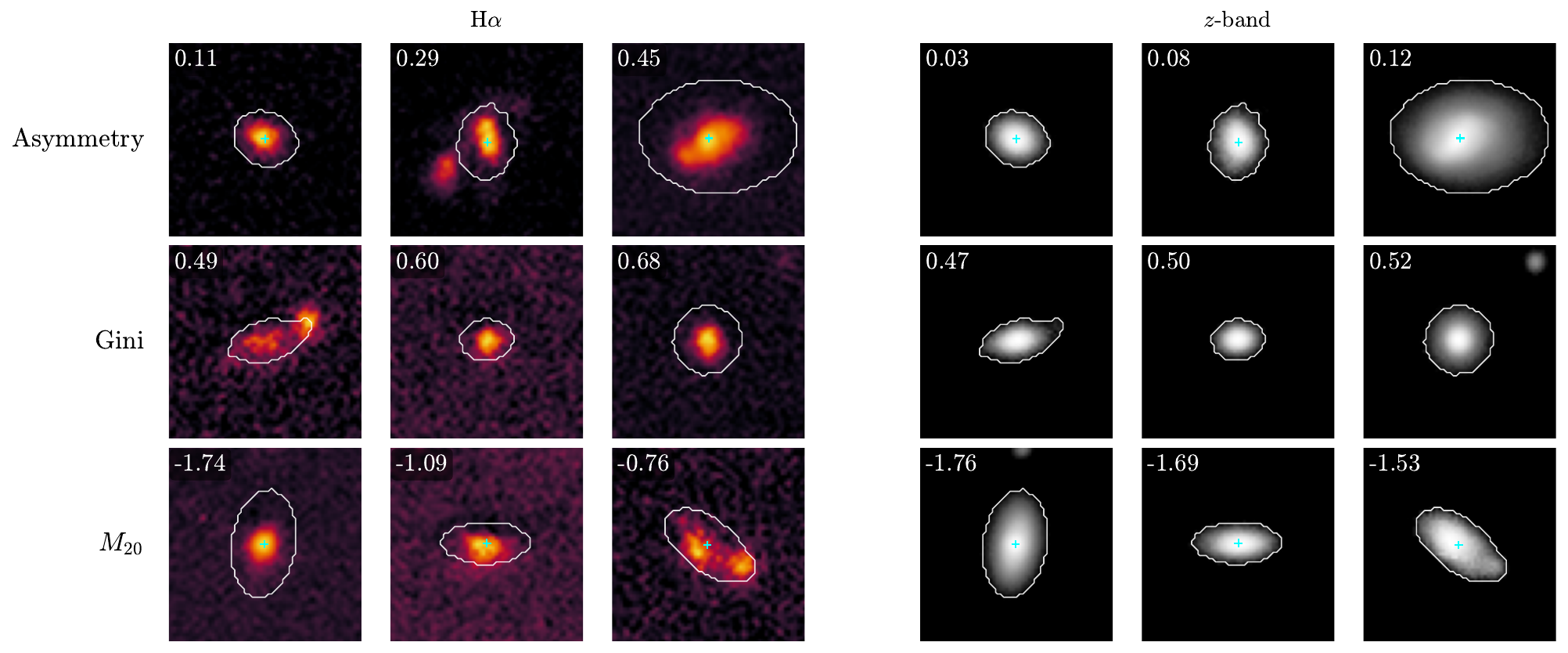}
    \caption{Example \ha emission maps (left) and HSC z-band images (right) for galaxies in our sample, illustrating the range of each non-parametric morphological statistic. Each row shows three galaxies with increasing values of asymmetry (top), Gini coefficient (middle), and \mtwenty\  statistic (bottom), with the measured value indicated in the upper left corner of each panel. The same galaxies are shown in \ha emission on the left and continuum emission on the right. White contours show the segmentation maps used in the morphological measurements and cyan points show the asymmetry center (top row) or \mtwenty\ center (bottom row). Higher asymmetry corresponds to more disturbed emission, higher Gini to less homogeneous flux distributions, and higher \mtwenty\ to bright regions that are more offset from the galaxy center.}
    \label{fig:imgrid}
\end{figure*}

Beyond the integrated estimates of \ha emission, the N708 medium band imaging provides an opportunity to construct two-dimensional estimates of \ha emission in the Merian galaxies. We generate the maps following a similar approach to \citet{Mintz2024}, but with small updates to the continuum method. The new continuum estimation method is more efficient and accounts for color variations over the face of the galaxies. We first convolve the \textit{r, i, z}, and N708 images to match the broadest PSF and reproject all of the images onto a common grid. To construct the continuum map, at each pixel a powerlaw is fit to the \textit{r, i,} and \textit{z} broadband photometry and then integrated through the N708 transmission curve, similar to the approach for the integrated photometry described in Section~\ref{subsec:ha_measure}. The continuum map is subtracted from the N708 images to obtain a map of the isolated H$\alpha$ emission. Examples of \ha and continuum images are shown in \autoref{fig:imgrid}, sorted by morphological parameters.

We measure non-parametric morphological parameters as in \citet{Mintz2024} using the \texttt{statmorph} code \citep{RodriguezGomez2019}. We make small adjustments to the code, allowing us to measure the morphological parameters of the broad-band \textit{z} image (convolved as described above to a common PSF) and the \ha map simultaneously, fixing the mask, segmentation map, and center of the \ha map to that of the broadband image. We use the \textit{z}-band to represent the stellar continuum emission of the galaxy as it contains no strong emission lines in the Merian redshift range and is therefore largely uncontaminated by nebular emission. 

We focus on three non-parametric statistics for the remainder of the analysis: the asymmetry (\asym), the Gini coefficient (\gini), and the M$_{20}$ statistic (\mtwenty). The asymmetry is calculated by comparing the original image to the same image rotated by 180$^\circ$. The Gini coefficient quantifies the homogeneity of a galaxy's flux and ranges from 0 to 1, with \gini=0 indicating a perfectly uniform flux distribution and \gini=1 indicating that all of the flux is concentrated in a single pixel. The \mtwenty\ statistic compares the second moment of the brightest 20\% of pixels to the second moment of the galaxy overall. A low \mtwenty\ indicates that the brightest regions are concentrated at the galaxy's center, while a high \mtwenty\ means that the brightest pixels are farther from the center of the galaxy. For additional details on the morphological statistics, we refer the reader to \citet{RodriguezGomez2019}.

As suggested in \citet{RodriguezGomez2019}, we only include \texttt{statmorph} results for sources with \texttt{flag} $\leq 1$, indicating reliable measurements, and sources whose smallest measured scale ($r_{20}$) is greater than half of the FWHM. We also require that the average signal-to-noise ratio per pixel exceed 2 for both the continuum and H$\alpha$ maps. Slightly over 2600 sources are removed, the majority for poor \ha morphology measurements, leaving 3473 sources with reliable \ha SFRs and morphologies. The morphology quality cuts preferentially remove galaxies with low sSFR. This is expected, as galaxies with weak \ha emission have low signal-to-noise \ha maps that fail the morphology quality criteria. The final sample is therefore slightly weighted toward actively star-forming systems, consistent with the selection effects discussed in Section~\ref{subsec:caveats}. But because our analysis measures correlations at fixed stellar mass and sSFR, this weighting does not bias the fitted trends unless the selection couples jointly to metallicity and morphology at fixed mass and sSFR, which we have no reason to expect from these cuts.

Additionally, we visually inspect all of the galaxies in the sample to remove sources with problematic images. We remove 15\% of sources, leaving 2943 sources that meet all selection criteria described above and have reliable morphology measurements. The majority of visual-inspection removals are due to image artifacts, which are prevalent in the Merian medium-band images. A small fraction are removed due to shredding, i.e. deblending issues in the reduction pipeline that mistakenly classify clumps of larger galaxies as small galaxies. The number of shredded galaxies is smaller for our sample than would be expected for a sample of Merian-only galaxies; the DESI crossmatching significantly increases the likelihood that the sources are true galaxies.

\begin{table*}[ht]
    \caption{Strong-line ratios used in each metallicity calibration.}
    \label{tab:lineratios}
    \centering
\begin{tabular}{llccccc}
    \hline
    \hline
    Ratio & Definition & C2024 & N2022 & S2024 & $N$ & \% \\
    (1) & (2) & (3) & (4) & (5) & (6) & (7) \\
    \hline
    R2         & [\ion{O}{2}]$\lambda\lambda3726,29\,/\,\mathrm{H}\beta$                                    & \checkmark & \checkmark &            &  5150 & 84.5 \\
    R3         & [\ion{O}{3}]$\lambda5007\,/\,\mathrm{H}\beta$                                              & \checkmark & \checkmark &            &  5580 & 91.6 \\
    O32        & [\ion{O}{3}]$\lambda5007\,/\,$[\ion{O}{2}]$\lambda\lambda3726,29$                          & \checkmark & \checkmark &            &  4963 & 81.5 \\
    R23        & ([\ion{O}{2}]$\lambda\lambda3726,29\,+\,$[\ion{O}{3}]$\lambda\lambda4959,5007$)$\,/\,\mathrm{H}\beta$ & \checkmark & \checkmark & \checkmark &  4962 & 81.5 \\
    N2         & [\ion{N}{2}]$\lambda6584\,/\,\mathrm{H}\alpha$                                             & \checkmark & \checkmark & \checkmark &  5579 & 91.6 \\
    S2         & [\ion{S}{2}]$\lambda\lambda6716,31\,/\,\mathrm{H}\alpha$                                   & \checkmark & \checkmark &            &  5544 & 91.0 \\
    O3N2       & ([\ion{O}{3}]$\lambda5007\,/\,\mathrm{H}\beta$)$\,/\,$([\ion{N}{2}]$\lambda6584\,/\,\mathrm{H}\alpha$) &            & \checkmark &            &  5106 & 83.8 \\
    $\hat{R}$  & $0.47\log R_2 + 0.88\log R_3$ \citep{Laseter2024}                                          & \checkmark &            &            &  4963 & 81.5 \\
    Ne3O2      & [\ion{Ne}{3}]$\lambda3869\,/\,$[\ion{O}{2}]$\lambda\lambda3726,29$                         & \checkmark &            &            &   910 & 14.9 \\
    \hline
\end{tabular}
    \tablecomments{The set of strong-line ratios we use to measure gas-phase oxygen abundances. Column (1): The name of the strong-line ratio. Column (2): The definition of the strong-line ratio. Column (3): Whether the strong-line ratio is included as part of the \citet{Curti2024} calibrations. Column (4): Whether the strong-line ratio is included as part of the \citet{Nakajima2022} calibrations. Column (5): Whether the strong-line ratio is included as part of the \citet{Scholte2024} calibrations. Column (6): The total number of galaxies in the sample with S/N $>3$ for each of the lines included in the strong-line ratio. Column (7): The percentage of the 6092 galaxies in the sample described in Section~\ref{sec:data} for which the strong-line ratio can be used reliably.}
\end{table*}

\begin{figure*}
    \centering
     \includegraphics[width=\linewidth]{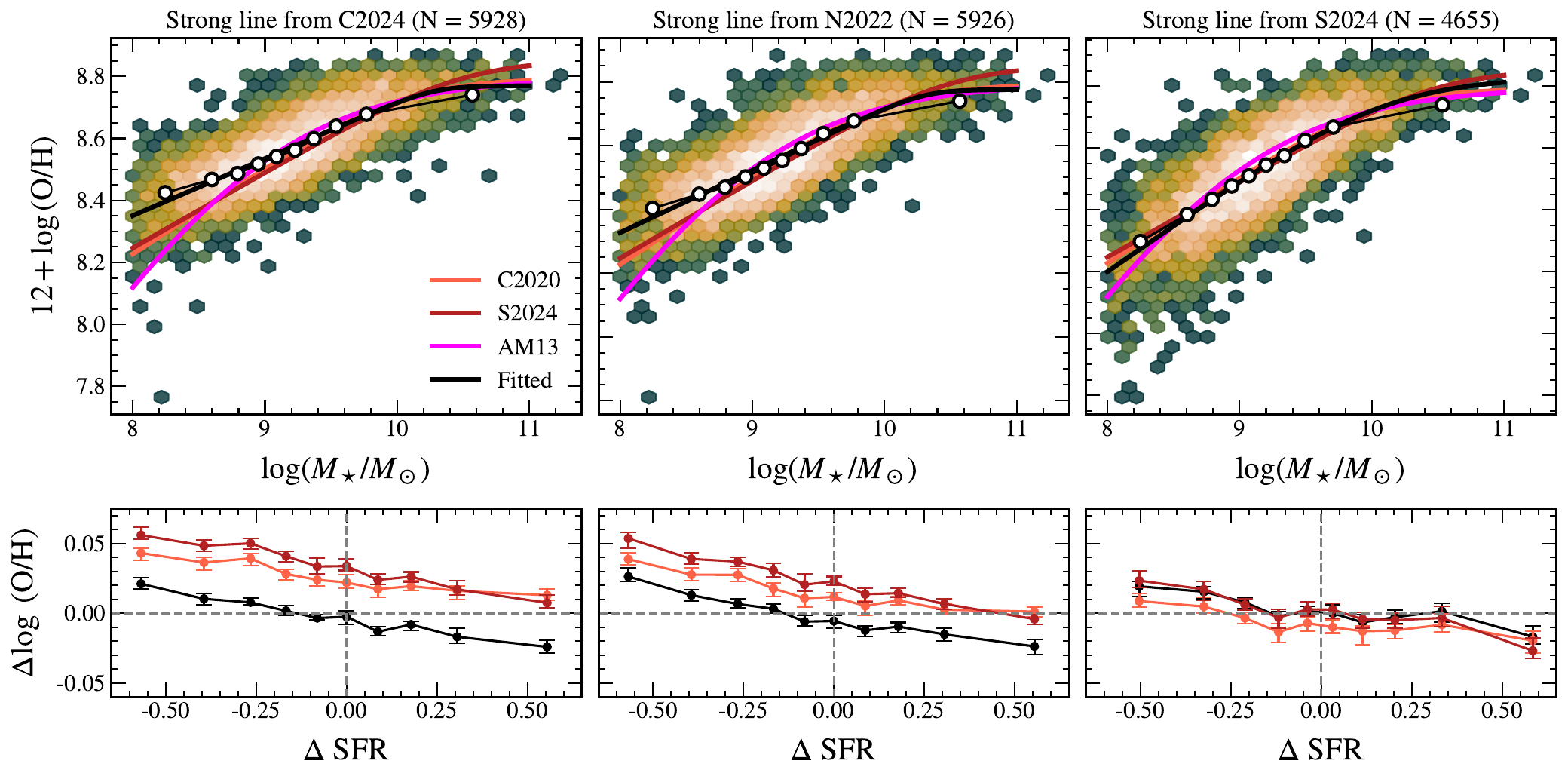}
    \caption{The mass-metallicity relation (MZR) for the galaxies in our Merian-DESI sample using metallicities derived from three different sets of strong-line calibrations, one shown in each column. Literature fits to the MZR are shown from \citet{Curti2020}, \citet{Scholte2024}, and \citet{Andrews2013}. Fits to our data using the functional form from \citet{Curti2017} are shown in black, with points indicating binned median values. The bottom row shows the metallicity offset ($\Delta \log$ (O/H)) versus SFR offset from the SFMS ($\Delta$ SFR). $\Delta \log$ (O/H) is measured from the MZR fits from \citet{Curti2020}, \citet{Scholte2024}, and the fit directly to the data and $\Delta$ SFR is taken with respect to the \citet{Popesso2023} SFMS at $z=0.08$ as shown in \autoref{fig:sfms}.}
    \label{fig:MZR}
\end{figure*}

\section{Metallicities}\label{sec:metallicities}

We estimate gas-phase oxygen abundances (O/H) for the galaxies in our sample (here, the sample of 6029 galaxies described in Section~\ref{sec:data} before the morphology cuts) using three sets of strong-line diagnostics that have all been calibrated to the low-metallicity regime. Low-mass galaxies (particularly those that are star forming) are expected to be more metal poor than massive galaxies, and so strong-line calibrations derived only from metal-rich, massive galaxies are not appropriate for our sample. Throughout, we use ``gas-phase oxygen abundances" and ``gas-phase metallicities" interchangeably as is common in the literature. We use three sets of strong-line calibrations:

\begin{enumerate}
    \item N2022: \citet{Nakajima2022} derived strong line calibrations using local SDSS galaxies and extremely metal-poor galaxies from the Subaru EMPRESS survey. 
    \item C2024: \citet{Curti2024} derived updated versions of their strong line calibrations from \citet{Curti2017} and \citet{Curti2020} using a sample of local metal-poor galaxies to extend their previous fits to lower metallicities. 
    \item S2024: \citet{Scholte2024} used a combination of the R23 calibration from \citet{Nakajima2022} and the N2 calibration from \citet{Denicolo2002} to derive gas-phase metallicities from DESI early data release (EDR) spectra. They compared their strong-line derived metallicities to metallicities measured using auroral lines and found them to be in agreement with a scatter of $\sim$ 0.3 dex. 
\end{enumerate}

A summary of the line ratios used in each method are included in \autoref{tab:lineratios}.

The gas-phase metallicity of each galaxy is found by minimizing the distance between the observed strong line ratios and the fitted calibrations for all line ratios whose component lines have S/N $>3$. In other words, we minimize

\begin{equation}
    \chi^2 = \sum\frac{(R_{\text{obs, }i} - R_{\text{cal, }i})^2}{\sigma^2_{\text{obs}, i} +\sigma^2_{\text{cal}, i}}
\end{equation}
where $R_{\text{obs, }i}$ is the observed value of line ratio \textit{i}, $R_{\text{cal, }i}$ is the calibration-predicted value of the ratio for a given metallicity, $\sigma^2_{\text{obs}}$ is the uncertainty on the observed ratio, and $\sigma^2_{\text{cal}}$ is the intrinsic dispersion of the calibration. For the N2022 and C2024 calibrations, we require at least two line ratios be detected with significance (each component line must have S/N $>3$) to measure a metallicity estimate. For S2024, we require that both of the included line ratios be significant. 

We successfully measure gas-phase oxygen abundances for 97\% of galaxies using the N2022 and C2024 calibrations and for 76\% of galaxies using the S2024 calibration. The N2022 and C2024 metallicities can  be measured from a variety of line combinations, while the S2024 calibrations require that all of [\ion{O}{2}]$\lambda${3726,3729},  [\ion{O}{3}]$\lambda${4959,5007}, and H$\beta$ be significantly detected. In \texttt{FastSpecFit}, the [OIII]$\lambda\lambda$4959,5007 doublet is fit with a tied 1:3 amplitude ratio, so [\ion{O}{3}]$\lambda$4959 is not an independent detection requirement. Instead, the difference between the C2024/N2022 and S2024 samples is driven primarily by the [\ion{O}{2}]$\lambda$3726,29 doublet: C2024 and N2022 have multiple line-ratio combinations to fall back on when [OII] is undetected, whereas the S2024 R23 requirement removes those sources entirely. Because [\ion{O}{2}] emission scales with star formation activity, the S2024 sample is preferentially biased toward higher-sSFR galaxies. (See \citet{Telford2016} for more discussion of systematic effects on metallicity measurements caused by various S/N requirements.)

\autoref{fig:MZR} shows the mass–metallicity relation for our sample under each of the three calibrations, with fits to our data using the functional form of \citet{Curti2017} shown in black. Literature fits from \citet{Curti2020}, \citet{Scholte2024} and \citet{Andrews2013} are plotted for comparison. The three calibrations produce relations of broadly similar shape above $10^9\,M_\odot$, offset by up to $\sim0.15$ dex at the low stellar mass end, reflecting systematic differences among strong-line calibrations as discussed above.

The discrepancies seen in \autoref{fig:MZR} at low mass between the MZRs fit to the C2024 metallicities in our Merian-DESI sample and the published MZRs are best explained by differences in our samples. The DESI spectra are deeper than the SDSS spectra that were used in \citet{Curti2020}, so we expect to have better coverage of galaxies with lower SFRs at low mass than were included in that sample. According to the general intuition of the FMR, we would therefore expect that our sample includes more low-mass galaxies with higher metallicities, which is precisely the observed disagreement. Similar logic can be applied when comparing with the \citet{Scholte2024} fit. When we use the same strong line calibrators from that study we recover the same MZR, but when we use the C2024 calibrations, we include more sources with weak [\ion{O}{2}] and so we recover additional weakly star-forming and high metallicity galaxies.

The bottom row of \autoref{fig:MZR} shows the metallicity offset from the MZR, $\Delta \log$(O/H), as a function of offset from the star-forming main sequence, $\Delta$SFR. We recover the anticorrelation expected from the FMR: at fixed stellar mass, galaxies with elevated SFR are systematically more metal-poor. The trend is present for all three calibrations. The S2024 relation is the shallowest of the three, again consistent with its more restrictive sample. Throughout the analysis below we compute $\Delta$FMR with respect to a fit to the samples shown in \autoref{fig:MZR} using the functional form presented in Equation 2 of \citet{Curti2020}. Our conclusions are unchanged if we instead use the reported FMR fit in \citet{Curti2020}.

\begin{figure*}
    \centering
    \includegraphics[width=\linewidth]{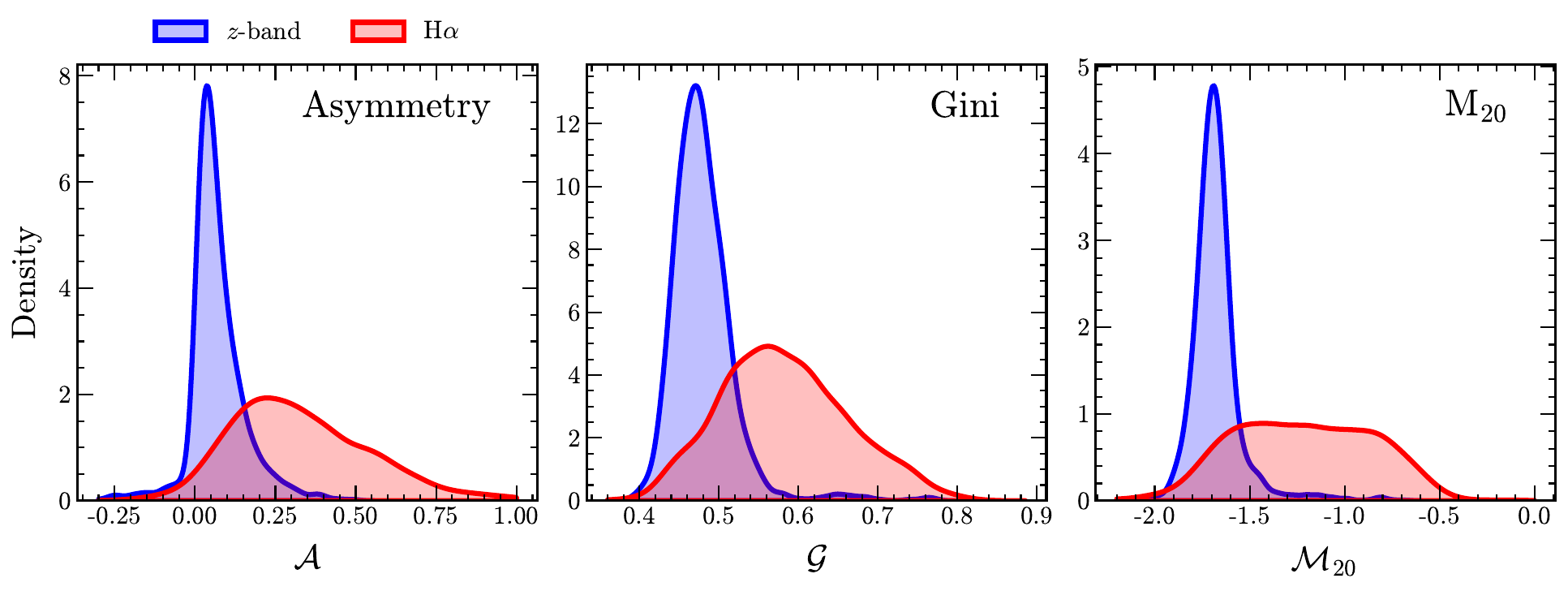}
    \caption{Distribution of non-parametric morphological statistics measured from HSC \textit{z}-band images (blue) and from \ha emission maps (red). The \ha emission is generally more asymmetric, less homogeneous, and less centralized than the stellar continuum emission. }
    \label{fig:morph_kdes}
\end{figure*}

\begin{figure*}
    \centering
    \includegraphics[width=1\linewidth]{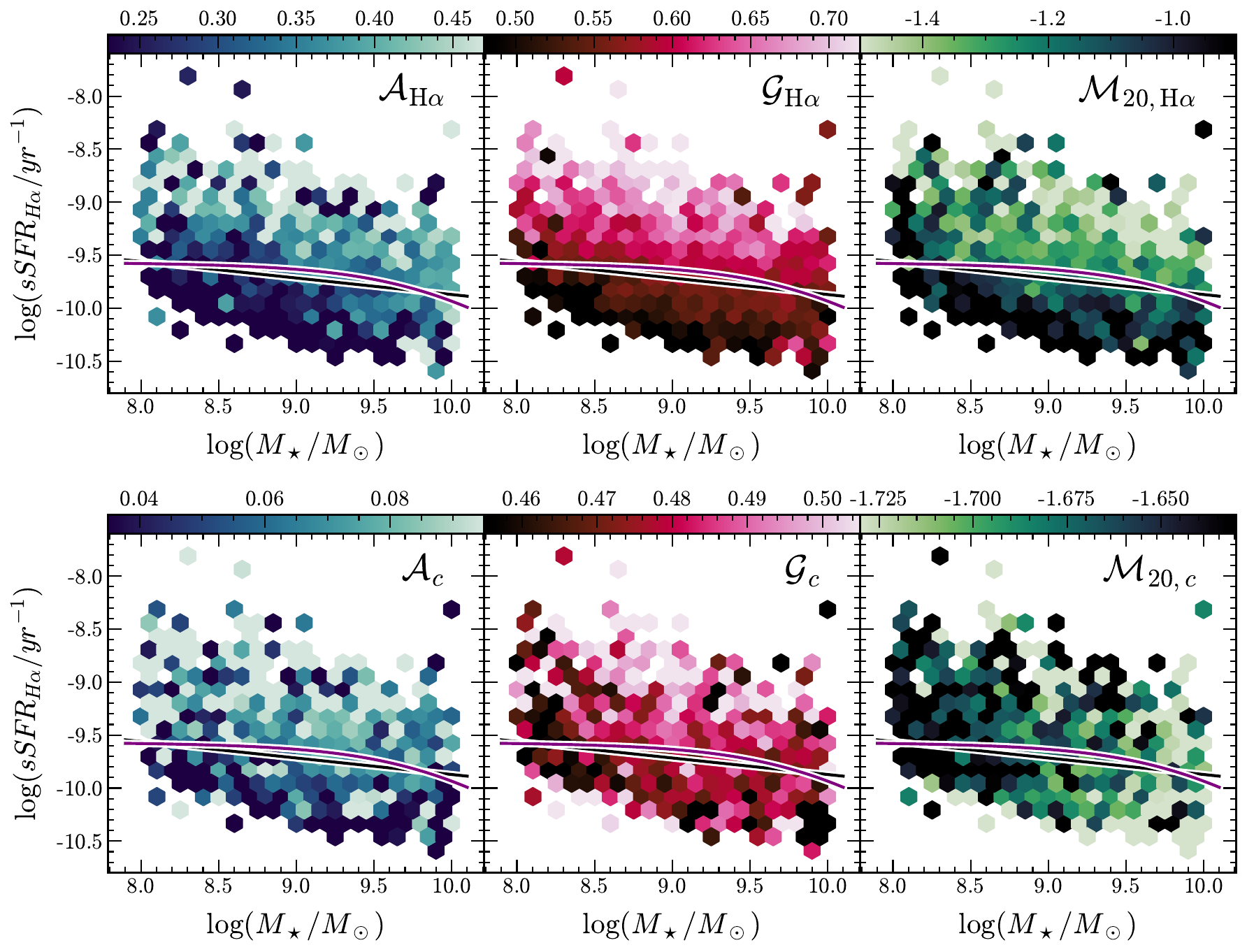}
    \caption{Specific star formation rate versus stellar mass for galaxies in the Merian-DESI sample with successful morphological measurements. sSFR values are derived from Merian photometry as described in Section~\ref{subsec:ha_measure}.  Points in each panel are colored by averages of different morphological parameters. The top row shows \ha morphology: asymmetry (left), Gini coefficient (middle), and M$_{20}$ statistic (right). The bottom row shows the same but for the stellar continuum morphology as measured from the \textit{z}-band. Note that while the color maps are the same for the three morphological parameters, the normalization is different for the \ha and continuum maps. The overplotted lines show the SFMS fits from \autoref{fig:sfms} (in black) and the \citet{Popesso2023} SFMS in purple.}
    \label{fig:sfms_morph}
\end{figure*}

\section{Results}\label{sec:results}

The distribution of a galaxy's light is tied to its stellar mass and star formation rate. Previous work has established this relationship for stellar continuum morphology, and has hinted at trends with morphology of line emission \citep{Conselice2003, Fossati2013, Boselli2015, Nersesian2023, Yao2023, Mintz2024}. In particular, using broadband and \ha maps from the Merian survey, \citet{Mintz2024} found that low-mass galaxies with high sSFR have high \giniha, low \mtwentyha, and high \asymha; they have clumpy \ha emission that is located close to, but slightly offset from, the galaxy's center. They also found that galaxies with higher sSFR have more asymmetric broadband morphologies, although the strength of the trend is weaker than that for the line emission. In general, the \ha maps were found to be more asymmetric, less homogeneous, and less centrally concentrated than the continuum emission and showed much larger spreads in all the morphological parameters considered, a result we reconfirm with our current sample in \autoref{fig:morph_kdes}. 

Here, with a significantly larger sample ($>2\times$), we conduct a more thorough analysis to robustly investigate the dependence of continuum and \ha emission morphology on $\log M_\star$ and sSFR. We then extend the analysis to include gas-phase metallicities, exploring whether this added covariate is independently tied to morphology. In the following analysis, we restrict the sample to $\log(M_\star/M_\odot)<10$, leaving 2648 galaxies. This cut serves two purposes. First, the processes we aim to probe -- bursty star formation and its triggering -- are specific to the low-mass regime. Second, the linear models adopted below are not expected to describe the high-mass population: the MZR flattens for $\log M_\star/M_\odot \gtrsim 10$ (see \autoref{fig:MZR}), so a metallicity term linear in stellar mass is not appropriate above that break point. The high-mass end of our sample is in any case unrepresentative, since the EW$_{\text{H}\alpha}$ cut of Section~\ref{subsec:ha_measure} preferentially removes massive galaxies with low sSFR. We clarify that all SFR and sSFR values used in the subsequent analysis are those measured from Merian photometry as described in Section~\ref{subsec:ha_measure}. The results are unchanged if the spectroscopically-derived DESI SFR values are used instead.

\begin{figure*}
    \centering
    \includegraphics[width=1\linewidth]{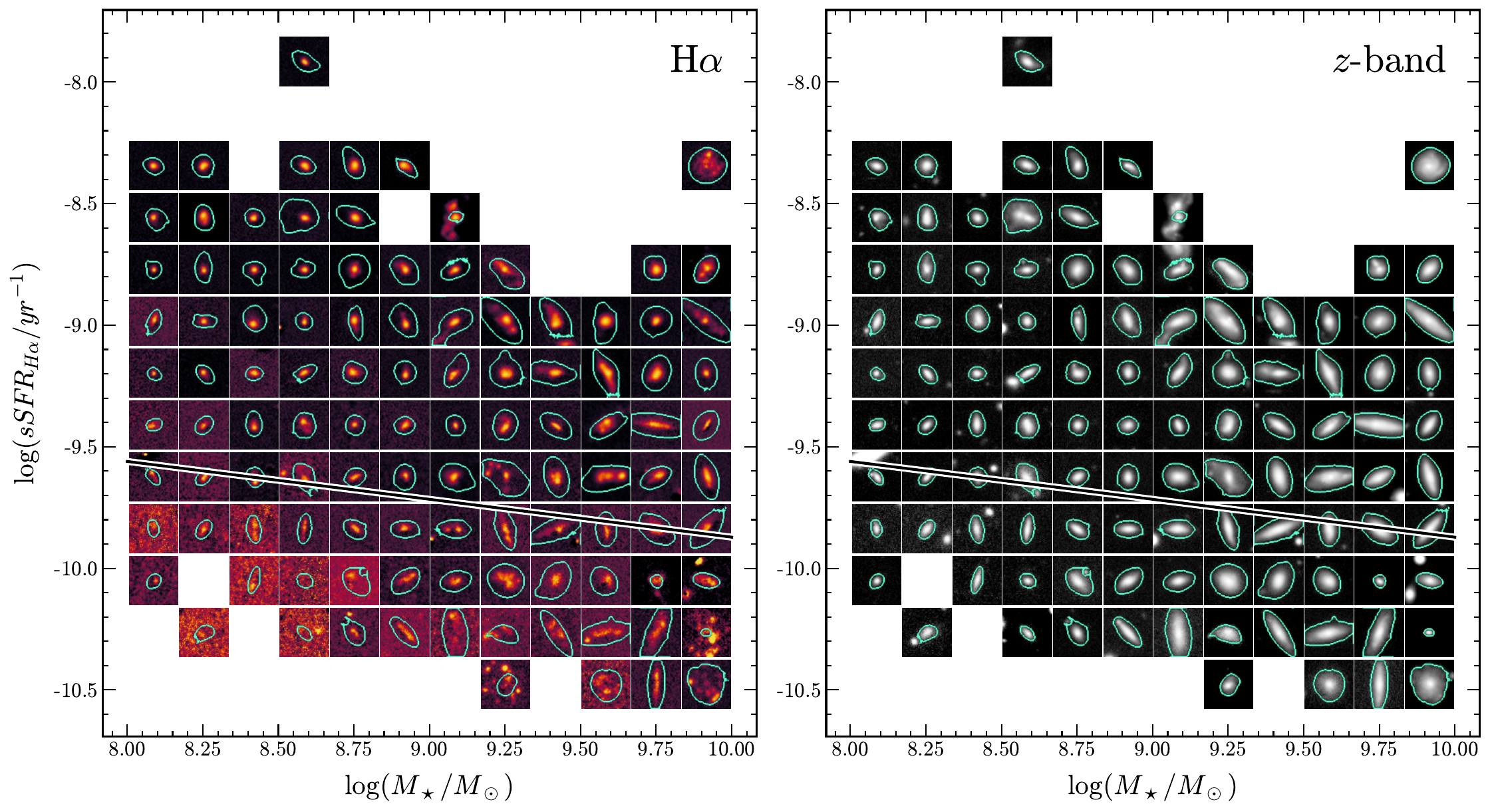}
    \caption{Representative \ha emission maps (left) and z-band images (right) for galaxies across the stellar mass - SFR plane. sSFR values are derived from Merian photometry as described in Section~\ref{subsec:ha_measure}. Each cutout shows a randomly selected galaxy from that bin of stellar mass and sSFR, with contours indicating the segmentation maps. The same galaxies are show in the \ha panel and the continuum panel. The overplotted lines show the SFMS fits from \autoref{fig:sfms} (in black) and the \citet{Popesso2023} SFMS in purple. The visual trends reinforce the statistical results of Section~\ref{subsec:morphtrends}: toward higher sSFR the \ha emission becomes visibly more asymmetric and clumpy, while the continuum morphology changes more subtly.}
    \label{fig:sfms_morph_ims}
\end{figure*}

\subsection{Dependence of \ha and Broadband Morphology on sSFR and $M_\star$}\label{subsec:morphtrends}

In \autoref{fig:sfms_morph}, we plot the galaxies in our sample in the $\log M_\star-\log$ sSFR space colored by binned average \asym, \gini, and \mtwenty\ for both the \ha and continuum maps. For the \ha morphologies, we see clear trends with stellar mass and sSFR, with higher mass galaxies having higher \asymha, higher \giniha, and lower \mtwentyha\ at fixed sSFR. At fixed mass, galaxies with higher sSFR also appear to have higher \asymha, higher \giniha, and lower \mtwentyha. In other words, at higher sSFR galaxies have \ha emission that is more asymmetric, less homogeneous, and more centralized than lower sSFR galaxies at fixed mass and the same is true for high mass galaxies at fixed sSFR. The trends in continuum morphology are more subtle, but by eye there is a clear increase in continuum \asym\ with sSFR for fixed mass and a decrease in $\mathcal{M}_{20,c}$\ with stellar mass for fixed sSFR. 

\begin{figure*}[th]
    \centering    \includegraphics[width=1\linewidth]{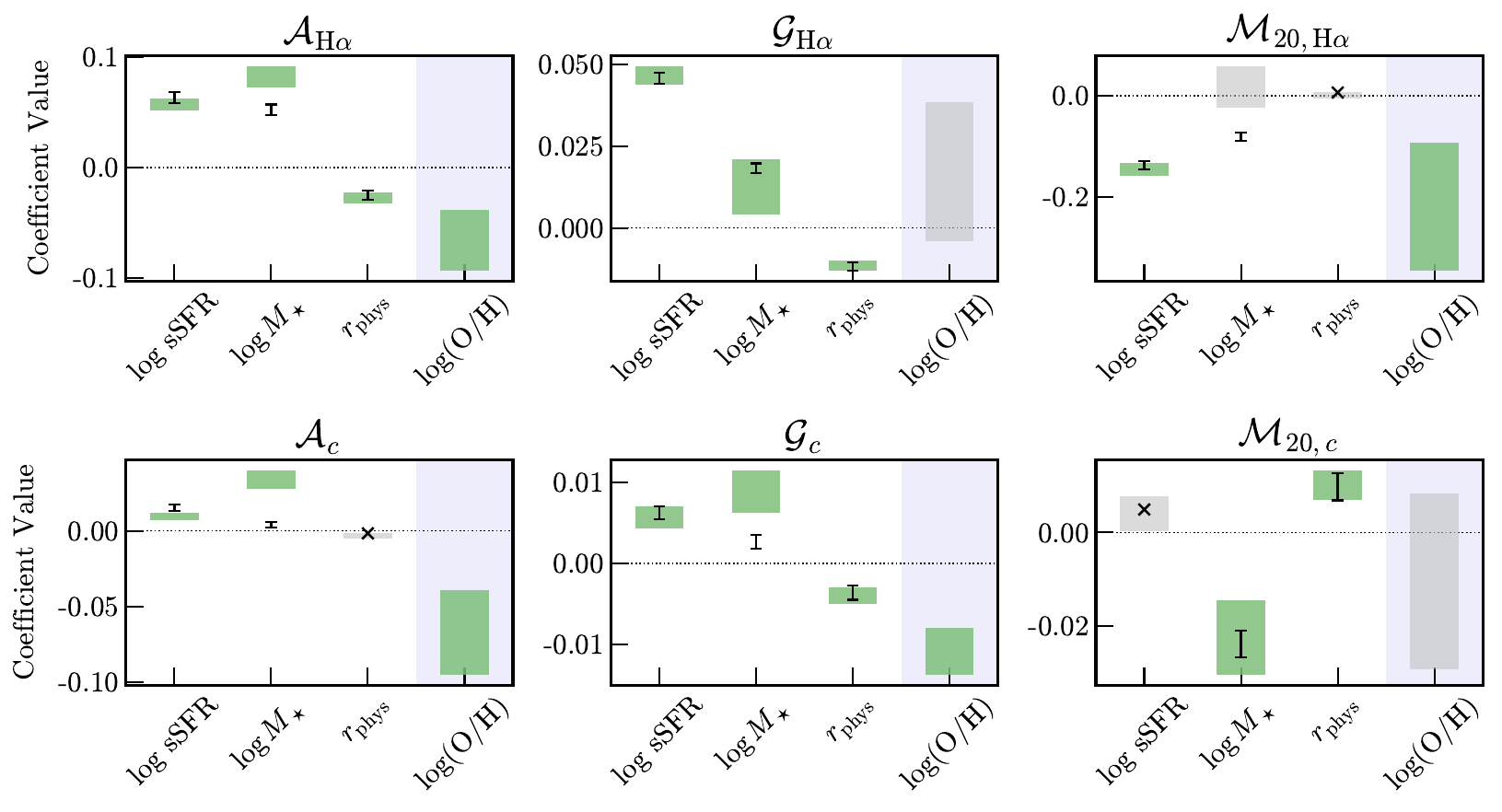}
    \caption{Coefficients from the linear model fits for the H$\alpha$ and continuum morphological parameters. The black points and errorbars show the fitted coefficients and standard errors for the models presented in Section~\ref{subsec:morphtrends} fit only to sSFR (derived from Merian photometry as described in Section~\ref{subsec:ha_measure}), $M_\star$, and $r_\text{phys}$. The `x' markers indicate an insignificant coefficient. The green and gray shaded rectangles show the range of fitted coefficients for the linear models presented in Section~\ref{subsec:morphmet}, which include metallicity as an additional predictor. The range covers the values derived from model fits to various strong-line derived metallicities, with a green shading indicating the coefficient was significant regardless of the choice of calibrations and a gray shading indicating that the coefficient was insignificant for at least one set of metallicities. Nearly all of the morphological parameters depend significantly on the specific star formation rate and stellar mass. The H$\alpha$ asymmetry (\asymha) and M$_{20}$ statistic (\mtwentyha) depend significantly on metallicity, as do the continuum asymmetry (\asym$_c$) and Gini coefficient (\gini$_c$).}
    \label{fig:coeffs}
\end{figure*}

To quantify the strength and significance of these observed trends, we next perform a linear regression analysis, using ordinary least squares to fit linear models predicting the morphological parameters as a function of stellar mass and sSFR. While we do not expect the effects of cosmic evolution to be significant over our narrow redshift range, the physical resolution of the images does vary over the sample and is correlated with redshift. To account for this, we calculate the physical resolution ($r_\text{phys}$) for each galaxy using the FWHM of the PSF and the galaxy's spectroscopic redshift and include this as an additional covariate. 
To summarize, our fitted linear models are of the form
\begin{equation}
    \mathcal{X} = \beta_0 + \beta_1 \times \log \frac{M\star}{M_\odot} + \beta_2\times \log \text{sSFR} + \beta_3 \times r_\text{phys} + \epsilon\label{eq:lm1}
\end{equation}
where all of the predictors are standardized to zero mean and unit standard deviation prior to fitting (in order to evaluate the relative contribution of each predictor) and $\mathcal{X}$ represents one of the morphological parameters. The fitted coefficients are reported in \autoref{tab:no_OH} in \autoref{app:models} and shown as black error bars (or x's for coefficients with $p>0.05$) in \autoref{fig:coeffs}.

The results of the linear regression analysis largely agree with the visually identified trends seen in \autoref{fig:sfms_morph}. \asymha\ and \giniha\ increase with sSFR and $\log M_\star$ and decrease for images with lower resolution (i.e. larger minimum resolved scale). For \asymha\ the size of the mass and sSFR effects are comparable to one another and larger than the resolution dependence. \giniha\ is more steeply dependent on sSFR than stellar mass and least dependent on the resolution. \mtwentyha\ decreases with increasing sSFR and stellar mass and is not dependent on the resolution.

The continuum morphology trends are much weaker than those for the \ha maps and the $R^2$ values for those fits are correspondingly lower; less of the variance in continuum morphology can be explained by the considered predictors. We find \asym$_{c}$ is positively correlated with sSFR and stellar mass.  \gini$_c$ increases significantly, though weakly, with sSFR and stellar mass. $\mathcal{M}_{20,c}$ does not depend significantly on sSFR and decreases strongly with increasing stellar mass. Galaxies with high stellar mass have continuum emission that is slightly less homogeneous and significantly more centrally concentrated while galaxies with high sSFR have continuum emission that is slightly more asymmetric and less homogeneous. 

The fitted trends are also apparent visually in \autoref{fig:sfms_morph_ims}, which shows \ha maps and continuum images for representative galaxies in the sSFR -- $M_\star$ plane.

\begin{figure*}
    \centering
    \includegraphics[width=1\linewidth]{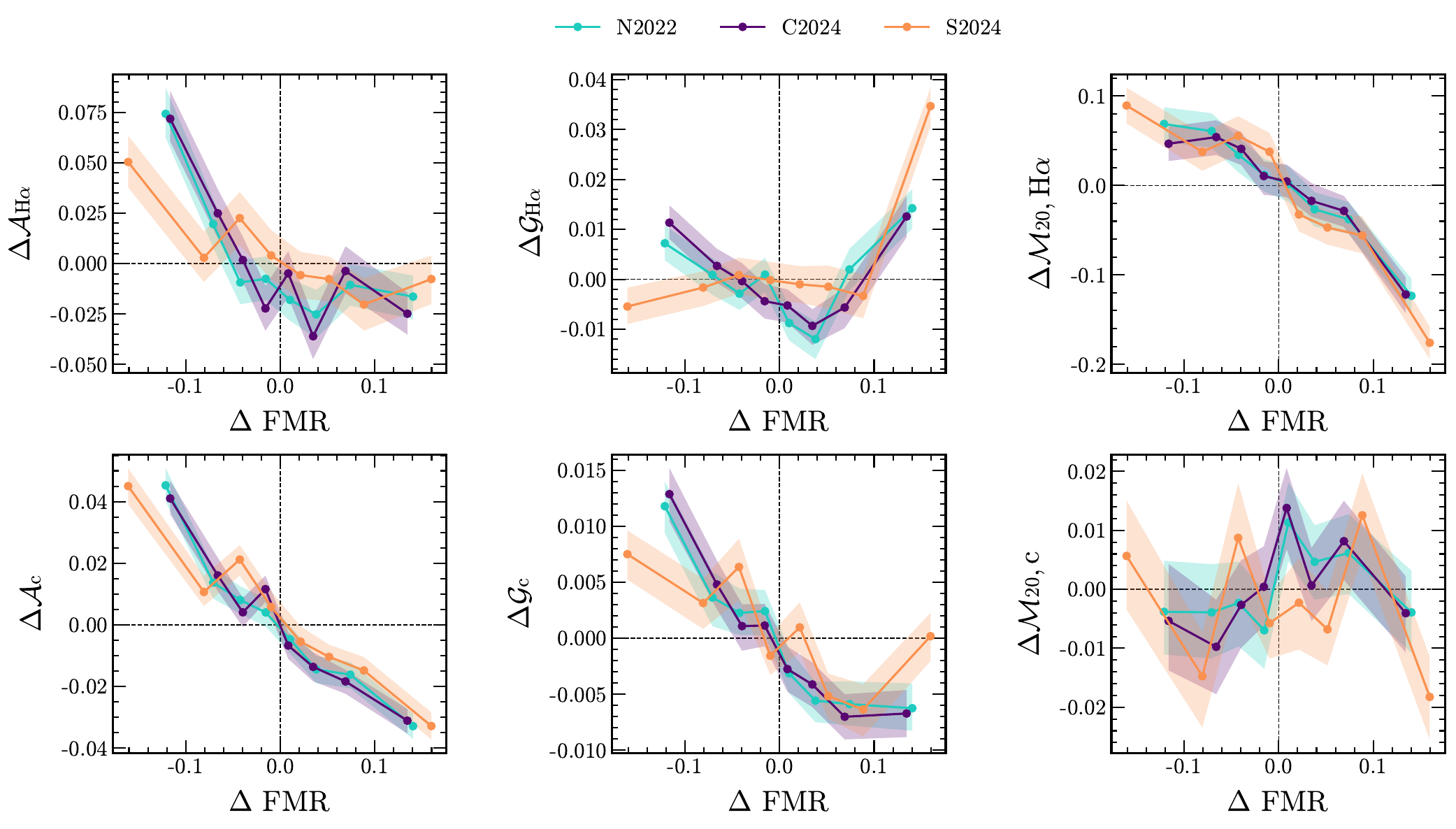}
    \caption{Each panel shows the residuals from the fitted metallicity-free morphological models (e.g. \asym$_{\text{H}\alpha,i} - \hat{\mathcal{A}_{\text{H}\alpha}}(\text{sSFR}_i,\ \log M_{\star, i}/M_\odot,\ r_{\text{phys}, i}$)) as a function of residuals from the fundamental metallicity relation (e.g. $\log(\text{O/H})_i - \text{FMR}(\text{sSFR}_i,\ \log M_{\star, i}/M_\odot)$). The points represent the binned means and the shaded regions show the 1$\sigma$ bootstrapped error on the means for the three different metallicity calibrations. We find significant trends for \asymha, \mtwentyha, \asym$_{c}$, and \gini$_{c}$. Galaxies that are particularly metal-poor for their mass and sSFR have \ha emission that is particularly asymmetric and offset from the center (higher \mtwentyha). They also have especially asymmetric and clumpy continuum emission.}
    \label{fig:offsets}
\end{figure*}

\subsection{A Weak, but Significant Morphological Dependence on Metallicity}\label{subsec:morphmet}

Having established that the morphology of stellar continuum and \ha emission depend on stellar mass and sSFR, we next repeat the above analysis with metallicity -- i.e., gas-phase oxygen abundance, 12 + $\log$(O/H) -- included as an additional predictor:
\begin{align}
    \mathcal{X} &= \beta_0 + \beta_1 \times \log \frac{M\star}{M_\odot} + \beta_2\times \log \text{sSFR} + \beta_3 \times r_\text{phys}\nonumber
    \\ 
    &+\beta_4\times [12 + \log(\text{O/H)}] + \epsilon.\label{eq:lm2}
\end{align}
We fit a linear model three times for each morphological parameter, once for each set of strong line calibrators described in Section~\ref{sec:metallicities} to assess the robustness of the results to small changes in measured metallicities. We report the full fitting results in \autoref{app:models} in Tables~\ref{tab:S24}, \ref{tab:C24}, and \ref{tab:N22}. We show the ranges covered by the fitted coefficients from the three metallicity calibrations in \autoref{fig:coeffs}, with green ranges indicating that the coefficient is significant ($p < 0.05$) across all three datasets and gray ranges indicating it is insignificant in at least one. 

We find that metallicity is a significant predictor of asymmetry, both for continuum and \ha emission. Galaxies that are especially metal poor for their mass and sSFR have the most asymmetric distribution of stars and recent star formation, though this added effect is weaker for the \ha emission; the increase in $R^2$ from adding metallicity as a predictor of \asymha\ is 0.01. For the continuum emission, the metallicity is more powerful with an increase in $R^2$ of $\sim$ 0.06--0.07, more than tripling the $R^2$ value from the no-metallicity model.

Additionally, \mtwentyha\ is negatively correlated with metallicity, which reduces the variance by an additional 2--6\%. At fixed sSFR and stellar mass, galaxies with lower metallicity have their brightest \ha regions less centralized. \gini$_c$ is negatively correlated with metallicity such that more diluted galaxies have clumpier continuum emission. No significant trends are found in \giniha\ or $\mathcal{M}_{20, c}$ with metallicity that are robust across all strong-line calibrators.

Because the effect of metallicity on the continuum and \ha morphology is weak, it is challenging to discern these trends visually. Grids of maps and images do not strikingly reinforce the derived statistical trends. Instead, in \autoref{fig:offsets}, we show the deviation of the morphological parameters from the no-metallicity linear models ($\Delta\mathcal{X}$) as a function of deviation from the fundamental metallicity relation ($\Delta$FMR), as discussed in Section~\ref{sec:metallicities}. 

Again, the results we discuss above from the linear modeling are apparent in the $\Delta\mathcal{X}$ -- $\Delta$FMR trends in \autoref{fig:offsets}. Galaxies that are particularly metal poor for their stellar mass and sSFR have \ha emission that is especially asymmetric and decentralized, but with no obvious difference in homogeneity. They also have continuum emission that is especially asymmetric and especially clumpy, but with no change in the degree of centralization.

Across the three calibrations, the fitted coefficients agree in sign and significance for \asymha, \mtwentyha, \asym$_{c}$, and \gini$_c$, but the S2024 metallicities consistently yield the strongest metallicity dependence: $\beta_\text{O/H}$ for \mtwentyha\ is roughly three times larger than for C2024 or N2022, and the $\Delta R^2$ for \asym$_c$ is the largest of the three. S2024 is also the only calibration for which \giniha\ shows a significant metallicity coefficient. This is consistent with the sample differences described in Section~\ref{sec:metallicities}: the R23 requirement removes galaxies without a significant [\ion{O}{2}] detection, leaving a subsample weighted toward higher sSFR. The S2024 coefficients can therefore be considered an upper bound on the strength of the metallicity dependence, but we note that the direction and significance of the trends are unchanged across calibrations.

\subsection{Caveats of Analysis}\label{subsec:caveats}
While the results presented above are robust to small changes in the sample construction and measurement of relevant statistics, we briefly discuss potential concerns regarding the analysis. Firstly, we recognize that our choice of linear modeling is simplistic and that the morphological trends are unlikely to be purely linear. In general, we find that the models are overall well fit to the data and the presence and direction of trends in morphology with stellar mass, sSFR, and metallicity can be recovered with simple binning of the sample (see \autoref{app:bin}). By constructing a low-metallicity sample and a control sample with average metallicity but similar mass and sSFR, we find that the metal-poor galaxies have the same significant differences in their morphologies as we found with the full linear modeling. While the models presented above should not necessarily be used to predict a galaxy's morphology, the general significance, strength, and direction of the trends are robust.

Secondly, our sample is subject to several layers of selection that limit its completeness. The DESI targeting is not complete at these magnitudes, and our emission-line, EW, and morphology quality cuts each preferentially remove galaxies with weak line emission or low surface brightness. Our sample is therefore not a complete census of low-mass galaxies, but is weighted toward actively star-forming systems with well-detected \hanospace. Importantly, however, our conclusions do not rest on the sample being volume-complete; the trends we measure are correlations at fixed stellar mass and sSFR within the observed population. Incompleteness would bias our results only if the selection acted jointly on metallicity and morphology at fixed mass and sSFR -- for example, if metal-poor, asymmetric galaxies were preferentially detected relative to metal-poor, symmetric ones. We are not aware of a selection effect in either the DESI targeting or our quality cuts that would couple to both quantities in this way.

Additionally, the DESI metallicities are measured through 1.5\arcsec\ fibers ($\sim\!2$ kpc for our redshift range), which cover only the central regions of our galaxies. Gas-phase metallicities are known to vary radially within galaxies \citep{Sanchez2014, Belfiore2017}, so our measurements should be interpreted as central, rather than global, metallicities. We note that if pristine gas inflows preferentially dilute central metallicities, the fiber measurements are most sensitive to exactly this region. Spatially resolved metallicity measurements, which are beyond the reach of single-fiber spectroscopy, could provide additional tests of the physical interpretations we present below.

Finally, dust attenuation affects both axes of our analysis. We correct the emission-line fluxes and integrated \ha measurements using the integrated Balmer decrement, but the maps cannot easily be corrected to account for spatial variations in the dust distribution across each galaxy. Inhomogeneous dust could artificially enhance the measured asymmetry and clumpiness of the \ha maps, and -- because dust content correlates with metallicity -- could in principle imprint a spurious metallicity dependence. We note, however, that this effect works against the observed trend: metal-poor galaxies are expected to be less dusty \citep{RemyRuyer2014}, and should therefore have less dust-induced structure in their maps, whereas we find they are the most asymmetric. Patchy attenuation is thus unlikely to produce the observed anticorrelation between metallicity and asymmetry, though it adds scatter to the morphological measurements.

\section{Discussion and Conclusions}\label{sec:discussion}

Combining optical medium-band imaging from the Merian Survey with optical spectroscopy from the first data release of DESI, we compiled a large sample of galaxies with measured strong-line metallicities, \ha SFRs, and resolved maps of \ha emission. We measured non-parametric morphological parameters from the \ha and continuum maps, finding strong trends in the \ha and continuum morphologies with sSFR and stellar mass. More actively star forming galaxies have asymmetric and clumpy continuum and \ha emission with the brightest regions located near but slightly offset from the continuum center. 

Incorporating metallicity as an additional predictor, we find weak but significant trends, demonstrating that metallicity is tied to galaxy morphology even at fixed stellar mass and star formation rate. Galaxies that are even more metal poor than the FMR predicts for their mass and sSFR have the most asymmetric \ha and continuum emission and have \ha emission that is slightly more offset from the continuum center. 

Stellar continuum emission and \ha emission are generated by distinct stellar populations, with continuum light tracing older generations (averaged over the past $\gtrsim1$Gyr) and \ha emission tracing the distribution of recent star formation only, averaged over the past $\sim$10 Myr \citep{Kennicutt2012, FloresVelazquez2021}. The metallicity dependence is substantially stronger for the continuum than for the \ha emission: including 12 + log(O/H) increases $R^2$ for \asym$_c$ by 0.06--0.07, more than twice the variance explained by stellar mass and sSFR together, while for \asymha\ the increase is $\sim$0.01, roughly a tenth of the mass and sSFR contribution. This contrast provides evidence to distinguish between the two triggering channels: internal feedback can perturb the ionized gas but not the collisionless stellar body on burst timescales, and the predicted structural response of dwarfs to feedback is a radial breathing mode \citep{El-Badry2016}, which would not substantially affect the continuum asymmetry. The dependence of the continuum asymmetry on metallicity therefore favors external triggering of the elevated star formation.

A dynamical disturbance in the galaxy would also account for the \ha trends, affecting the preexisting stellar population, increasing the observed continuum asymmetry and simultaneously funneling gas towards the galaxy center \citep{Bustamante2018, Yesuf2021, Bottrell2024}. The abrupt central concentration of the gas could lead to a burst of star formation, located near but not perfectly at the continuum center, causing an elevated SFR and a central metallicity dilution \citep{Ceverino2016} -- to which our fiber-based metallicities are particularly sensitive (see Section~\ref{subsec:caveats}) --as well as the observed increase in \mtwentyha\ with decreasing metallicity. The most strongly star-forming galaxies still have generally centralized \ha emission (low \mtwentyha), but within bins of fixed mass and sSFR, the most metal-poor galaxies have their brightest \ha clumps slightly off-center (higher \mtwentyha).

The diversity of metallicity and \ha centrality in the highest sSFR population (some galaxies are more metal rich and centralized than others) could reflect a diversity of triggering mechanisms or alternatively, could be explained by temporal evolution in the \ha morphology and metallicity over the duration of the burst. If the bursts are sufficiently long lived  ($\gtrsim10$Myr), the SFR would stay elevated while the \ha morphology becomes more centralized and the nuclear metallicity increases with ongoing star formation and mixing. To further constrain the duration of the bursts and the possibility of morphological evolution over their duration, additional data constraining intermediate timescales would be required, UV imaging being the most plausible addition.

This work demonstrates the value of resolved star-formation tracers to constrain the drivers of bursts in low-mass galaxies. Characterizing the environment of dwarfs can be challenging and directly identifying very low mass companions outside of the Local Volume is beyond current observational capabilities. Here, we have shown how resolved maps of \ha emission combined with gas-phase metallicity measurements can provide additional insight into the environmental triggers of elevated activity. This type of analysis can be extended to even larger datasets with future releases of Merian and DESI and can be applied to other cosmic epochs where medium-bands are available to measure \ha emission -- for example, from JWST NIRCam.

\begin{acknowledgments}
A.M. acknowledges support from the National Science Foundation Graduate Research Fellowship under Grant No. 2039656. J.G. and A.L. are supported by a National Science Foundation Astronomy and Astrophysics Research Grant under Grant Nos. 1007052 and 2106839 respectively. AL is supported by the National Science Foundation under Grant No. 2510898.

The Hyper Suprime-Cam (HSC) collaboration includes the astronomical communities of Japan and Taiwan and Princeton University. The HSC instrumentation and software were developed by the National Astronomical Observatory of Japan (NAOJ), Kavli Institute for the Physics and Mathematics of the Universe (Kavli IPMU), University of Tokyo, High Energy Accelerator Research Organization (KEK), Academia Sinica Institute for Astronomy and Astrophysics in Taiwan (ASIAA), and Princeton University. Funding was contributed by the FIRST program from the Japanese Cabinet Office, Ministry of Education, Culture, Sports, Science and Technology (MEXT), Japan Society for the Promotion of Science (JSPS), Japan Science and Technology Agency (JST), Toray Science Foundation, NAOJ, Kavli IPMU, KEK, ASIAA, and Princeton University.

This project used data obtained with the Dark Energy Camera (DECam), which was constructed by the Dark Energy Survey (DES) collaboration. Funding for the DES Projects has been provided by the US Department of Energy, the US National Science Foundation, the Ministry of Science and Education of Spain, the Science and Technology Facilities Council of the United Kingdom, the Higher Education Funding Council for England, the National Center for Supercomputing Applications at the University of Illinois at Urbana-Champaign, the Kavli Institute for Cosmological Physics at the University of Chicago, Center for Cosmology and Astro-Particle Physics at the Ohio State University, the Mitchell Institute for Fundamental Physics and Astronomy at Texas A\&M University, Financiadora de Estudos e Projetos, Fundação Carlos Chagas Filho de Amparo à Pesquisa do Estado do Rio de Janeiro, Conselho Nacional de Desenvolvimento Científico e Tecnológico and the Ministério da Ciência, Tecnologia e Inovação, the Deutsche Forschungsgemeinschaft and the Collaborating Institutions in the Dark Energy Survey.

The Collaborating Institutions are Argonne National Laboratory, the University of California at Santa Cruz, the University of Cambridge, Centro de Investigaciones Enérgeticas, Medioambientales y Tecnológicas–Madrid, the University of Chicago, University College London, the DES-Brazil Consortium, the University of Edinburgh, the Eidgenössische Technische Hochschule (ETH) Zürich, Fermi National Accelerator Laboratory, the University of Illinois at Urbana-Champaign, the Institut de Ciències de l’Espai (IEEC/CSIC), the Institut de Física d’Altes Energies, Lawrence Berkeley National Laboratory, the Ludwig-Maximilians Universität München and the associated Excellence Cluster Universe, the University of Michigan, NSF’s NOIRLab, the University of Nottingham, the Ohio State University, the OzDES Membership Consortium, the University of Pennsylvania, the University of Portsmouth, SLAC National Accelerator Laboratory, Stanford University, the University of Sussex, and Texas A\&M University.

The authors are pleased to acknowledge that the work reported on in this paper was substantially performed using the Princeton Research Computing resources at Princeton University which is consortium of groups led by the Princeton Institute for Computational Science and Engineering (PICSciE) and Office of Information Technology's Research Computing.

This research used data obtained with the Dark Energy Spectroscopic Instrument (DESI). DESI construction and operations is managed by the Lawrence Berkeley National Laboratory. This material is based upon work supported by the U.S. Department of Energy, Office of Science, Office of High-Energy Physics, under Contract No. DE–AC02–05CH11231, and by the National Energy Research Scientific Computing Center, a DOE Office of Science User Facility under the same contract. Additional support for DESI was provided by the U.S. National Science Foundation (NSF), Division of Astronomical Sciences under Contract No. AST-0950945 to the NSF’s National Optical-Infrared Astronomy Research Laboratory; the Science and Technology Facilities Council of the United Kingdom; the Gordon and Betty Moore Foundation; the Heising-Simons Foundation; the French Alternative Energies and Atomic Energy Commission (CEA); the National Council of Humanities, Science and Technology of Mexico (CONAHCYT); the Ministry of Science and Innovation of Spain (MICINN), and by the DESI Member Institutions: www.desi.lbl.gov/collaborating-institutions. The DESI collaboration is honored to be permitted to conduct scientific research on I’oligam Du’ag (Kitt Peak), a mountain with particular significance to the Tohono O’odham Nation. Any opinions, findings, and conclusions or recommendations expressed in this material are those of the author(s) and do not necessarily reflect the views of the U.S. National Science Foundation, the U.S. Department of Energy, or any of the listed funding agencies.

\textit{Software}: \texttt{NumPy} \citep[\url{https://numpy.org},][]{Harris2020}, \texttt{Astropy}  \citep[\url{https://astropy.org},][]{Astropy2013, Astropy2018, Astropy2022}, \texttt{Matplotlib} \citep[\url{https://matplotlib.org},][]{Hunter2007}, \texttt{SciPy} \citep[\url{https://scipy.org},][]{Virtanen2020}, \texttt{statmorph}  \citep[\url{https://statmorph.readthedocs.io/en/latest/},][]{RodriguezGomez2019}, \texttt{SEP}  \citep[\url{https://sep.readthedocs.io/en/v1.1.x/},][]{Bertin1996, Barbary2016}, \texttt{statsmodels} \citep{statsmodel}, \texttt{FastSpecFit} \citep{fastspecfit}.
\end{acknowledgments}

\bibliography{met}

@ARTICLE{Nakajima2022,
       author = {{Nakajima}, Kimihiko and {Ouchi}, Masami and {Xu}, Yi and {Rauch}, Michael and {Harikane}, Yuichi and {Nishigaki}, Moka and {Isobe}, Yuki and {Kusakabe}, Haruka and {Nagao}, Tohru and {Ono}, Yoshiaki and {Onodera}, Masato and {Sugahara}, Yuma and {Kim}, Ji Hoon and {Komiyama}, Yutaka and {Lee}, Chien-Hsiu and {Zahedy}, Fakhri S.},
        title = "{EMPRESS. V. Metallicity Diagnostics of Galaxies over 12 + log(O/H) ≃ 6.9-8.9 Established by a Local Galaxy Census: Preparing for JWST Spectroscopy}",
      journal = {\apjs},
         year = 2022,
        month = sep,
       volume = {262},
       number = {1},
          eid = {3},
        pages = {3},
          doi = {10.3847/1538-4365/ac7710},
archivePrefix = {arXiv},
       eprint = {2206.02824},
 primaryClass = {astro-ph.GA},
       adsurl = {https://ui.adsabs.harvard.edu/abs/2022ApJS..262....3N}
}

@ARTICLE{Curti2024,
       author = {{Curti}, Mirko and {Maiolino}, Roberto and {Curtis-Lake}, Emma and {Chevallard}, Jacopo and {Carniani}, Stefano and {D'Eugenio}, Francesco and {Looser}, Tobias J. and {Scholtz}, Jan and {Charlot}, Stephane and {Cameron}, Alex and {{\"U}bler}, Hannah and {Witstok}, Joris and {Boyett}, Kristian and {Laseter}, Isaac and {Sandles}, Lester and {Arribas}, Santiago and {Bunker}, Andrew and {Giardino}, Giovanna and {Maseda}, Michael V. and {Rawle}, Tim and {Rodr{\'\i}guez Del Pino}, Bruno and {Smit}, Renske and {Willott}, Chris J. and {Eisenstein}, Daniel J. and {Hausen}, Ryan and {Johnson}, Benjamin and {Rieke}, Marcia and {Robertson}, Brant and {Tacchella}, Sandro and {Williams}, Christina C. and {Willmer}, Christopher and {Baker}, William M. and {Bhatawdekar}, Rachana and {Egami}, Eiichi and {Helton}, Jakob M. and {Ji}, Zhiyuan and {Kumari}, Nimisha and {Perna}, Michele and {Shivaei}, Irene and {Sun}, Fengwu},
        title = "{JADES: Insights into the low-mass end of the mass-metallicity-SFR relation at 3 < z < 10 from deep JWST/NIRSpec spectroscopy}",
      journal = {\aap},
         year = 2024,
        month = apr,
       volume = {684},
          eid = {A75},
        pages = {A75},
          doi = {10.1051/0004-6361/202346698},
archivePrefix = {arXiv},
       eprint = {2304.08516},
 primaryClass = {astro-ph.GA},
       adsurl = {https://ui.adsabs.harvard.edu/abs/2024A&A...684A..75C}
}

@ARTICLE{Curti2017,
       author = {{Curti}, M. and {Cresci}, G. and {Mannucci}, F. and {Marconi}, A. and {Maiolino}, R. and {Esposito}, S.},
        title = "{New fully empirical calibrations of strong-line metallicity indicators in star-forming galaxies}",
      journal = {\mnras},
         year = 2017,
        month = feb,
       volume = {465},
       number = {2},
        pages = {1384-1400},
          doi = {10.1093/mnras/stw2766},
archivePrefix = {arXiv},
       eprint = {1610.06939},
 primaryClass = {astro-ph.GA},
       adsurl = {https://ui.adsabs.harvard.edu/abs/2017MNRAS.465.1384C}
}

@ARTICLE{Curti2020,
       author = {{Curti}, Mirko and {Mannucci}, Filippo and {Cresci}, Giovanni and {Maiolino}, Roberto},
        title = "{The mass-metallicity and the fundamental metallicity relation revisited on a fully T$_{e}$-based abundance scale for galaxies}",
      journal = {\mnras},
         year = 2020,
        month = jan,
       volume = {491},
       number = {1},
        pages = {944-964},
          doi = {10.1093/mnras/stz2910},
archivePrefix = {arXiv},
       eprint = {1910.00597},
 primaryClass = {astro-ph.GA},
       adsurl = {https://ui.adsabs.harvard.edu/abs/2020MNRAS.491..944C}
}

@ARTICLE{Andrews2013,
       author = {{Andrews}, Brett H. and {Martini}, Paul},
        title = "{The Mass-Metallicity Relation with the Direct Method on Stacked Spectra of SDSS Galaxies}",
      journal = {\apj},
         year = 2013,
        month = mar,
       volume = {765},
       number = {2},
          eid = {140},
        pages = {140},
          doi = {10.1088/0004-637X/765/2/140},
archivePrefix = {arXiv},
       eprint = {1211.3418},
 primaryClass = {astro-ph.CO},
       adsurl = {https://ui.adsabs.harvard.edu/abs/2013ApJ...765..140A}
}

@ARTICLE{Scholte2024,
       author = {{Scholte}, Dirk and {Saintonge}, Am{\'e}lie and {Moustakas}, John and {Catinella}, Barbara and {Zou}, Hu and {Dey}, Biprateep and {Aguilar}, J. and {Ahlen}, S. and {Anand}, A. and {Blum}, R. and {Brooks}, D. and {Circosta}, C. and {Claybaugh}, T. and {de la Macorra}, A. and {Doel}, P. and {Font-Ribera}, A. and {F{\"o}rster}, P.~U. and {Forero-Romero}, J.~E. and {Gazta{\~n}aga}, E. and {Gontcho A Gontcho}, S. and {Juneau}, S. and {Kehoe}, R. and {Kisner}, T. and {Koposov}, S.~E. and {Kremin}, A. and {Lambert}, A. and {Landriau}, M. and {Maraston}, C. and {Martini}, P. and {Meisner}, A. and {Mighty}, A.~S. and {Miquel}, R. and {Myers}, A.~D. and {Nie}, J. and {Poppett}, C. and {Prada}, F. and {Rezaie}, M. and {Rossi}, G. and {Sanchez}, E. and {Schubnell}, M. and {Silber}, J. and {Sprayberry}, D. and {Siudek}, M. and {Speranza}, F. and {Tarl{\'e}}, G. and {Tojeiro}, R. and {Weaver}, B.~A.},
        title = "{The atomic gas sequence and mass-metallicity relation from dwarfs to massive galaxies}",
      journal = {\mnras},
         year = 2024,
        month = dec,
       volume = {535},
       number = {3},
        pages = {2341-2356},
          doi = {10.1093/mnras/stae2477},
archivePrefix = {arXiv},
       eprint = {2408.03996},
 primaryClass = {astro-ph.GA},
       adsurl = {https://ui.adsabs.harvard.edu/abs/2024MNRAS.535.2341S}
}

@ARTICLE{Denicolo2002,
       author = {{Denicol{\'o}}, Glenda and {Terlevich}, Roberto and {Terlevich}, Elena},
        title = "{New light on the search for low-metallicity galaxies - I. The N2 calibrator}",
      journal = {\mnras},
         year = 2002,
        month = feb,
       volume = {330},
       number = {1},
        pages = {69-74},
          doi = {10.1046/j.1365-8711.2002.05041.x},
archivePrefix = {arXiv},
       eprint = {astro-ph/0110356},
 primaryClass = {astro-ph},
       adsurl = {https://ui.adsabs.harvard.edu/abs/2002MNRAS.330...69D}
}

@ARTICLE{Laseter2024,
       author = {{Laseter}, Isaac H. and {Maseda}, Michael V. and {Curti}, Mirko and {Maiolino}, Roberto and {D'Eugenio}, Francesco and {Cameron}, Alex J. and {Looser}, Tobias J. and {Arribas}, Santiago and {Baker}, William M. and {Bhatawdekar}, Rachana and {Boyett}, Kristan and {Bunker}, Andrew J. and {Carniani}, Stefano and {Charlot}, Stephane and {Chevallard}, Jacopo and {Curtis-lake}, Emma and {Egami}, Eiichi and {Eisenstein}, Daniel J. and {Hainline}, Kevin and {Hausen}, Ryan and {Ji}, Zhiyuan and {Kumari}, Nimisha and {Perna}, Michele and {Rawle}, Tim and {Rix}, Hans-Walter and {Robertson}, Brant and {Rodr{\'\i}guez Del Pino}, Bruno and {Sandles}, Lester and {Scholtz}, Jan and {Smit}, Renske and {Tacchella}, Sandro and {{\"U}bler}, Hannah and {Williams}, Christina C. and {Willott}, Chris and {Witstok}, Joris},
        title = "{JADES: Detecting [OIII]{\ensuremath{\lambda}}4363 emitters and testing strong line calibrations in the high-z Universe with ultra-deep JWST/NIRSpec spectroscopy up to z {\ensuremath{\sim}} 9.5}",
      journal = {\aap},
         year = 2024,
        month = jan,
       volume = {681},
          eid = {A70},
        pages = {A70},
          doi = {10.1051/0004-6361/202347133},
archivePrefix = {arXiv},
       eprint = {2306.03120},
 primaryClass = {astro-ph.GA},
       adsurl = {https://ui.adsabs.harvard.edu/abs/2024A&A...681A..70L}
}

@ARTICLE{Danieli2025,
       author = {{Danieli}, Shany and {Kado-Fong}, Erin and {Huang}, Song and {Luo}, Yifei and {Li}, Ting S. and {Kelvin}, Lee S. and {Leauthaud}, Alexie and {Greene}, Jenny E. and {Mintz}, Abby and {Lin}, Xiaojing and {Li}, Jiaxuan and {Baldassare}, Vivienne and {Banerjee}, Arka and {Bhattacharyya}, Joy and {Blanco}, Diana and {Brooks}, Alyson and {Cai}, Zheng and {Chen}, Xinjun and {Cruz}, Akaxia and {Geda}, Robel and {Guan}, Runquan and {Johnson}, Sean and {Kannawadi}, Arun and {Kim}, Stacy Y. and {Li}, Mingyu and {Lupton}, Robert and {Mace}, Charlie and {Medina}, Gustavo E. and {Pan}, Yue and {Peter}, Annika H.~G. and {Read}, Justin I. and {C{\'o}rdova Rosado}, Rodrigo and {Seifert}, Allen and {Wasleske}, Erik J. and {Wick}, Joseph},
        title = "{First Data Release of the Merian Survey: A Wide-field Imaging Survey of Dwarf Galaxies at z {\ensuremath{\sim}} 0.06─0.10}",
      journal = {\apj},
         year = 2025,
        month = nov,
       volume = {993},
       number = {1},
          eid = {110},
        pages = {110},
          doi = {10.3847/1538-4357/ae003e},
archivePrefix = {arXiv},
       eprint = {2410.01884},
 primaryClass = {astro-ph.GA},
       adsurl = {https://ui.adsabs.harvard.edu/abs/2025ApJ...993..110D}
}

@ARTICLE{DESIdr1,
       author = {{DESI Collaboration} and {Abdul-Karim}, M. and {Adame}, A.~G. and {Aguado}, D. and {Aguilar}, J. and {Ahlen}, S. and {Alam}, S. and {Aldering}, G. and {Alexander}, D.~M. and {Alfarsy}, R. and {others}},
        title = "{Data Release 1 of the Dark Energy Spectroscopic Instrument}",
      journal = {arXiv e-prints},
         year = 2025,
        month = mar,
          eid = {arXiv:2503.14745},
        pages = {arXiv:2503.14745},
          doi = {10.48550/arXiv.2503.14745},
archivePrefix = {arXiv},
       eprint = {2503.14745},
 primaryClass = {astro-ph.CO},
       adsurl = {https://ui.adsabs.harvard.edu/abs/2025arXiv250314745D}
}

@ARTICLE{Mao2024,
       author = {{Mao}, Yao-Yuan and {Geha}, Marla and {Wechsler}, Risa H. and {Asali}, Yasmeen and {Wang}, Yunchong and {Kado-Fong}, Erin and {Kallivayalil}, Nitya and {Nadler}, Ethan O. and {Tollerud}, Erik J. and {Weiner}, Benjamin and {de los Reyes}, Mithi A.~C. and {Wu}, John F.},
        title = "{The SAGA Survey. III. A Census of 101 Satellite Systems around Milky Way{\textendash}mass Galaxies}",
      journal = {\apj},
         year = 2024,
        month = nov,
       volume = {976},
       number = {1},
          eid = {117},
        pages = {117},
          doi = {10.3847/1538-4357/ad64c4},
archivePrefix = {arXiv},
       eprint = {2404.14498},
 primaryClass = {astro-ph.GA},
       adsurl = {https://ui.adsabs.harvard.edu/abs/2024ApJ...976..117M}
}

@ARTICLE{Cardelli1989,
       author = {{Cardelli}, Jason A. and {Clayton}, Geoffrey C. and {Mathis}, John S.},
        title = "{The Relationship between Infrared, Optical, and Ultraviolet Extinction}",
      journal = {\apj},
         year = 1989,
        month = oct,
       volume = {345},
        pages = {245},
          doi = {10.1086/167900},
       adsurl = {https://ui.adsabs.harvard.edu/abs/1989ApJ...345..245C}
}

@BOOK{osterbrock2006,
       author = {{Osterbrock}, Donald E. and {Ferland}, Gary J.},
        title = "{Astrophysics of gaseous nebulae and active galactic nuclei}",
         year = 2006,
       adsurl = {https://ui.adsabs.harvard.edu/abs/2006agna.book.....O}
}

@ARTICLE{Hummer1987,
       author = {{Hummer}, D.~G. and {Storey}, P.~J.},
        title = "{Recombination-line intensities for hydrogenic ions - I. Case B calculations for H I and He II.}",
      journal = {\mnras},
         year = 1987,
        month = feb,
       volume = {224},
        pages = {801-820},
          doi = {10.1093/mnras/224.3.801},
       adsurl = {https://ui.adsabs.harvard.edu/abs/1987MNRAS.224..801H}
}

@ARTICLE{Kauffmann2003,
       author = {{Kauffmann}, Guinevere and {Heckman}, Timothy M. and {Tremonti}, Christy and {Brinchmann}, Jarle and {Charlot}, St{\'e}phane and {White}, Simon D.~M. and {Ridgway}, Susan E. and {Brinkmann}, Jon and {Fukugita}, Masataka and {Hall}, Patrick B. and {Ivezi{\'c}}, {\v{Z}}eljko and {Richards}, Gordon T. and {Schneider}, Donald P.},
        title = "{The host galaxies of active galactic nuclei}",
      journal = {\mnras},
         year = 2003,
        month = dec,
       volume = {346},
       number = {4},
        pages = {1055-1077},
          doi = {10.1111/j.1365-2966.2003.07154.x},
archivePrefix = {arXiv},
       eprint = {astro-ph/0304239},
 primaryClass = {astro-ph},
       adsurl = {https://ui.adsabs.harvard.edu/abs/2003MNRAS.346.1055K}
}

@ARTICLE{Mintz2024,
       author = {{Mintz}, Abby and {Greene}, Jenny E. and {Kado-Fong}, Erin and {Danieli}, Shany and {Li}, Jiaxuan and {Luo}, Yifei and {Leauthaud}, Alexie and {Baldassare}, Vivienne and {Huang}, Song and {Peter}, Annika H.~G. and {Bhattacharyya}, Joy and {Li}, Mingyu and {Pan}, Yue},
        title = "{A Nonparametric Morphological Analysis of H{\ensuremath{\alpha}} Emission in Bright Dwarfs Using the Merian Survey}",
      journal = {\apj},
         year = 2024,
        month = oct,
       volume = {974},
       number = {2},
          eid = {273},
        pages = {273},
          doi = {10.3847/1538-4357/ad6861},
archivePrefix = {arXiv},
       eprint = {2410.01886},
 primaryClass = {astro-ph.GA},
       adsurl = {https://ui.adsabs.harvard.edu/abs/2024ApJ...974..273M}
}

@ARTICLE{RodriguezGomez2019,
       author = {{Rodriguez-Gomez}, Vicente and {Snyder}, Gregory F. and {Lotz}, Jennifer M. and {Nelson}, Dylan and {Pillepich}, Annalisa and {Springel}, Volker and {Genel}, Shy and {Weinberger}, Rainer and {Tacchella}, Sandro and {Pakmor}, R{\"u}diger and {Torrey}, Paul and {Marinacci}, Federico and {Vogelsberger}, Mark and {Hernquist}, Lars and {Thilker}, David A.},
        title = "{The optical morphologies of galaxies in the IllustrisTNG simulation: a comparison to Pan-STARRS observations}",
      journal = {\mnras},
         year = 2019,
        month = mar,
       volume = {483},
       number = {3},
        pages = {4140-4159},
          doi = {10.1093/mnras/sty3345},
archivePrefix = {arXiv},
       eprint = {1809.08239},
 primaryClass = {astro-ph.GA},
       adsurl = {https://ui.adsabs.harvard.edu/abs/2019MNRAS.483.4140R}
}

@ARTICLE{Telford2016,
       author = {{Telford}, O. Grace and {Dalcanton}, Julianne J. and {Skillman}, Evan D. and {Conroy}, Charlie},
        title = "{Exploring Systematic Effects in the Relation Between Stellar Mass, Gas Phase Metallicity, and Star Formation Rate}",
      journal = {\apj},
         year = 2016,
        month = aug,
       volume = {827},
       number = {1},
          eid = {35},
        pages = {35},
          doi = {10.3847/0004-637X/827/1/35},
archivePrefix = {arXiv},
       eprint = {1606.08850},
 primaryClass = {astro-ph.GA},
       adsurl = {https://ui.adsabs.harvard.edu/abs/2016ApJ...827...35T}
}

@ARTICLE{Yao2023,
       author = {{Yao}, Yao and {Song}, Jie and {Kong}, Xu and {Fang}, Guanwen and {Zhang}, Hong-Xin and {Chen}, Xinkai},
        title = "{Evolution of Nonparametric Morphology of Galaxies in the JWST CEERS Field at z ≃ 0.8-3.0}",
      journal = {\apj},
         year = 2023,
        month = sep,
       volume = {954},
       number = {2},
          eid = {113},
        pages = {113},
          doi = {10.3847/1538-4357/ace7b5},
archivePrefix = {arXiv},
       eprint = {2307.13975},
 primaryClass = {astro-ph.GA},
       adsurl = {https://ui.adsabs.harvard.edu/abs/2023ApJ...954..113Y}
}

@ARTICLE{Nersesian2023,
       author = {{Nersesian}, Angelos and {Zibetti}, Stefano and {D'Eugenio}, Francesco and {Baes}, Maarten},
        title = "{Non-parametric galaxy morphology from stellar and nebular emission with the CALIFA sample}",
      journal = {\aap},
         year = 2023,
        month = may,
       volume = {673},
          eid = {A63},
        pages = {A63},
          doi = {10.1051/0004-6361/202345962},
archivePrefix = {arXiv},
       eprint = {2303.01907},
 primaryClass = {astro-ph.GA},
       adsurl = {https://ui.adsabs.harvard.edu/abs/2023A&A...673A..63N}
}

@ARTICLE{Conselice2003,
       author = {{Conselice}, Christopher J.},
        title = "{The Relationship between Stellar Light Distributions of Galaxies and Their Formation Histories}",
      journal = {\apjs},
         year = 2003,
        month = jul,
       volume = {147},
       number = {1},
        pages = {1-28},
          doi = {10.1086/375001},
archivePrefix = {arXiv},
       eprint = {astro-ph/0303065},
 primaryClass = {astro-ph},
       adsurl = {https://ui.adsabs.harvard.edu/abs/2003ApJS..147....1C}
}

@ARTICLE{Fossati2013,
       author = {{Fossati}, M. and {Gavazzi}, G. and {Savorgnan}, G. and {Fumagalli}, M. and {Boselli}, A. and {Guti{\'e}rrez}, L. and {Hern{\'a}ndez Toledo}, H. and {Giovanelli}, R. and {Haynes}, M.~P.},
        title = "{H{\ensuremath{\alpha}}3: an H{\ensuremath{\alpha}} imaging survey of HI selected galaxies from ALFALFA. IV. Structure of galaxies in the Local and Coma superclusters}",
      journal = {\aap},
         year = 2013,
        month = may,
       volume = {553},
          eid = {A91},
        pages = {A91},
          doi = {10.1051/0004-6361/201220915},
archivePrefix = {arXiv},
       eprint = {1303.0840},
 primaryClass = {astro-ph.CO},
       adsurl = {https://ui.adsabs.harvard.edu/abs/2013A&A...553A..91F}
}

@ARTICLE{Boselli2015,
       author = {{Boselli}, A. and {Fossati}, M. and {Gavazzi}, G. and {Ciesla}, L. and {Buat}, V. and {Boissier}, S. and {Hughes}, T.~M.},
        title = "{H{\ensuremath{\alpha}} imaging of the Herschel Reference Survey. The star formation properties of a volume-limited, K-band-selected sample of nearby late-type galaxies}",
      journal = {\aap},
         year = 2015,
        month = jul,
       volume = {579},
          eid = {A102},
        pages = {A102},
          doi = {10.1051/0004-6361/201525712},
archivePrefix = {arXiv},
       eprint = {1504.01876},
 primaryClass = {astro-ph.GA},
       adsurl = {https://ui.adsabs.harvard.edu/abs/2015A&A...579A.102B}
}

@ARTICLE{Bottrell2024,
       author = {{Bottrell}, Connor and {Yesuf}, Hassen M. and {Popping}, Gerg{\"o} and {Omori}, Kiyoaki Christopher and {Tang}, Shenli and {Ding}, Xuheng and {Pillepich}, Annalisa and {Nelson}, Dylan and {Eisert}, Lukas and {Gao}, Hua and {Goulding}, Andy D. and {Kalita}, Boris S. and {Luo}, Wentao and {Greene}, Jenny E. and {Shi}, Jingjing and {Silverman}, John D.},
        title = "{IllustrisTNG in the HSC-SSP: image data release and the major role of mini mergers as drivers of asymmetry and star formation}",
      journal = {\mnras},
         year = 2024,
        month = jan,
       volume = {527},
       number = {3},
        pages = {6506-6539},
          doi = {10.1093/mnras/stad2971},
archivePrefix = {arXiv},
       eprint = {2308.14793},
 primaryClass = {astro-ph.GA},
       adsurl = {https://ui.adsabs.harvard.edu/abs/2024MNRAS.527.6506B}
}

@ARTICLE{Yesuf2021,
       author = {{Yesuf}, Hassen M. and {Ho}, Luis C. and {Faber}, S.~M.},
        title = "{What is Important? Morphological Asymmetries are Useful Predictors of Star Formation Rates of Star-forming Galaxies in SDSS Stripe 82}",
      journal = {\apj},
         year = 2021,
        month = dec,
       volume = {923},
       number = {2},
          eid = {205},
        pages = {205},
          doi = {10.3847/1538-4357/ac27a7},
archivePrefix = {arXiv},
       eprint = {2109.08882},
 primaryClass = {astro-ph.GA},
       adsurl = {https://ui.adsabs.harvard.edu/abs/2021ApJ...923..205Y}
}

@ARTICLE{Schlafly2011,
       author = {{Schlafly}, Edward F. and {Finkbeiner}, Douglas P.},
        title = "{Measuring Reddening with Sloan Digital Sky Survey Stellar Spectra and Recalibrating SFD}",
      journal = {\apj},
         year = 2011,
        month = aug,
       volume = {737},
       number = {2},
          eid = {103},
        pages = {103},
          doi = {10.1088/0004-637X/737/2/103},
archivePrefix = {arXiv},
       eprint = {1012.4804},
 primaryClass = {astro-ph.GA},
       adsurl = {https://ui.adsabs.harvard.edu/abs/2011ApJ...737..103S}
}

@ARTICLE{Bustamante2018,
       author = {{Bustamante}, Sebasti{\'a}n and {Sparre}, Martin and {Springel}, Volker and {Grand}, Robert J.~J.},
        title = "{Merger-induced metallicity dilution in cosmological galaxy formation simulations}",
      journal = {\mnras},
         year = 2018,
        month = sep,
       volume = {479},
       number = {3},
        pages = {3381-3392},
          doi = {10.1093/mnras/sty1692},
archivePrefix = {arXiv},
       eprint = {1712.03250},
 primaryClass = {astro-ph.GA},
       adsurl = {https://ui.adsabs.harvard.edu/abs/2018MNRAS.479.3381B}
}

@ARTICLE{Ellison2008,
       author = {{Ellison}, Sara L. and {Patton}, David R. and {Simard}, Luc and {McConnachie}, Alan W.},
        title = "{Galaxy Pairs in the Sloan Digital Sky Survey. I. Star Formation, Active Galactic Nucleus Fraction, and the Mass-Metallicity Relation}",
      journal = {\aj},
         year = 2008,
        month = may,
       volume = {135},
       number = {5},
        pages = {1877-1899},
          doi = {10.1088/0004-6256/135/5/1877},
archivePrefix = {arXiv},
       eprint = {0803.0161},
 primaryClass = {astro-ph},
       adsurl = {https://ui.adsabs.harvard.edu/abs/2008AJ....135.1877E}
}

@ARTICLE{Hopkins2014,
       author = {{Hopkins}, Philip F. and {Kere{\v{s}}}, Du{\v{s}}an and {O{\~n}orbe}, Jos{\'e} and {Faucher-Gigu{\`e}re}, Claude-Andr{\'e} and {Quataert}, Eliot and {Murray}, Norman and {Bullock}, James S.},
        title = "{Galaxies on FIRE (Feedback In Realistic Environments): stellar feedback explains cosmologically inefficient star formation}",
      journal = {\mnras},
         year = 2014,
        month = nov,
       volume = {445},
       number = {1},
        pages = {581-603},
          doi = {10.1093/mnras/stu1738},
archivePrefix = {arXiv},
       eprint = {1311.2073},
 primaryClass = {astro-ph.CO},
       adsurl = {https://ui.adsabs.harvard.edu/abs/2014MNRAS.445..581H}
}

@ARTICLE{Benavides2025,
       author = {{Benavides}, Jos{\'e} A. and {Sales}, Laura V. and {Wetzel}, Andrew and {Moreno}, Jorge and {Feldmann}, Robert and {Mercado}, Francisco J. and {Bullock}, James S. and {Hopkins}, Philip F. and {Faucher-Gigu{\'e}re}, Claude-Andr{\'e} and {Stern}, Jonathan and {Wheeler}, Coral and {Kere{\v{s}}}, Du{\v{s}}an},
        title = "{Disks no more: the morphology of low-mass simulated galaxies in FIREbox}",
      journal = {\mnras},
         year = 2025,
        month = oct,
       volume = {544},
       number = {4},
        pages = {4651-4664},
          doi = {10.1093/mnras/staf1847},
archivePrefix = {arXiv},
       eprint = {2508.00991},
 primaryClass = {astro-ph.GA},
       adsurl = {https://ui.adsabs.harvard.edu/abs/2025MNRAS.544.4651B}
}

@ARTICLE{Read2016,
       author = {{Read}, J.~I. and {Agertz}, O. and {Collins}, M.~L.~M.},
        title = "{Dark matter cores all the way down}",
      journal = {\mnras},
         year = 2016,
        month = jul,
       volume = {459},
       number = {3},
        pages = {2573-2590},
          doi = {10.1093/mnras/stw713},
archivePrefix = {arXiv},
       eprint = {1508.04143},
 primaryClass = {astro-ph.GA},
       adsurl = {https://ui.adsabs.harvard.edu/abs/2016MNRAS.459.2573R}
}

@ARTICLE{Sales2022,
       author = {{Sales}, Laura V. and {Wetzel}, Andrew and {Fattahi}, Azadeh},
        title = "{Baryonic solutions and challenges for cosmological models of dwarf galaxies}",
      journal = {Nature Astronomy},
         year = 2022,
        month = jun,
       volume = {6},
        pages = {897-910},
          doi = {10.1038/s41550-022-01689-w},
archivePrefix = {arXiv},
       eprint = {2206.05295},
 primaryClass = {astro-ph.GA},
       adsurl = {https://ui.adsabs.harvard.edu/abs/2022NatAs...6..897S}
}

@ARTICLE{Popesso2023,
       author = {{Popesso}, P. and {Concas}, A. and {Cresci}, G. and {Belli}, S. and {Rodighiero}, G. and {Inami}, H. and {Dickinson}, M. and {Ilbert}, O. and {Pannella}, M. and {Elbaz}, D.},
        title = "{The main sequence of star-forming galaxies across cosmic times}",
      journal = {\mnras},
         year = 2023,
        month = feb,
       volume = {519},
       number = {1},
        pages = {1526-1544},
          doi = {10.1093/mnras/stac3214},
archivePrefix = {arXiv},
       eprint = {2203.10487},
 primaryClass = {astro-ph.GA},
       adsurl = {https://ui.adsabs.harvard.edu/abs/2023MNRAS.519.1526P}
}

@ARTICLE{FloresVelazquez2021,
       author = {{Flores Vel{\'a}zquez}, Jos{\'e} A. and {Gurvich}, Alexander B. and {Faucher-Gigu{\`e}re}, Claude-Andr{\'e} and {Bullock}, James S. and {Starkenburg}, Tjitske K. and {Moreno}, Jorge and {Lazar}, Alexandres and {Mercado}, Francisco J. and {Stern}, Jonathan and {Sparre}, Martin and {Hayward}, Christopher C. and {Wetzel}, Andrew and {El-Badry}, Kareem},
        title = "{The time-scales probed by star formation rate indicators for realistic, bursty star formation histories from the FIRE simulations}",
      journal = {\mnras},
         year = 2021,
        month = mar,
       volume = {501},
       number = {4},
        pages = {4812-4824},
          doi = {10.1093/mnras/staa3893},
archivePrefix = {arXiv},
       eprint = {2008.08582},
 primaryClass = {astro-ph.GA},
       adsurl = {https://ui.adsabs.harvard.edu/abs/2021MNRAS.501.4812F}
}

@ARTICLE{Stierwalt2015,
       author = {{Stierwalt}, S. and {Besla}, G. and {Patton}, D. and {Johnson}, K. and {Kallivayalil}, N. and {Putman}, M. and {Privon}, G. and {Ross}, G.},
        title = "{TiNy Titans: The Role of Dwarf-Dwarf Interactions in Low-mass Galaxy Evolution}",
      journal = {\apj},
         year = 2015,
        month = may,
       volume = {805},
       number = {1},
          eid = {2},
        pages = {2},
          doi = {10.1088/0004-637X/805/1/2},
archivePrefix = {arXiv},
       eprint = {1412.4796},
 primaryClass = {astro-ph.GA},
       adsurl = {https://ui.adsabs.harvard.edu/abs/2015ApJ...805....2S}
}

@ARTICLE{KadoFong2024b,
       author = {{Kado-Fong}, Erin and {Geha}, Marla and {Mao}, Yao-Yuan and {de los Reyes}, Mithi A.~C. and {Wechsler}, Risa H. and {Weiner}, Benjamin and {Asali}, Yasmeen and {Kallivayalil}, Nitya and {Nadler}, Ethan O. and {Tollerud}, Erik J. and {Wang}, Yunchong},
        title = "{SAGAbg. II. The Low-mass Star-forming Sequence Evolves Significantly between 0.05 < z < 0.21}",
      journal = {\apj},
         year = 2024,
        month = nov,
       volume = {976},
       number = {1},
          eid = {83},
        pages = {83},
          doi = {10.3847/1538-4357/ad8137},
archivePrefix = {arXiv},
       eprint = {2409.12221},
 primaryClass = {astro-ph.GA},
       adsurl = {https://ui.adsabs.harvard.edu/abs/2024ApJ...976...83K}
}

@ARTICLE{Lee2009,
       author = {{Lee}, Janice C. and {Kennicutt}, Jr., Robert C. and {Funes}, S.~J. Jos{\'e} G. and {Sakai}, Shoko and {Akiyama}, Sanae},
        title = "{Dwarf Galaxy Starburst Statistics in the Local Volume}",
      journal = {\apj},
         year = 2009,
        month = feb,
       volume = {692},
       number = {2},
        pages = {1305-1320},
          doi = {10.1088/0004-637X/692/2/1305},
archivePrefix = {arXiv},
       eprint = {0810.5132},
 primaryClass = {astro-ph},
       adsurl = {https://ui.adsabs.harvard.edu/abs/2009ApJ...692.1305L}
}

@ARTICLE{Wang2024,
       author = {{Wang}, Bingjie and {Leja}, Joel and {Labb{\'e}}, Ivo and {Bezanson}, Rachel and {Whitaker}, Katherine E. and {Brammer}, Gabriel and {Furtak}, Lukas J. and {Weaver}, John R. and {Price}, Sedona H. and {Zitrin}, Adi and {Atek}, Hakim and {Coe}, Dan and {Cutler}, Sam E. and {Dayal}, Pratika and {van Dokkum}, Pieter and {Feldmann}, Robert and {Marchesini}, Danilo and {Franx}, Marijn and {F{\"o}rster Schreiber}, Natascha and {Fujimoto}, Seiji and {Geha}, Marla and {Glazebrook}, Karl and {de Graaff}, Anna and {Greene}, Jenny E. and {Juneau}, St{\'e}phanie and {Kassin}, Susan and {Kriek}, Mariska and {Khullar}, Gourav and {Maseda}, Michael and {Mowla}, Lamiya A. and {Muzzin}, Adam and {Nanayakkara}, Themiya and {Nelson}, Erica J. and {Oesch}, Pascal A. and {Pacifici}, Camilla and {Pan}, Richard and {Papovich}, Casey and {Setton}, David J. and {Shapley}, Alice E. and {Smit}, Renske and {Stefanon}, Mauro and {Suess}, Katherine A. and {Taylor}, Edward N. and {Williams}, Christina C.},
        title = "{The UNCOVER Survey: A First-look HST+JWST Catalog of Galaxy Redshifts and Stellar Population Properties Spanning 0.2 {\ensuremath{\lesssim}} z {\ensuremath{\lesssim}} 15}",
      journal = {\apjs},
         year = 2024,
        month = jan,
       volume = {270},
       number = {1},
          eid = {12},
        pages = {12},
          doi = {10.3847/1538-4365/ad0846},
archivePrefix = {arXiv},
       eprint = {2310.01276},
 primaryClass = {astro-ph.GA},
       adsurl = {https://ui.adsabs.harvard.edu/abs/2024ApJS..270...12W}
}

@ARTICLE{Kennicutt2012,
       author = {{Kennicutt}, Robert C. and {Evans}, Neal J.},
        title = "{Star Formation in the Milky Way and Nearby Galaxies}",
      journal = {\araa},
         year = 2012,
        month = sep,
       volume = {50},
        pages = {531-608},
          doi = {10.1146/annurev-astro-081811-125610},
archivePrefix = {arXiv},
       eprint = {1204.3552},
 primaryClass = {astro-ph.GA},
       adsurl = {https://ui.adsabs.harvard.edu/abs/2012ARA&A..50..531K}
}

@ARTICLE{McQuinn2010a,
       author = {{McQuinn}, Kristen B.~W. and {Skillman}, Evan D. and {Cannon}, John M. and {Dalcanton}, Julianne and {Dolphin}, Andrew and {Hidalgo-Rodr{\'\i}guez}, Sebastian and {Holtzman}, Jon and {Stark}, David and {Weisz}, Daniel and {Williams}, Benjamin},
        title = "{The Nature of Starbursts. I. The Star Formation Histories of Eighteen Nearby Starburst Dwarf Galaxies}",
      journal = {\apj},
         year = 2010,
        month = sep,
       volume = {721},
       number = {1},
        pages = {297-317},
          doi = {10.1088/0004-637X/721/1/297},
archivePrefix = {arXiv},
       eprint = {1008.1589},
 primaryClass = {astro-ph.CO},
       adsurl = {https://ui.adsabs.harvard.edu/abs/2010ApJ...721..297M}
}

@ARTICLE{McQuinn2010b,
       author = {{McQuinn}, Kristen B.~W. and {Skillman}, Evan D. and {Cannon}, John M. and {Dalcanton}, Julianne and {Dolphin}, Andrew and {Hidalgo-Rodr{\'\i}guez}, Sebastian and {Holtzman}, Jon and {Stark}, David and {Weisz}, Daniel and {Williams}, Benjamin},
        title = "{The Nature of Starbursts. II. The Duration of Starbursts in Dwarf Galaxies}",
      journal = {\apj},
         year = 2010,
        month = nov,
       volume = {724},
       number = {1},
        pages = {49-58},
          doi = {10.1088/0004-637X/724/1/49},
archivePrefix = {arXiv},
       eprint = {1009.2940},
 primaryClass = {astro-ph.CO},
       adsurl = {https://ui.adsabs.harvard.edu/abs/2010ApJ...724...49M}
}

@ARTICLE{Kauffmann2014,
       author = {{Kauffmann}, Guinevere},
        title = "{Quantitative constraints on starburst cycles in galaxies with stellar masses in the range {}10$^{8}$-{}10$^{10}$ M$_{{\ensuremath{\odot}}}$}",
      journal = {\mnras},
         year = 2014,
        month = jul,
       volume = {441},
       number = {3},
        pages = {2717-2724},
          doi = {10.1093/mnras/stu752},
       adsurl = {https://ui.adsabs.harvard.edu/abs/2014MNRAS.441.2717K}
}

@ARTICLE{Virtanen2020,
       author = {{Virtanen}, Pauli and {Gommers}, Ralf and {Oliphant}, Travis E. and {Haberland}, Matt and {Reddy}, Tyler and {Cournapeau}, David and {Burovski}, Evgeni and {Peterson}, Pearu and {Weckesser}, Warren and {Bright}, Jonathan and {van der Walt}, St{\'e}fan J. and {Brett}, Matthew and {Wilson}, Joshua and {Millman}, K. Jarrod and {Mayorov}, Nikolay and {Nelson}, Andrew R.~J. and {Jones}, Eric and {Kern}, Robert and {Larson}, Eric and {Carey}, C.~J. and {Polat}, {\.I}lhan and {Feng}, Yu and {Moore}, Eric W. and {VanderPlas}, Jake and {Laxalde}, Denis and {Perktold}, Josef and {Cimrman}, Robert and {Henriksen}, Ian and {Quintero}, E.~A. and {Harris}, Charles R. and {Archibald}, Anne M. and {Ribeiro}, Ant{\^o}nio H. and {Pedregosa}, Fabian and {van Mulbregt}, Paul and {SciPy 1. 0 Contributors}},
        title = "{SciPy 1.0: fundamental algorithms for scientific computing in Python}",
      journal = {Nature Methods},
         year = 2020,
        month = feb,
       volume = {17},
        pages = {261-272},
          doi = {10.1038/s41592-019-0686-2},
archivePrefix = {arXiv},
       eprint = {1907.10121},
 primaryClass = {cs.MS},
       adsurl = {https://ui.adsabs.harvard.edu/abs/2020NatMe..17..261V}
}

@ARTICLE{Pontzen2012,
       author = {{Pontzen}, Andrew and {Governato}, Fabio},
        title = "{How supernova feedback turns dark matter cusps into cores}",
      journal = {\mnras},
         year = 2012,
        month = apr,
       volume = {421},
       number = {4},
        pages = {3464-3471},
          doi = {10.1111/j.1365-2966.2012.20571.x},
archivePrefix = {arXiv},
       eprint = {1106.0499},
 primaryClass = {astro-ph.CO},
       adsurl = {https://ui.adsabs.harvard.edu/abs/2012MNRAS.421.3464P}
}

@ARTICLE{KadoFong2024c,
       author = {{Kado-Fong}, Erin and {Robinson}, Azia and {Nyland}, Kristina and {Greene}, Jenny E. and {Suess}, Katherine A. and {Stierwalt}, Sabrina and {Beaton}, Rachael},
        title = "{Dwarf─Dwarf Interactions Can Both Trigger and Quench Star Formation}",
      journal = {\apj},
         year = 2024,
        month = mar,
       volume = {963},
       number = {1},
          eid = {37},
        pages = {37},
          doi = {10.3847/1538-4357/ad18cb},
archivePrefix = {arXiv},
       eprint = {2311.09280},
 primaryClass = {astro-ph.GA},
       adsurl = {https://ui.adsabs.harvard.edu/abs/2024ApJ...963...37K}
}

@ARTICLE{Lelli2014,
       author = {{Lelli}, Federico and {Verheijen}, Marc and {Fraternali}, Filippo},
        title = "{The triggering of starbursts in low-mass galaxies}",
      journal = {\mnras},
         year = 2014,
        month = dec,
       volume = {445},
       number = {2},
        pages = {1694-1712},
          doi = {10.1093/mnras/stu1804},
archivePrefix = {arXiv},
       eprint = {1409.1239},
 primaryClass = {astro-ph.GA},
       adsurl = {https://ui.adsabs.harvard.edu/abs/2014MNRAS.445.1694L}
}

@ARTICLE{SanchezAlmeida2015,
       author = {{S{\'a}nchez Almeida}, J. and {Elmegreen}, B.~G. and {Mu{\~n}oz-Tu{\~n}{\'o}n}, C. and {Elmegreen}, D.~M. and {P{\'e}rez-Montero}, E. and {Amor{\'\i}n}, R. and {Filho}, M.~E. and {Ascasibar}, Y. and {Papaderos}, P. and {V{\'\i}lchez}, J.~M.},
        title = "{Localized Starbursts in Dwarf Galaxies Produced by the Impact of Low-metallicity Cosmic Gas Clouds}",
      journal = {\apjl},
         year = 2015,
        month = sep,
       volume = {810},
       number = {2},
          eid = {L15},
        pages = {L15},
          doi = {10.1088/2041-8205/810/2/L15},
archivePrefix = {arXiv},
       eprint = {1509.00180},
 primaryClass = {astro-ph.GA},
       adsurl = {https://ui.adsabs.harvard.edu/abs/2015ApJ...810L..15S}
}

@ARTICLE{Mannucci2010,
       author = {{Mannucci}, F. and {Cresci}, G. and {Maiolino}, R. and {Marconi}, A. and {Gnerucci}, A.},
        title = "{A fundamental relation between mass, star formation rate and metallicity in local and high-redshift galaxies}",
      journal = {\mnras},
         year = 2010,
        month = nov,
       volume = {408},
       number = {4},
        pages = {2115-2127},
          doi = {10.1111/j.1365-2966.2010.17291.x},
archivePrefix = {arXiv},
       eprint = {1005.0006},
 primaryClass = {astro-ph.CO},
       adsurl = {https://ui.adsabs.harvard.edu/abs/2010MNRAS.408.2115M}
}

@ARTICLE{Dave2012,
       author = {{Dav{\'e}}, Romeel and {Finlator}, Kristian and {Oppenheimer}, Benjamin D.},
        title = "{An analytic model for the evolution of the stellar, gas and metal content of galaxies}",
      journal = {\mnras},
         year = 2012,
        month = mar,
       volume = {421},
       number = {1},
        pages = {98-107},
          doi = {10.1111/j.1365-2966.2011.20148.x},
archivePrefix = {arXiv},
       eprint = {1108.0426},
 primaryClass = {astro-ph.CO},
       adsurl = {https://ui.adsabs.harvard.edu/abs/2012MNRAS.421...98D}
}

@ARTICLE{Lilly2013,
       author = {{Lilly}, Simon J. and {Carollo}, C. Marcella and {Pipino}, Antonio and {Renzini}, Alvio and {Peng}, Yingjie},
        title = "{Gas Regulation of Galaxies: The Evolution of the Cosmic Specific Star Formation Rate, the Metallicity-Mass-Star-formation Rate Relation, and the Stellar Content of Halos}",
      journal = {\apj},
         year = 2013,
        month = aug,
       volume = {772},
       number = {2},
          eid = {119},
        pages = {119},
          doi = {10.1088/0004-637X/772/2/119},
archivePrefix = {arXiv},
       eprint = {1303.5059},
 primaryClass = {astro-ph.CO},
       adsurl = {https://ui.adsabs.harvard.edu/abs/2013ApJ...772..119L}
}

@ARTICLE{Luo2024,
       author = {{Luo}, Yifei and {Leauthaud}, Alexie and {Greene}, Jenny and {Huang}, Song and {Kado-Fong}, Erin and {Danieli}, Shany and {Li}, Ting S. and {Li}, Jiaxuan and {Blanco}, Diana and {Wasleske}, Erik J. and {Wick}, Joseph and {Mintz}, Abby and {Guan}, Runquan and {Peter}, Annika H.~G. and {Baldassare}, Vivienne and {Brooks}, Alyson and {Banerjee}, Arka and {Bhattacharyya}, Joy and {Cai}, Zheng and {Chen}, Xinjun and {Gunn}, Jim and {Johnson}, Sean D. and {Kelvin}, Lee S. and {Li}, Mingyu and {Lin}, Xiaojing and {Lupton}, Robert and {Mace}, Charlie and {Medina}, Gustavo E. and {Read}, Justin and {Rosado}, Rodrigo C{\'o}rdova and {Seifert}, Allen},
        title = "{The Merian survey: design, construction, and characterization of a filter set optimized to find dwarf galaxies and measure their dark matter halo properties with weak lensing}",
      journal = {\mnras},
         year = 2024,
        month = jun,
       volume = {530},
       number = {4},
        pages = {4988-5005},
          doi = {10.1093/mnras/stae925},
archivePrefix = {arXiv},
       eprint = {2305.19310},
 primaryClass = {astro-ph.GA},
       adsurl = {https://ui.adsabs.harvard.edu/abs/2024MNRAS.530.4988L}
}

@ARTICLE{Aihara2018,
       author = {{Aihara}, Hiroaki and {Arimoto}, Nobuo and {Armstrong}, Robert and {Arnouts}, St{\'e}phane and {Bahcall}, Neta A. and {Bickerton}, Steven and {Bosch}, James and {Bundy}, Kevin and {Capak}, Peter L. and {Chan}, James H.~H. and {Chiba}, Masashi and {Coupon}, Jean and {Egami}, Eiichi and {Enoki}, Motohiro and {Finet}, Francois and {Fujimori}, Hiroki and {Fujimoto}, Seiji and {Furusawa}, Hisanori and {Furusawa}, Junko and {Goto}, Tomotsugu and {Goulding}, Andy and {Greco}, Johnny P. and {Greene}, Jenny E. and {Gunn}, James E. and {Hamana}, Takashi and {Harikane}, Yuichi and {Hashimoto}, Yasuhiro and {Hattori}, Takashi and {Hayashi}, Masao and {Hayashi}, Yusuke and {He{\l}miniak}, Krzysztof G. and {Higuchi}, Ryo and {Hikage}, Chiaki and {Ho}, Paul T.~P. and {Hsieh}, Bau-Ching and {Huang}, Kuiyun and {Huang}, Song and {Ikeda}, Hiroyuki and {Imanishi}, Masatoshi and {Inoue}, Akio K. and {Iwasawa}, Kazushi and {Iwata}, Ikuru and {Jaelani}, Anton T. and {Jian}, Hung-Yu and {Kamata}, Yukiko and {Karoji}, Hiroshi and {Kashikawa}, Nobunari and {Katayama}, Nobuhiko and {Kawanomoto}, Satoshi and {Kayo}, Issha and {Koda}, Jin and {Koike}, Michitaro and {Kojima}, Takashi and {Komiyama}, Yutaka and {Konno}, Akira and {Koshida}, Shintaro and {Koyama}, Yusei and {Kusakabe}, Haruka and {Leauthaud}, Alexie and {Lee}, Chien-Hsiu and {Lin}, Lihwai and {Lin}, Yen-Ting and {Lupton}, Robert H. and {Mandelbaum}, Rachel and {Matsuoka}, Yoshiki and {Medezinski}, Elinor and {Mineo}, Sogo and {Miyama}, Shoken and {Miyatake}, Hironao and {Miyazaki}, Satoshi and {Momose}, Rieko and {More}, Anupreeta and {More}, Surhud and {Moritani}, Yuki and {Moriya}, Takashi J. and {Morokuma}, Tomoki and {Mukae}, Shiro and {Murata}, Ryoma and {Murayama}, Hitoshi and {Nagao}, Tohru and {Nakata}, Fumiaki and {Niida}, Mana and {Niikura}, Hiroko and {Nishizawa}, Atsushi J. and {Obuchi}, Yoshiyuki and {Oguri}, Masamune and {Oishi}, Yukie and {Okabe}, Nobuhiro and {Okamoto}, Sakurako and {Okura}, Yuki and {Ono}, Yoshiaki and {Onodera}, Masato and {Onoue}, Masafusa and {Osato}, Ken and {Ouchi}, Masami and {Price}, Paul A. and {Pyo}, Tae-Soo and {Sako}, Masao and {Sawicki}, Marcin and {Shibuya}, Takatoshi and {Shimasaku}, Kazuhiro and {Shimono}, Atsushi and {Shirasaki}, Masato and {Silverman}, John D. and {Simet}, Melanie and {Speagle}, Joshua and {Spergel}, David N. and {Strauss}, Michael A. and {Sugahara}, Yuma and {Sugiyama}, Naoshi and {Suto}, Yasushi and {Suyu}, Sherry H. and {Suzuki}, Nao and {Tait}, Philip J. and {Takada}, Masahiro and {Takata}, Tadafumi and {Tamura}, Naoyuki and {Tanaka}, Manobu M. and {Tanaka}, Masaomi and {Tanaka}, Masayuki and {Tanaka}, Yoko and {Terai}, Tsuyoshi and {Terashima}, Yuichi and {Toba}, Yoshiki and {Tominaga}, Nozomu and {Toshikawa}, Jun and {Turner}, Edwin L. and {Uchida}, Tomohisa and {Uchiyama}, Hisakazu and {Umetsu}, Keiichi and {Uraguchi}, Fumihiro and {Urata}, Yuji and {Usuda}, Tomonori and {Utsumi}, Yousuke and {Wang}, Shiang-Yu and {Wang}, Wei-Hao and {Wong}, Kenneth C. and {Yabe}, Kiyoto and {Yamada}, Yoshihiko and {Yamanoi}, Hitomi and {Yasuda}, Naoki and {Yeh}, Sherry and {Yonehara}, Atsunori and {Yuma}, Suraphong},
        title = "{The Hyper Suprime-Cam SSP Survey: Overview and survey design}",
      journal = {\pasj},
         year = 2018,
        month = jan,
       volume = {70},
          eid = {S4},
        pages = {S4},
          doi = {10.1093/pasj/psx066},
archivePrefix = {arXiv},
       eprint = {1704.05858},
 primaryClass = {astro-ph.IM},
       adsurl = {https://ui.adsabs.harvard.edu/abs/2018PASJ...70S...4A}
}

@ARTICLE{DESI2016,
       author = {{DESI Collaboration} and {Aghamousa}, Amir and {Aguilar}, Jessica and {Ahlen}, Steve and {Alam}, Shadab and {Allen}, Lori E. and {Allende Prieto}, Carlos and {Annis}, James and {Bailey}, Stephen and {Balland}, Christophe and {Ballester}, Otger and {Baltay}, Charles and {Beaufore}, Lucas and {Bebek}, Chris and {Beers}, Timothy C. and {Bell}, Eric F. and {Bernal}, Jos{\'e} Luis and {Besuner}, Robert and {Beutler}, Florian and {Blake}, Chris and {Bleuler}, Hannes and {Blomqvist}, Michael and {Blum}, Robert and {Bolton}, Adam S. and {Briceno}, Cesar and {Brooks}, David and {Brownstein}, Joel R. and {Buckley-Geer}, Elizabeth and {Burden}, Angela and {Burtin}, Etienne and {Busca}, Nicolas G. and {Cahn}, Robert N. and {Cai}, Yan-Chuan and {Cardiel-Sas}, Laia and {Carlberg}, Raymond G. and {Carton}, Pierre-Henri and {Casas}, Ricard and {Castander}, Francisco J. and {Cervantes-Cota}, Jorge L. and {Claybaugh}, Todd M. and {Close}, Madeline and {Coker}, Carl T. and {Cole}, Shaun and {Comparat}, Johan and {Cooper}, Andrew P. and {Cousinou}, M.-C. and {Crocce}, Martin and {Cuby}, Jean-Gabriel and {Cunningham}, Daniel P. and {Davis}, Tamara M. and {Dawson}, Kyle S. and {de la Macorra}, Axel and {De Vicente}, Juan and {Delubac}, Timoth{\'e}e and {Derwent}, Mark and {Dey}, Arjun and {Dhungana}, Govinda and {Ding}, Zhejie and {Doel}, Peter and {Duan}, Yutong T. and {Ealet}, Anne and {Edelstein}, Jerry and {Eftekharzadeh}, Sarah and {Eisenstein}, Daniel J. and {Elliott}, Ann and {Escoffier}, St{\'e}phanie and {Evatt}, Matthew and {Fagrelius}, Parker and {Fan}, Xiaohui and {Fanning}, Kevin and {Farahi}, Arya and {Farihi}, Jay and {Favole}, Ginevra and {Feng}, Yu and {Fernandez}, Enrique and {Findlay}, Joseph R. and {Finkbeiner}, Douglas P. and {Fitzpatrick}, Michael J. and {Flaugher}, Brenna and {Flender}, Samuel and {Font-Ribera}, Andreu and {Forero-Romero}, Jaime E. and {Fosalba}, Pablo and {Frenk}, Carlos S. and {Fumagalli}, Michele and {Gaensicke}, Boris T. and {Gallo}, Giuseppe and {Garcia-Bellido}, Juan and {Gaztanaga}, Enrique and {Pietro Gentile Fusillo}, Nicola and {Gerard}, Terry and {Gershkovich}, Irena and {Giannantonio}, Tommaso and {Gillet}, Denis and {Gonzalez-de-Rivera}, Guillermo and {Gonzalez-Perez}, Violeta and {Gott}, Shelby and {Graur}, Or and {Gutierrez}, Gaston and {Guy}, Julien and {Habib}, Salman and {Heetderks}, Henry and {Heetderks}, Ian and {Heitmann}, Katrin and {Hellwing}, Wojciech A. and {Herrera}, David A. and {Ho}, Shirley and {Holland}, Stephen and {Honscheid}, Klaus and {Huff}, Eric and {Hutchinson}, Timothy A. and {Huterer}, Dragan and {Hwang}, Ho Seong and {Illa Laguna}, Joseph Maria and {Ishikawa}, Yuzo and {Jacobs}, Dianna and {Jeffrey}, Niall and {Jelinsky}, Patrick and {Jennings}, Elise and {Jiang}, Linhua and {Jimenez}, Jorge and {Johnson}, Jennifer and {Joyce}, Richard and {Jullo}, Eric and {Juneau}, St{\'e}phanie and {Kama}, Sami and {Karcher}, Armin and {Karkar}, Sonia and {Kehoe}, Robert and {Kennamer}, Noble and {Kent}, Stephen and {Kilbinger}, Martin and {Kim}, Alex G. and {Kirkby}, David and {Kisner}, Theodore and {Kitanidis}, Ellie and {Kneib}, Jean-Paul and {Koposov}, Sergey and {Kovacs}, Eve and {Koyama}, Kazuya and {Kremin}, Anthony and {Kron}, Richard and {Kronig}, Luzius and {Kueter-Young}, Andrea and {Lacey}, Cedric G. and {Lafever}, Robin and {Lahav}, Ofer and {Lambert}, Andrew and {Lampton}, Michael and {Landriau}, Martin and {Lang}, Dustin and {Lauer}, Tod R. and {Le Goff}, Jean-Marc and {Le Guillou}, Laurent and {Le Van Suu}, Auguste and {Lee}, Jae Hyeon and {Lee}, Su-Jeong and {Leitner}, Daniela and {Lesser}, Michael and {Levi}, Michael E. and {L'Huillier}, Benjamin and {Li}, Baojiu and {Liang}, Ming and {Lin}, Huan and {Linder}, Eric and {Loebman}, Sarah R. and {Luki{\'c}}, Zarija and {Ma}, Jun and {MacCrann}, Niall and {Magneville}, Christophe and {Makarem}, Laleh and {Manera}, Marc and {Manser}, Christopher J. and {Marshall}, Robert and {Martini}, Paul and {Massey}, Richard and {Matheson}, Thomas and {McCauley}, Jeremy and {McDonald}, Patrick and {McGreer}, Ian D. and {Meisner}, Aaron and {Metcalfe}, Nigel and {Miller}, Timothy N. and {Miquel}, Ramon and {Moustakas}, John and {Myers}, Adam and {Naik}, Milind and {Newman}, Jeffrey A. and {Nichol}, Robert C. and {Nicola}, Andrina and {Nicolati da Costa}, Luiz and {Nie}, Jundan and {Niz}, Gustavo and {Norberg}, Peder and {Nord}, Brian and {Norman}, Dara and {Nugent}, Peter and {O'Brien}, Thomas and {Oh}, Minji and {Olsen}, Knut A.~G.},
        title = "{The DESI Experiment Part II: Instrument Design}",
      journal = {arXiv e-prints},
         year = 2016,
        month = oct,
          eid = {arXiv:1611.00037},
        pages = {arXiv:1611.00037},
          doi = {10.48550/arXiv.1611.00037},
archivePrefix = {arXiv},
       eprint = {1611.00037},
 primaryClass = {astro-ph.IM},
       adsurl = {https://ui.adsabs.harvard.edu/abs/2016arXiv161100037D}
}

@ARTICLE{DESI2022,
       author = {{DESI Collaboration} and {Abareshi}, B. and {Aguilar}, J. and {Ahlen}, S. and {Alam}, Shadab and {Alexander}, David M. and {Alfarsy}, R. and {Allen}, L. and {Allende Prieto}, C. and {Alves}, O. and {Ameel}, J. and {Armengaud}, E. and {Asorey}, J. and {Aviles}, Alejandro and {Bailey}, S. and {Balaguera-Antol{\'\i}nez}, A. and {Ballester}, O. and {Baltay}, C. and {Bault}, A. and {Beltran}, S.~F. and {Benavides}, B. and {BenZvi}, S. and {Berti}, A. and {Besuner}, R. and {Beutler}, Florian and {Bianchi}, D. and {Blake}, C. and {Blanc}, P. and {Blum}, R. and {Bolton}, A. and {Bose}, S. and {Bramall}, D. and {Brieden}, S. and {Brodzeller}, A. and {Brooks}, D. and {Brownewell}, C. and {Buckley-Geer}, E. and {Cahn}, R.~N. and {Cai}, Z. and {Canning}, R. and {Capasso}, R. and {Carnero Rosell}, A. and {Carton}, P. and {Casas}, R. and {Castander}, F.~J. and {Cervantes-Cota}, J.~L. and {Chabanier}, S. and {Chaussidon}, E. and {Chuang}, C. and {Circosta}, C. and {Cole}, S. and {Cooper}, A.~P. and {da Costa}, L. and {Cousinou}, M.-C. and {Cuceu}, A. and {Davis}, T.~M. and {Dawson}, K. and {de la Cruz-Noriega}, R. and {de la Macorra}, A. and {de Mattia}, A. and {Della Costa}, J. and {Demmer}, P. and {Derwent}, M. and {Dey}, A. and {Dey}, B. and {Dhungana}, G. and {Ding}, Z. and {Dobson}, C. and {Doel}, P. and {Donald-McCann}, J. and {Donaldson}, J. and {Douglass}, K. and {Duan}, Y. and {Dunlop}, P. and {Edelstein}, J. and {Eftekharzadeh}, S. and {Eisenstein}, D.~J. and {Enriquez-Vargas}, M. and {Escoffier}, S. and {Evatt}, M. and {Fagrelius}, P. and {Fan}, X. and {Fanning}, K. and {Fawcett}, V.~A. and {Ferraro}, S. and {Ereza}, J. and {Flaugher}, B. and {Font-Ribera}, A. and {Forero-Romero}, J.~E. and {Frenk}, C.~S. and {Fromenteau}, S. and {G{\"a}nsicke}, B.~T. and {Garcia-Quintero}, C. and {Garrison}, L. and {Gazta{\~n}aga}, E. and {Gerardi}, F. and {Gil-Mar{\'\i}n}, H. and {Gontcho A Gontcho}, S. and {Gonzalez-Morales}, Alma X. and {Gonzalez-de-Rivera}, G. and {Gonzalez-Perez}, V. and {Gordon}, C. and {Graur}, O. and {Green}, D. and {Grove}, C. and {Gruen}, D. and {Gutierrez}, G. and {Guy}, J. and {Hahn}, C. and {Harris}, S. and {Herrera}, D. and {Herrera-Alcantar}, Hiram K. and {Honscheid}, K. and {Howlett}, C. and {Huterer}, D. and {Ir{\v{s}}i{\v{c}}}, V. and {Ishak}, M. and {Jelinsky}, P. and {Jiang}, L. and {Jimenez}, J. and {Jing}, Y.~P. and {Joyce}, R. and {Jullo}, E. and {Juneau}, S. and {Kara{\c{c}}ayl{\i}}, N.~G. and {Karamanis}, M. and {Karcher}, A. and {Karim}, T. and {Kehoe}, R. and {Kent}, S. and {Kirkby}, D. and {Kisner}, T. and {Kitaura}, F. and {Koposov}, S.~E. and {Kov{\'a}cs}, A. and {Kremin}, A. and {Krolewski}, Alex and {L'Huillier}, B. and {Lahav}, O. and {Lambert}, A. and {Lamman}, C. and {Lan}, Ting-Wen and {Landriau}, M. and {Lane}, S. and {Lang}, D. and {Lange}, J.~U. and {Lasker}, J. and {Le Guillou}, L. and {Leauthaud}, A. and {Le Van Suu}, A. and {Levi}, Michael E. and {Li}, T.~S. and {Magneville}, C. and {Manera}, M. and {Manser}, Christopher J. and {Marshall}, B. and {Martini}, Paul and {McCollam}, W. and {McDonald}, P. and {Meisner}, Aaron M. and {Mena-Fern{\'a}ndez}, J. and {Meneses-Rizo}, J. and {Mezcua}, M. and {Miller}, T. and {Miquel}, R. and {Montero-Camacho}, P. and {Moon}, J. and {Moustakas}, J. and {Mueller}, E. and {Mu{\~n}oz-Guti{\'e}rrez}, Andrea and {Myers}, Adam D. and {Nadathur}, S. and {Najita}, J. and {Napolitano}, L. and {Neilsen}, E. and {Newman}, Jeffrey A. and {Nie}, J.~D. and {Ning}, Y. and {Niz}, G. and {Norberg}, P. and {Noriega}, Hern{\'a}n E. and {O'Brien}, T. and {Obuljen}, A. and {Palanque-Delabrouille}, N. and {Palmese}, A. and {Zhiwei}, P. and {Pappalardo}, D. and {PENG}, X. and {Percival}, W.~J. and {Perruchot}, S. and {Pogge}, R. and {Poppett}, C. and {Porredon}, A. and {Prada}, F. and {Prochaska}, J. and {Pucha}, R. and {P{\'e}rez-Fern{\'a}ndez}, A. and {P{\'e}rez-R{\`a}fols}, I. and {Rabinowitz}, D. and {Raichoor}, A.},
        title = "{Overview of the Instrumentation for the Dark Energy Spectroscopic Instrument}",
      journal = {\aj},
         year = 2022,
        month = nov,
       volume = {164},
       number = {5},
          eid = {207},
        pages = {207},
          doi = {10.3847/1538-3881/ac882b},
archivePrefix = {arXiv},
       eprint = {2205.10939},
 primaryClass = {astro-ph.IM},
       adsurl = {https://ui.adsabs.harvard.edu/abs/2022AJ....164..207D}
}

@ARTICLE{DESI2024,
       author = {{DESI Collaboration} and {Adame}, A.~G. and {Aguilar}, J. and {Ahlen}, S. and {Alam}, S. and {Aldering}, G. and {Alexander}, D.~M. and {Alfarsy}, R. and {Allende Prieto}, C. and {Alvarez}, M. and {Alves}, O. and {Anand}, A. and {Andrade-Oliveira}, F. and {Armengaud}, E. and {Asorey}, J. and {Avila}, S. and {Aviles}, A. and {Bailey}, S. and {Balaguera-Antol{\'\i}nez}, A. and {Ballester}, O. and {Baltay}, C. and {Bault}, A. and {Bautista}, J. and {Behera}, J. and {Beltran}, S.~F. and {BenZvi}, S. and {Beraldo e Silva}, L. and {Bermejo-Climent}, J.~R. and {Berti}, A. and {Besuner}, R. and {Beutler}, F. and {Bianchi}, D. and {Blake}, C. and {Blum}, R. and {Bolton}, A.~S. and {Brieden}, S. and {Brodzeller}, A. and {Brooks}, D. and {Brown}, Z. and {Buckley-Geer}, E. and {Burtin}, E. and {Cabayol-Garcia}, L. and {Cai}, Z. and {Canning}, R. and {Cardiel-Sas}, L. and {Carnero Rosell}, A. and {Castander}, F.~J. and {Cervantes-Cota}, J.~L. and {Chabanier}, S. and {Chaussidon}, E. and {Chaves-Montero}, J. and {Chen}, S. and {Chen}, X. and {Chuang}, C. and {Claybaugh}, T. and {Cole}, S. and {Cooper}, A.~P. and {Cuceu}, A. and {Davis}, T.~M. and {Dawson}, K. and {de Belsunce}, R. and {de la Cruz}, R. and {de la Macorra}, A. and {de Mattia}, A. and {Demina}, R. and {Demirbozan}, U. and {DeRose}, J. and {Dey}, A. and {Dey}, B. and {Dhungana}, G. and {Ding}, J. and {Ding}, Z. and {Doel}, P. and {Doshi}, R. and {Douglass}, K. and {Edge}, A. and {Eftekharzadeh}, S. and {Eisenstein}, D.~J. and {Elliott}, A. and {Escoffier}, S. and {Fagrelius}, P. and {Fan}, X. and {Fanning}, K. and {Fawcett}, V.~A. and {Ferraro}, S. and {Ereza}, J. and {Flaugher}, B. and {Font-Ribera}, A. and {Forero-S{\'a}nchez}, D. and {Forero-Romero}, J.~E. and {Frenk}, C.~S. and {G{\"a}nsicke}, B.~T. and {Garc{\'\i}a}, L. {\'A}. and {Garc{\'\i}a-Bellido}, J. and {Garcia-Quintero}, C. and {Garrison}, L.~H. and {Gil-Mar{\'\i}n}, H. and {Golden-Marx}, J. and {Gontcho A Gontcho}, S. and {Gonzalez-Morales}, A.~X. and {Gonzalez-Perez}, V. and {Gordon}, C. and {Graur}, O. and {Green}, D. and {Gruen}, D. and {Guy}, J. and {Hadzhiyska}, B. and {Hahn}, C. and {Han}, J.~J. and {Hanif}, M.~M.~S. and {Herrera-Alcantar}, H.~K. and {Honscheid}, K. and {Hou}, J. and {Howlett}, C. and {Huterer}, D. and {Ir{\v{s}}i{\v{c}}}, V. and {Ishak}, M. and {Jana}, A. and {Jiang}, L. and {Jimenez}, J. and {Jing}, Y.~P. and {Joudaki}, S. and {Jullo}, E. and {Joyce}, R. and {Juneau}, S. and {Kizhuprakkat}, N. and {Kara{\c{c}}ayl{\i}}, N.~G. and {Karim}, T. and {Kehoe}, R. and {Kent}, S. and {Khederlarian}, A. and {Kim}, S. and {Kirkby}, D. and {Kisner}, T. and {Kitaura}, F. and {Kneib}, J. and {Koposov}, S.~E. and {Kov{\'a}cs}, A. and {Kremin}, A. and {Krolewski}, A. and {L'Huillier}, B. and {Lahav}, O. and {Lambert}, A. and {Lamman}, C. and {Lan}, T.-W. and {Landriau}, M. and {Lang}, D. and {Lange}, J.~U. and {Lasker}, J. and {Le Guillou}, L. and {Leauthaud}, A. and {Levi}, M.~E. and {Li}, T.~S. and {Linder}, E. and {Lyons}, A. and {Magneville}, C. and {Manera}, M. and {Manser}, C.~J. and {Margala}, D. and {Martini}, P. and {McDonald}, P. and {Medina}, G.~E. and {Medina-Varela}, L. and {Meisner}, A. and {Mena-Fern{\'a}ndez}, J. and {Meneses-Rizo}, J. and {Mezcua}, M. and {Miquel}, R. and {Montero-Camacho}, P. and {Moon}, J. and {Moore}, S. and {Moustakas}, J. and {Mueller}, E. and {Mundet}, J. and {Mu{\~n}oz-Guti{\'e}rrez}, A. and {Myers}, A.~D. and {Nadathur}, S. and {Napolitano}, L. and {Neveux}, R. and {Newman}, J.~A. and {Nie}, J. and {Niz}, G. and {Norberg}, P. and {Noriega}, H.~E. and {Paillas}, E. and {Palanque-Delabrouille}, N. and {Palmese}, A. and {Zhiwei}, P. and {Parkinson}, D. and {Penmetsa}, S. and {Percival}, W.~J. and {P{\'e}rez-Fern{\'a}ndez}, A. and {P{\'e}rez-R{\`a}fols}, I. and {Pieri}, M. and {Poppett}, C. and {Porredon}, A. and {Prada}, F. and {Pucha}, R. and {Raichoor}, A. and {Ram{\'\i}rez-P{\'e}rez}, C.},
        title = "{Validation of the Scientific Program for the Dark Energy Spectroscopic Instrument}",
      journal = {\aj},
         year = 2024,
        month = feb,
       volume = {167},
       number = {2},
          eid = {62},
        pages = {62},
          doi = {10.3847/1538-3881/ad0b08},
archivePrefix = {arXiv},
       eprint = {2306.06307},
 primaryClass = {astro-ph.CO},
       adsurl = {https://ui.adsabs.harvard.edu/abs/2024AJ....167...62D}
}

@ARTICLE{Hahn2023,
       author = {{Hahn}, ChangHoon and {Wilson}, Michael J. and {Ruiz-Macias}, Omar and {Cole}, Shaun and {Weinberg}, David H. and {Moustakas}, John and {Kremin}, Anthony and {Tinker}, Jeremy L. and {Smith}, Alex and {Wechsler}, Risa H. and {Ahlen}, Steven and {Alam}, Shadab and {Bailey}, Stephen and {Brooks}, David and {Cooper}, Andrew P. and {Davis}, Tamara M. and {Dawson}, Kyle and {Dey}, Arjun and {Dey}, Biprateep and {Eftekharzadeh}, Sarah and {Eisenstein}, Daniel J. and {Fanning}, Kevin and {Forero-Romero}, Jaime E. and {Frenk}, Carlos S. and {Gazta{\~n}aga}, Enrique and {A Gontcho}, Satya Gontcho and {Guy}, Julien and {Honscheid}, Klaus and {Ishak}, Mustapha and {Juneau}, St{\'e}phanie and {Kehoe}, Robert and {Kisner}, Theodore and {Lan}, Ting-Wen and {Landriau}, Martin and {Le Guillou}, Laurent and {Levi}, Michael E. and {Magneville}, Christophe and {Martini}, Paul and {Meisner}, Aaron and {Myers}, Adam D. and {Nie}, Jundan and {Norberg}, Peder and {Palanque-Delabrouille}, Nathalie and {Percival}, Will J. and {Poppett}, Claire and {Prada}, Francisco and {Raichoor}, Anand and {Ross}, Ashley J. and {Gaines}, Sasha and {Saulder}, Christoph and {Schlafly}, Eddie and {Schlegel}, David and {Sierra-Porta}, David and {Tarle}, Gregory and {Weaver}, Benjamin A. and {Y{\`e}che}, Christophe and {Zarrouk}, Pauline and {Zhou}, Rongpu and {Zhou}, Zhimin and {Zou}, Hu},
        title = "{The DESI Bright Galaxy Survey: Final Target Selection, Design, and Validation}",
      journal = {\aj},
         year = 2023,
        month = jun,
       volume = {165},
       number = {6},
          eid = {253},
        pages = {253},
          doi = {10.3847/1538-3881/accff8},
archivePrefix = {arXiv},
       eprint = {2208.08512},
 primaryClass = {astro-ph.CO},
       adsurl = {https://ui.adsabs.harvard.edu/abs/2023AJ....165..253H}
}

@ARTICLE{DiCintio2014,
       author = {{Di Cintio}, Arianna and {Brook}, Chris B. and {Macci{\`o}}, Andrea V. and {Stinson}, Greg S. and {Knebe}, Alexander and {Dutton}, Aaron A. and {Wadsley}, James},
        title = "{The dependence of dark matter profiles on the stellar-to-halo mass ratio: a prediction for cusps versus cores}",
      journal = {\mnras},
         year = 2014,
        month = jan,
       volume = {437},
       number = {1},
        pages = {415-423},
          doi = {10.1093/mnras/stt1891},
archivePrefix = {arXiv},
       eprint = {1306.0898},
 primaryClass = {astro-ph.CO},
       adsurl = {https://ui.adsabs.harvard.edu/abs/2014MNRAS.437..415D}
}

@ARTICLE{Dutton2019,
       author = {{Dutton}, Aaron A. and {Macci{\`o}}, Andrea V. and {Buck}, Tobias and {Dixon}, Keri L. and {Blank}, Marvin and {Obreja}, Aura},
        title = "{NIHAO XX: the impact of the star formation threshold on the cusp-core transformation of cold dark matter haloes}",
      journal = {\mnras},
         year = 2019,
        month = jun,
       volume = {486},
       number = {1},
        pages = {655-671},
          doi = {10.1093/mnras/stz889},
archivePrefix = {arXiv},
       eprint = {1811.10625},
 primaryClass = {astro-ph.GA},
       adsurl = {https://ui.adsabs.harvard.edu/abs/2019MNRAS.486..655D}
}

@ARTICLE{Barbary2016,
       author = {{Barbary}, Kyle},
        title = "{SEP: Source Extractor as a library}",
      journal = {The Journal of Open Source Software},
         year = 2016,
        month = oct,
       volume = {1},
       number = {6},
          eid = {58},
        pages = {58},
          doi = {10.21105/joss.00058},
       adsurl = {https://ui.adsabs.harvard.edu/abs/2016JOSS....1...58B}
}

@ARTICLE{Bertin1996,
       author = {{Bertin}, E. and {Arnouts}, S.},
        title = "{SExtractor: Software for source extraction.}",
      journal = {\aaps},
         year = 1996,
        month = jun,
       volume = {117},
        pages = {393-404},
          doi = {10.1051/aas:1996164},
       adsurl = {https://ui.adsabs.harvard.edu/abs/1996A&AS..117..393B}
}

@inproceedings{statsmodel,
  title={statsmodels: Econometric and statistical modeling with python},
  author={Seabold, Skipper and Perktold, Josef},
  booktitle={9th Python in Science Conference},
  year={2010},
}

@misc{fastspecfit,
       author = {{Moustakas}, John and {Buhler}, Jeremy and {Scholte}, Dirk and {Dey}, Biprateep and {Khederlarian}, Ashod},
        title = "{FastSpecFit: Fast spectral synthesis and emission-line fitting of DESI spectra}",
 howpublished = {Astrophysics Source Code Library, record ascl:2308.005},
         year = 2023,
        month = aug,
          eid = {ascl:2308.005},
archivePrefix = {ascl},
       eprint = {2308.005},
       adsurl = {https://ui.adsabs.harvard.edu/abs/2023ascl.soft08005M}
}

@ARTICLE{GildePaz2003,
       author = {{Gil de Paz}, A. and {Madore}, B.~F. and {Pevunova}, O.},
        title = "{Palomar/Las Campanas Imaging Atlas of Blue Compact Dwarf Galaxies. I. Images and Integrated Photometry}",
      journal = {\apjs},
         year = 2003,
        month = jul,
       volume = {147},
       number = {1},
        pages = {29-59},
          doi = {10.1086/374737},
archivePrefix = {arXiv},
       eprint = {astro-ph/0302221},
 primaryClass = {astro-ph},
       adsurl = {https://ui.adsabs.harvard.edu/abs/2003ApJS..147...29G}
}

@ARTICLE{Gavazzi2012,
       author = {{Gavazzi}, G. and {Fumagalli}, M. and {Galardo}, V. and {Grossetti}, F. and {Boselli}, A. and {Giovanelli}, R. and {Haynes}, M.~P. and {Fabello}, S.},
        title = "{H{\ensuremath{\alpha}}3: an H{\ensuremath{\alpha}} imaging survey of HI selected galaxies from ALFALFA. I. Catalogue in the Local Supercluster}",
      journal = {\aap},
         year = 2012,
        month = sep,
       volume = {545},
          eid = {A16},
        pages = {A16},
          doi = {10.1051/0004-6361/201218788},
archivePrefix = {arXiv},
       eprint = {1206.0061},
 primaryClass = {astro-ph.CO},
       adsurl = {https://ui.adsabs.harvard.edu/abs/2012A&A...545A..16G}
}

@ARTICLE{Hopkins2013,
       author = {{Hopkins}, A.~M. and {Driver}, S.~P. and {Brough}, S. and {Owers}, M.~S. and {Bauer}, A.~E. and {Gunawardhana}, M.~L.~P. and {Cluver}, M.~E. and {Colless}, M. and {Foster}, C. and {Lara-L{\'o}pez}, M.~A. and {Roseboom}, I. and {Sharp}, R. and {Steele}, O. and {Thomas}, D. and {Baldry}, I.~K. and {Brown}, M.~J.~I. and {Liske}, J. and {Norberg}, P. and {Robotham}, A.~S.~G. and {Bamford}, S. and {Bland-Hawthorn}, J. and {Drinkwater}, M.~J. and {Loveday}, J. and {Meyer}, M. and {Peacock}, J.~A. and {Tuffs}, R. and {Agius}, N. and {Alpaslan}, M. and {Andrae}, E. and {Cameron}, E. and {Cole}, S. and {Ching}, J.~H.~Y. and {Christodoulou}, L. and {Conselice}, C. and {Croom}, S. and {Cross}, N.~J.~G. and {De Propris}, R. and {Delhaize}, J. and {Dunne}, L. and {Eales}, S. and {Ellis}, S. and {Frenk}, C.~S. and {Graham}, Alister W. and {Grootes}, M.~W. and {H{\"a}u{\ss}ler}, B. and {Heymans}, C. and {Hill}, D. and {Hoyle}, B. and {Hudson}, M. and {Jarvis}, M. and {Johansson}, J. and {Jones}, D.~H. and {van Kampen}, E. and {Kelvin}, L. and {Kuijken}, K. and {L{\'o}pez-S{\'a}nchez}, {\'A}. and {Maddox}, S. and {Madore}, B. and {Maraston}, C. and {McNaught-Roberts}, T. and {Nichol}, R.~C. and {Oliver}, S. and {Parkinson}, H. and {Penny}, S. and {Phillipps}, S. and {Pimbblet}, K.~A. and {Ponman}, T. and {Popescu}, C.~C. and {Prescott}, M. and {Proctor}, R. and {Sadler}, E.~M. and {Sansom}, A.~E. and {Seibert}, M. and {Staveley-Smith}, L. and {Sutherland}, W. and {Taylor}, E. and {Van Waerbeke}, L. and {V{\'a}zquez-Mata}, J.~A. and {Warren}, S. and {Wijesinghe}, D.~B. and {Wild}, V. and {Wilkins}, S.},
        title = "{Galaxy And Mass Assembly (GAMA): spectroscopic analysis}",
      journal = {\mnras},
         year = 2013,
        month = apr,
       volume = {430},
       number = {3},
        pages = {2047-2066},
          doi = {10.1093/mnras/stt030},
archivePrefix = {arXiv},
       eprint = {1301.7127},
 primaryClass = {astro-ph.CO},
       adsurl = {https://ui.adsabs.harvard.edu/abs/2013MNRAS.430.2047H}
}

@ARTICLE{Emami2019,
       author = {{Emami}, Najmeh and {Siana}, Brian and {Weisz}, Daniel R. and {Johnson}, Benjamin D. and {Ma}, Xiangcheng and {El-Badry}, Kareem},
        title = "{A Closer Look at Bursty Star Formation with L $_{H{\ensuremath{\alpha}} }$ and L $_{UV}$ Distributions}",
      journal = {\apj},
         year = 2019,
        month = aug,
       volume = {881},
       number = {1},
          eid = {71},
        pages = {71},
          doi = {10.3847/1538-4357/ab211a},
archivePrefix = {arXiv},
       eprint = {1809.06380},
 primaryClass = {astro-ph.GA},
       adsurl = {https://ui.adsabs.harvard.edu/abs/2019ApJ...881...71E}
}

@ARTICLE{Weisz2012,
       author = {{Weisz}, Daniel R. and {Johnson}, Benjamin D. and {Johnson}, L. Clifton and {Skillman}, Evan D. and {Lee}, Janice C. and {Kennicutt}, Robert C. and {Calzetti}, Daniela and {van Zee}, Liese and {Bothwell}, Matthew S. and {Dalcanton}, Julianne J. and {Dale}, Daniel A. and {Williams}, Benjamin F.},
        title = "{Modeling the Effects of Star Formation Histories on H{\ensuremath{\alpha}} and Ultraviolet Fluxes in nearby Dwarf Galaxies}",
      journal = {\apj},
         year = 2012,
        month = jan,
       volume = {744},
       number = {1},
          eid = {44},
        pages = {44},
          doi = {10.1088/0004-637X/744/1/44},
archivePrefix = {arXiv},
       eprint = {1109.2905},
 primaryClass = {astro-ph.CO},
       adsurl = {https://ui.adsabs.harvard.edu/abs/2012ApJ...744...44W}
}

@ARTICLE{Collins2022,
       author = {{Collins}, Michelle L.~M. and {Read}, Justin I.},
        title = "{Observational constraints on stellar feedback in dwarf galaxies}",
      journal = {Nature Astronomy},
         year = 2022,
        month = may,
       volume = {6},
        pages = {647-658},
          doi = {10.1038/s41550-022-01657-4},
archivePrefix = {arXiv},
       eprint = {2205.06825},
 primaryClass = {astro-ph.GA},
       adsurl = {https://ui.adsabs.harvard.edu/abs/2022NatAs...6..647C}
}

@ARTICLE{Sparre2017,
       author = {{Sparre}, Martin and {Hayward}, Christopher C. and {Feldmann}, Robert and {Faucher-Gigu{\`e}re}, Claude-Andr{\'e} and {Muratov}, Alexander L. and {Kere{\v{s}}}, Du{\v{s}}an and {Hopkins}, Philip F.},
        title = "{(Star)bursts of FIRE: observational signatures of bursty star formation in galaxies}",
      journal = {\mnras},
         year = 2017,
        month = apr,
       volume = {466},
       number = {1},
        pages = {88-104},
          doi = {10.1093/mnras/stw3011},
archivePrefix = {arXiv},
       eprint = {1510.03869},
 primaryClass = {astro-ph.GA},
       adsurl = {https://ui.adsabs.harvard.edu/abs/2017MNRAS.466...88S}
}

@ARTICLE{El-Badry2016,
       author = {{El-Badry}, Kareem and {Wetzel}, Andrew and {Geha}, Marla and {Hopkins}, Philip F. and {Kere{\v{s}}}, Dusan and {Chan}, T.~K. and {Faucher-Gigu{\`e}re}, Claude-Andr{\'e}},
        title = "{Breathing FIRE: How Stellar Feedback Drives Radial Migration, Rapid Size Fluctuations, and Population Gradients in Low-mass Galaxies}",
      journal = {\apj},
         year = 2016,
        month = apr,
       volume = {820},
       number = {2},
          eid = {131},
        pages = {131},
          doi = {10.3847/0004-637X/820/2/131},
archivePrefix = {arXiv},
       eprint = {1512.01235},
 primaryClass = {astro-ph.GA},
       adsurl = {https://ui.adsabs.harvard.edu/abs/2016ApJ...820..131E}
}

@ARTICLE{Teyssier2013,
       author = {{Teyssier}, Romain and {Pontzen}, Andrew and {Dubois}, Yohan and {Read}, Justin I.},
        title = "{Cusp-core transformations in dwarf galaxies: observational predictions}",
      journal = {\mnras},
         year = 2013,
        month = mar,
       volume = {429},
       number = {4},
        pages = {3068-3078},
          doi = {10.1093/mnras/sts563},
archivePrefix = {arXiv},
       eprint = {1206.4895},
 primaryClass = {astro-ph.CO},
       adsurl = {https://ui.adsabs.harvard.edu/abs/2013MNRAS.429.3068T}
}

@ARTICLE{Cenci2024,
       author = {{Cenci}, Elia and {Feldmann}, Robert and {Gensior}, Jindra and {Moreno}, Jorge and {Bassini}, Luigi and {Bernardini}, Mauro},
        title = "{Starbursts driven by central gas compaction}",
      journal = {\mnras},
         year = 2024,
        month = jan,
       volume = {527},
       number = {3},
        pages = {7871-7890},
          doi = {10.1093/mnras/stad3709},
archivePrefix = {arXiv},
       eprint = {2309.09046},
 primaryClass = {astro-ph.GA},
       adsurl = {https://ui.adsabs.harvard.edu/abs/2024MNRAS.527.7871C}
}

@ARTICLE{Hinshaw2013,
       author = {{Hinshaw}, G. and {Larson}, D. and {Komatsu}, E. and {Spergel}, D.~N. and {Bennett}, C.~L. and {Dunkley}, J. and {Nolta}, M.~R. and {Halpern}, M. and {Hill}, R.~S. and {Odegard}, N. and {Page}, L. and {Smith}, K.~M. and {Weiland}, J.~L. and {Gold}, B. and {Jarosik}, N. and {Kogut}, A. and {Limon}, M. and {Meyer}, S.~S. and {Tucker}, G.~S. and {Wollack}, E. and {Wright}, E.~L.},
        title = "{Nine-year Wilkinson Microwave Anisotropy Probe (WMAP) Observations: Cosmological Parameter Results}",
      journal = {\apjs},
         year = 2013,
        month = oct,
       volume = {208},
       number = {2},
          eid = {19},
        pages = {19},
          doi = {10.1088/0067-0049/208/2/19},
archivePrefix = {arXiv},
       eprint = {1212.5226},
 primaryClass = {astro-ph.CO},
       adsurl = {https://ui.adsabs.harvard.edu/abs/2013ApJS..208...19H}
}

@ARTICLE{Oke1983,
       author = {{Oke}, J.~B. and {Gunn}, J.~E.},
        title = "{Secondary standard stars for absolute spectrophotometry.}",
      journal = {\apj},
         year = 1983,
        month = mar,
       volume = {266},
        pages = {713-717},
          doi = {10.1086/160817},
       adsurl = {https://ui.adsabs.harvard.edu/abs/1983ApJ...266..713O}
}

@ARTICLE{Harris2020,
       author = {{Harris}, Charles R. and {Millman}, K. Jarrod and {van der Walt}, St{\'e}fan J. and {Gommers}, Ralf and {Virtanen}, Pauli and {Cournapeau}, David and {Wieser}, Eric and {Taylor}, Julian and {Berg}, Sebastian and {Smith}, Nathaniel J. and {Kern}, Robert and {Picus}, Matti and {Hoyer}, Stephan and {van Kerkwijk}, Marten H. and {Brett}, Matthew and {Haldane}, Allan and {del R{\'\i}o}, Jaime Fern{\'a}ndez and {Wiebe}, Mark and {Peterson}, Pearu and {G{\'e}rard-Marchant}, Pierre and {Sheppard}, Kevin and {Reddy}, Tyler and {Weckesser}, Warren and {Abbasi}, Hameer and {Gohlke}, Christoph and {Oliphant}, Travis E.},
        title = "{Array programming with NumPy}",
      journal = {\nat},
         year = 2020,
        month = sep,
       volume = {585},
       number = {7825},
        pages = {357-362},
          doi = {10.1038/s41586-020-2649-2},
archivePrefix = {arXiv},
       eprint = {2006.10256},
 primaryClass = {cs.MS},
       adsurl = {https://ui.adsabs.harvard.edu/abs/2020Natur.585..357H}
}

@ARTICLE{Astropy2018,
       author = {{Astropy Collaboration} and {Price-Whelan}, A.~M. and {Sip{\H{o}}cz}, B.~M. and {G{\"u}nther}, H.~M. and {Lim}, P.~L. and {Crawford}, S.~M. and {Conseil}, S. and {Shupe}, D.~L. and {Craig}, M.~W. and {Dencheva}, N. and {Ginsburg}, A. and {VanderPlas}, J.~T. and {Bradley}, L.~D. and {P{\'e}rez-Su{\'a}rez}, D. and {de Val-Borro}, M. and {Aldcroft}, T.~L. and {Cruz}, K.~L. and {Robitaille}, T.~P. and {Tollerud}, E.~J. and {Ardelean}, C. and {Babej}, T. and {Bach}, Y.~P. and {Bachetti}, M. and {Bakanov}, A.~V. and {Bamford}, S.~P. and {Barentsen}, G. and {Barmby}, P. and {Baumbach}, A. and {Berry}, K.~L. and {Biscani}, F. and {Boquien}, M. and {Bostroem}, K.~A. and {Bouma}, L.~G. and {Brammer}, G.~B. and {Bray}, E.~M. and {Breytenbach}, H. and {Buddelmeijer}, H. and {Burke}, D.~J. and {Calderone}, G. and {Cano Rodr{\'\i}guez}, J.~L. and {Cara}, M. and {Cardoso}, J.~V.~M. and {Cheedella}, S. and {Copin}, Y. and {Corrales}, L. and {Crichton}, D. and {D'Avella}, D. and {Deil}, C. and {Depagne}, {\'E}. and {Dietrich}, J.~P. and {Donath}, A. and {Droettboom}, M. and {Earl}, N. and {Erben}, T. and {Fabbro}, S. and {Ferreira}, L.~A. and {Finethy}, T. and {Fox}, R.~T. and {Garrison}, L.~H. and {Gibbons}, S.~L.~J. and {Goldstein}, D.~A. and {Gommers}, R. and {Greco}, J.~P. and {Greenfield}, P. and {Groener}, A.~M. and {Grollier}, F. and {Hagen}, A. and {Hirst}, P. and {Homeier}, D. and {Horton}, A.~J. and {Hosseinzadeh}, G. and {Hu}, L. and {Hunkeler}, J.~S. and {Ivezi{\'c}}, {\v{Z}}. and {Jain}, A. and {Jenness}, T. and {Kanarek}, G. and {Kendrew}, S. and {Kern}, N.~S. and {Kerzendorf}, W.~E. and {Khvalko}, A. and {King}, J. and {Kirkby}, D. and {Kulkarni}, A.~M. and {Kumar}, A. and {Lee}, A. and {Lenz}, D. and {Littlefair}, S.~P. and {Ma}, Z. and {Macleod}, D.~M. and {Mastropietro}, M. and {McCully}, C. and {Montagnac}, S. and {Morris}, B.~M. and {Mueller}, M. and {Mumford}, S.~J. and {Muna}, D. and {Murphy}, N.~A. and {Nelson}, S. and {Nguyen}, G.~H. and {Ninan}, J.~P. and {N{\"o}the}, M. and {Ogaz}, S. and {Oh}, S. and {Parejko}, J.~K. and {Parley}, N. and {Pascual}, S. and {Patil}, R. and {Patil}, A.~A. and {Plunkett}, A.~L. and {Prochaska}, J.~X. and {Rastogi}, T. and {Reddy Janga}, V. and {Sabater}, J. and {Sakurikar}, P. and {Seifert}, M. and {Sherbert}, L.~E. and {Sherwood-Taylor}, H. and {Shih}, A.~Y. and {Sick}, J. and {Silbiger}, M.~T. and {Singanamalla}, S. and {Singer}, L.~P. and {Sladen}, P.~H. and {Sooley}, K.~A. and {Sornarajah}, S. and {Streicher}, O. and {Teuben}, P. and {Thomas}, S.~W. and {Tremblay}, G.~R. and {Turner}, J.~E.~H. and {Terr{\'o}n}, V. and {van Kerkwijk}, M.~H. and {de la Vega}, A. and {Watkins}, L.~L. and {Weaver}, B.~A. and {Whitmore}, J.~B. and {Woillez}, J. and {Zabalza}, V. and {Astropy Contributors}},
        title = "{The Astropy Project: Building an Open-science Project and Status of the v2.0 Core Package}",
      journal = {\aj},
         year = 2018,
        month = sep,
       volume = {156},
       number = {3},
          eid = {123},
        pages = {123},
          doi = {10.3847/1538-3881/aabc4f},
archivePrefix = {arXiv},
       eprint = {1801.02634},
 primaryClass = {astro-ph.IM},
       adsurl = {https://ui.adsabs.harvard.edu/abs/2018AJ....156..123A}
}

@ARTICLE{Astropy2022,
       author = {{Astropy Collaboration} and {Price-Whelan}, Adrian M. and {Lim}, Pey Lian and {Earl}, Nicholas and {Starkman}, Nathaniel and {Bradley}, Larry and {Shupe}, David L. and {Patil}, Aarya A. and {Corrales}, Lia and {Brasseur}, C.~E. and {N{\"o}the}, Maximilian and {Donath}, Axel and {Tollerud}, Erik and {Morris}, Brett M. and {Ginsburg}, Adam and {Vaher}, Eero and {Weaver}, Benjamin A. and {Tocknell}, James and {Jamieson}, William and {van Kerkwijk}, Marten H. and {Robitaille}, Thomas P. and {Merry}, Bruce and {Bachetti}, Matteo and {G{\"u}nther}, H. Moritz and {Aldcroft}, Thomas L. and {Alvarado-Montes}, Jaime A. and {Archibald}, Anne M. and {B{\'o}di}, Attila and {Bapat}, Shreyas and {Barentsen}, Geert and {Baz{\'a}n}, Juanjo and {Biswas}, Manish and {Boquien}, M{\'e}d{\'e}ric and {Burke}, D.~J. and {Cara}, Daria and {Cara}, Mihai and {Conroy}, Kyle E. and {Conseil}, Simon and {Craig}, Matthew W. and {Cross}, Robert M. and {Cruz}, Kelle L. and {D'Eugenio}, Francesco and {Dencheva}, Nadia and {Devillepoix}, Hadrien A.~R. and {Dietrich}, J{\"o}rg P. and {Eigenbrot}, Arthur Davis and {Erben}, Thomas and {Ferreira}, Leonardo and {Foreman-Mackey}, Daniel and {Fox}, Ryan and {Freij}, Nabil and {Garg}, Suyog and {Geda}, Robel and {Glattly}, Lauren and {Gondhalekar}, Yash and {Gordon}, Karl D. and {Grant}, David and {Greenfield}, Perry and {Groener}, Austen M. and {Guest}, Steve and {Gurovich}, Sebastian and {Handberg}, Rasmus and {Hart}, Akeem and {Hatfield-Dodds}, Zac and {Homeier}, Derek and {Hosseinzadeh}, Griffin and {Jenness}, Tim and {Jones}, Craig K. and {Joseph}, Prajwel and {Kalmbach}, J. Bryce and {Karamehmetoglu}, Emir and {Ka{\l}uszy{\'n}ski}, Miko{\l}aj and {Kelley}, Michael S.~P. and {Kern}, Nicholas and {Kerzendorf}, Wolfgang E. and {Koch}, Eric W. and {Kulumani}, Shankar and {Lee}, Antony and {Ly}, Chun and {Ma}, Zhiyuan and {MacBride}, Conor and {Maljaars}, Jakob M. and {Muna}, Demitri and {Murphy}, N.~A. and {Norman}, Henrik and {O'Steen}, Richard and {Oman}, Kyle A. and {Pacifici}, Camilla and {Pascual}, Sergio and {Pascual-Granado}, J. and {Patil}, Rohit R. and {Perren}, Gabriel I. and {Pickering}, Timothy E. and {Rastogi}, Tanuj and {Roulston}, Benjamin R. and {Ryan}, Daniel F. and {Rykoff}, Eli S. and {Sabater}, Jose and {Sakurikar}, Parikshit and {Salgado}, Jes{\'u}s and {Sanghi}, Aniket and {Saunders}, Nicholas and {Savchenko}, Volodymyr and {Schwardt}, Ludwig and {Seifert-Eckert}, Michael and {Shih}, Albert Y. and {Jain}, Anany Shrey and {Shukla}, Gyanendra and {Sick}, Jonathan and {Simpson}, Chris and {Singanamalla}, Sudheesh and {Singer}, Leo P. and {Singhal}, Jaladh and {Sinha}, Manodeep and {Sip{\H{o}}cz}, Brigitta M. and {Spitler}, Lee R. and {Stansby}, David and {Streicher}, Ole and {{\v{S}}umak}, Jani and {Swinbank}, John D. and {Taranu}, Dan S. and {Tewary}, Nikita and {Tremblay}, Grant R. and {de Val-Borro}, Miguel and {Van Kooten}, Samuel J. and {Vasovi{\'c}}, Zlatan and {Verma}, Shresth and {de Miranda Cardoso}, Jos{\'e} Vin{\'\i}cius and {Williams}, Peter K.~G. and {Wilson}, Tom J. and {Winkel}, Benjamin and {Wood-Vasey}, W.~M. and {Xue}, Rui and {Yoachim}, Peter and {Zhang}, Chen and {Zonca}, Andrea and {Astropy Project Contributors}},
        title = "{The Astropy Project: Sustaining and Growing a Community-oriented Open-source Project and the Latest Major Release (v5.0) of the Core Package}",
      journal = {\apj},
         year = 2022,
        month = aug,
       volume = {935},
       number = {2},
          eid = {167},
        pages = {167},
          doi = {10.3847/1538-4357/ac7c74},
archivePrefix = {arXiv},
       eprint = {2206.14220},
 primaryClass = {astro-ph.IM},
       adsurl = {https://ui.adsabs.harvard.edu/abs/2022ApJ...935..167A}
}

@ARTICLE{Astropy2013,
       author = {{Astropy Collaboration} and {Robitaille}, Thomas P. and {Tollerud}, Erik J. and {Greenfield}, Perry and {Droettboom}, Michael and {Bray}, Erik and {Aldcroft}, Tom and {Davis}, Matt and {Ginsburg}, Adam and {Price-Whelan}, Adrian M. and {Kerzendorf}, Wolfgang E. and {Conley}, Alexander and {Crighton}, Neil and {Barbary}, Kyle and {Muna}, Demitri and {Ferguson}, Henry and {Grollier}, Fr{\'e}d{\'e}ric and {Parikh}, Madhura M. and {Nair}, Prasanth H. and {Unther}, Hans M. and {Deil}, Christoph and {Woillez}, Julien and {Conseil}, Simon and {Kramer}, Roban and {Turner}, James E.~H. and {Singer}, Leo and {Fox}, Ryan and {Weaver}, Benjamin A. and {Zabalza}, Victor and {Edwards}, Zachary I. and {Azalee Bostroem}, K. and {Burke}, D.~J. and {Casey}, Andrew R. and {Crawford}, Steven M. and {Dencheva}, Nadia and {Ely}, Justin and {Jenness}, Tim and {Labrie}, Kathleen and {Lim}, Pey Lian and {Pierfederici}, Francesco and {Pontzen}, Andrew and {Ptak}, Andy and {Refsdal}, Brian and {Servillat}, Mathieu and {Streicher}, Ole},
        title = "{Astropy: A community Python package for astronomy}",
      journal = {\aap},
         year = 2013,
        month = oct,
       volume = {558},
          eid = {A33},
        pages = {A33},
          doi = {10.1051/0004-6361/201322068},
archivePrefix = {arXiv},
       eprint = {1307.6212},
 primaryClass = {astro-ph.IM},
       adsurl = {https://ui.adsabs.harvard.edu/abs/2013A\&A...558A..33A}
}

@ARTICLE{Hunter2007,
       author = {{Hunter}, John D.},
        title = "{Matplotlib: A 2D Graphics Environment}",
      journal = {Computing in Science and Engineering},
         year = 2007,
        month = may,
       volume = {9},
       number = {3},
        pages = {90-95},
          doi = {10.1109/MCSE.2007.55},
       adsurl = {https://ui.adsabs.harvard.edu/abs/2007CSE.....9...90H}
}
\bibliographystyle{aasjournal}

\appendix

\section{Contaminating line correction}\label{app:lines}
The Merian N708 medium-band filter ($\lambda_\text{eff} = 7080$ \AA, FWHM $= 275$ \AA) captures not only \ha but also the neighboring [\ion{N}{2}]$\lambda\lambda$6548,6583 and [\ion{S}{2}]$\lambda\lambda$6716,6731 emission lines, whose contribution must be subtracted to recover the pure \ha flux.

We derive an average empirical multiplicative correction factor for each galaxy using the full sample of Merian-matched DESI DR1 spectra. For each galaxy with a DESI spectrum, we fit a power-law continuum to the rest-frame 6900–7300 $\AA$ window of the dust-corrected spectrum, masking the  \ha+[\ion{N}{2}]  and [\ion{S}{2}] regions. We then construct a synthetic \hanospace-only spectrum by replacing the observed flux outside a narrow \ha window (rest 6555–6575 $\AA$) with the fitted continuum, preserving only the \ha line itself. Both the full line spectrum and the \hanospace-only spectrum are integrated through the N708 transmission curve and the correction factor for each galaxy is taken to be the ratio of the full flux through the filter to the \hanospace-only flux.

Because different lines fall in the N708 filter at different redshifts, the correction varies systematically with $z$. It also varies with stellar mass, since higher-mass galaxies tend to have higher metallicities and therefore higher [\ion{N}{2}]/\ha ratios. We therefore bin the sample by stellar mass (four bins: $\log M_\star/M_\odot < 9.0, \in(9.0, 9.5), \in (9.5, 10), >10$) and measure the median correction in the low-redshift regime where the [\ion{S}{2}] doublet is fully covered by N708 ($z < 0.074$) and a high-redshift regime where the [\ion{S}{2}] doublet falls outside of the filter ($z > 0.083$). The correction within the transition range $z \in [0.074, 0.083]$ is linearly interpolated between the two medians. This calibration was performed on the over 6000 Merian galaxies with DESI spectra.

The multiplicative correction is computed according to the following parameterization.

\begin{equation}
c(z) = 
\left\{
    \begin{array}{lr}
        c_1 & \text{if } z < 0.074 \\
        \frac{c_2 - c_1}{0.009}(z-0.074) + c_1 & \text{if } 0.074 < z < 0.083 \\
        c_2 & \text{if } 0.083 < z
    \end{array}
\right.
\end{equation}

with mass-dependent coefficients,
\begin{equation}
\left\{
    \begin{array}{lr}
        c_1= 1.50,c_2= 1.09& \text{if } \log (M_\star/M_\odot) < 9.0 \\
        c_1= 1.71,c_2= 1.24& \text{if } 9.0 < \log (M_\star/M_\odot) < 9.5 \\
        c_1= 1.86,c_2= 1.43& \text{if } 9.5 < \log (M_\star/M_\odot) < 10.0 \\
        c_1= 1.96,c_2= 1.64& \text{if } 10 < \log (M_\star/M_\odot)
    \end{array}
    \right.
\end{equation}

such that ${\rm EW}_{{\rm H}\alpha+[\text{N II}]} = c \cdot {\rm EW}_{{\rm H}\alpha}$
\section{Results of Linear Regressions}\label{app:models}
Here we present the full results of the linear regressions described in Section~\ref{sec:results}. \autoref{tab:no_OH} includes the results of fitting \autoref{eq:lm1} to the sample - i.e. of fitting the morphological parameters as a function of $\log$ sSFR, $\log M_\star/M_\odot$, and $r_\text{phys}$. Tables \ref{tab:S24}, \ref{tab:C24}, and \ref{tab:N22} include the results of fitting \autoref{eq:lm2} using metallicities measured with strong line calibrations from \citet{Scholte2024}, \citet{Curti2024}, and \citet{Nakajima2022}, respectively.

\begin{table*}[h]
    \caption{Linear regression results for fits to morphological parameters without metallicity.}\label{tab:no_OH}
    \hspace{5in}
    \centering
\begin{tabular}{rrrrr}
\hline\hline
      & $\beta_\text{sSFR}$ & $\beta_{M_\star}$ & $\beta_{r_\text{phys}}$ & $R^2$\\
      & (1) & (2) & (3) & (4)\\
     \hline\\
$\mathcal{A}_{\text{H}\alpha}$ &$0.063^{***}$ & $0.052^{***}$ & $-0.025^{***}$ & 0.090 \\
$\mathcal{G}_{\text{H}\alpha}$ &$0.046^{***}$ & $0.018^{***}$ & $-0.012^{***}$ & 0.290 \\
$\mathcal{M}_{20, \text{H}\alpha}$ &$-0.138^{***}$ & $-0.081^{***}$ & $0.006^{\ \ \ }$ & 0.126 \\
$\mathcal{A}_{c}$ &$0.015^{***}$ & $0.004^{*\ \ }$ & $-0.002^{\ \ \ }$ & 0.027 \\
$\mathcal{G}_{c}$ &$0.006^{***}$ & $0.003^{**\ }$ & $-0.004^{***}$ & 0.026 \\
$\mathcal{M}_{20, c}$ &$0.005^{\ \ \ }$ & $-0.024^{***}$ & $0.010^{***}$ & 0.039 \\
\hline
\end{tabular}

    \hspace{5in}
    \tablecomments{Results of fitting the linear model presented in \autoref{eq:lm1} with results for each morphological parameter presented in each row. Fitted coefficients for $\log$ sSFR, $\log M_\star/M_\odot$, and $r_\text{phys}$ are presented in columns (1), (2), and (3) respectively with significance indicated by * for $p < 0.05$, ** for $p < 0.01$, and *** for $p<0.001$. Column (4) includes the value of $R^2$ for the fit, a measure of the fraction of variance in the dependent variable explained by the predictors.}
\end{table*}

\begin{table*}[ht]
    \caption{Linear regression results for fits to morphological parameters with metallicity included for S2024-calibrated metallicities.}\label{tab:S24}
    \hspace{5in}
    \centering
\begin{tabular}{rrrrrr}
\hline\hline
      & $\beta_\text{sSFR}$ & $\beta_{M_\star}$ & $\beta_{r_\text{phys}}$ & $\beta_{\text{O/H}}$ & $\Delta R^2$\\
   & (1) & (2) & (3) & (4) & (5)\\
     \hline\\
$\mathcal{A}_{\text{H}\alpha}$ &$0.057^{***}$ & $0.080^{***}$ & $-0.028^{***}$ & $-0.077^{***}$ & 0.009 \\
$\mathcal{G}_{\text{H}\alpha}$ &$0.048^{***}$ & $0.007^{**\ }$ & $-0.011^{***}$ & $0.033^{***}$ & 0.012 \\
$\mathcal{M}_{20, \text{H}\alpha}$ &$-0.141^{***}$ & $0.046^{***}$ & $0.001^{\ \ \ }$ & $-0.321^{***}$ & 0.056 \\
$\mathcal{A}_{c}$ &$0.009^{***}$ & $0.037^{***}$ & $-0.004^{*\ \ }$ & $-0.088^{***}$ & 0.073 \\
$\mathcal{G}_{c}$ &$0.006^{***}$ & $0.008^{***}$ & $-0.004^{***}$ & $-0.011^{***}$ & 0.005 \\
$\mathcal{M}_{20, c}$ &$0.004^{\ \ \ }$ & $-0.019^{***}$ & $0.010^{**\ }$ & $-0.018^{\ \ \ }$ & 0.001 \\
\hline
\end{tabular}

    \hspace{5in}
    \tablecomments{Results of fitting the linear model presented in \autoref{eq:lm2} to the sample with metallicities measured using strong line calibrations from \citet{Scholte2024} (N=2324). Fitted coefficients for $\log$ sSFR, $\log M_\star/M_\odot$, $r_\text{phys}$, and $12 + \log(\text{O/H})$ are presented in columns (1), (2), (3), and (4) respectively with significance indicated as in \autoref{tab:no_OH}. Column (5) indicates the increase in $R^2$ as compared to a linear model fit without metallicity to the same sample.}
\end{table*}

\begin{table*}[ht]
    \caption{Linear regression results for fits to morphological parameters with metallicity included for C2024-calibrated metallicities.}\label{tab:C24}
    \hspace{5in}
    \centering
\begin{tabular}{rrrrrr}
\hline\hline
      & $\beta_\text{sSFR}$ & $\beta_{M_\star}$ & $\beta_{r_\text{phys}}$ & $\beta_{\text{O/H}}$ & $\Delta R^2$\\
      & (1) & (2) & (3) & (4) & (5)\\
     \hline\\
$\mathcal{A}_{\text{H}\alpha}$ &$0.056^{***}$ & $0.084^{***}$ & $-0.027^{***}$ & $-0.052^{***}$ & 0.014 \\
$\mathcal{G}_{\text{H}\alpha}$ &$0.045^{***}$ & $0.019^{***}$ & $-0.012^{***}$ & $-0.001^{\ \ \ }$ & 0.000 \\
$\mathcal{M}_{20, \text{H}\alpha}$ &$-0.148^{***}$ & $-0.012^{\ \ \ }$ & $0.001^{\ \ \ }$ & $-0.107^{***}$ & 0.021 \\
$\mathcal{A}_{c}$ &$0.010^{***}$ & $0.031^{***}$ & $-0.003^{\ \ \ }$ & $-0.043^{***}$ & 0.060 \\
$\mathcal{G}_{c}$ &$0.005^{***}$ & $0.010^{***}$ & $-0.004^{***}$ & $-0.012^{***}$ & 0.021 \\
$\mathcal{M}_{20, c}$ &$0.005^{\ \ \ }$ & $-0.026^{***}$ & $0.010^{***}$ & $0.003^{\ \ \ }$ & 0.000 \\
\hline
\end{tabular}

    \hspace{5in}
    \tablecomments{Same as \autoref{tab:S24}, but for metallicities calculated using strong line calibrations from \citet{Curti2024} (N=2632). }
\end{table*}

\begin{table*}[ht]
    \caption{Linear regression results for fits to morphological parameters with metallicity included for N2022-calibrated metallicities.}\label{tab:N22}
    \hspace{5in}
    \centering
\begin{tabular}{rrrrrr}
\hline\hline
      & $\beta_\text{sSFR}$ & $\beta_{M_\star}$ & $\beta_{r_\text{phys}}$ & $\beta_{\text{O/H}}$ & $\Delta R^2$\\
       & (1) & (2) & (3) & (4) & (5)\\
\hline\\
$\mathcal{A}_{\text{H}\alpha}$ &$0.056^{***}$ & $0.082^{***}$ & $-0.027^{***}$ & $-0.047^{***}$ & 0.012 \\
$\mathcal{G}_{\text{H}\alpha}$ &$0.046^{***}$ & $0.016^{***}$ & $-0.012^{***}$ & $0.003^{\ \ \ }$ & 0.000 \\
$\mathcal{M}_{20, \text{H}\alpha}$ &$-0.150^{***}$ & $-0.003^{\ \ \ }$ & $0.001^{\ \ \ }$ & $-0.120^{***}$ & 0.027 \\
$\mathcal{A}_{c}$ &$0.010^{***}$ & $0.032^{***}$ & $-0.003^{\ \ \ }$ & $-0.044^{***}$ & 0.065 \\
$\mathcal{G}_{c}$ &$0.005^{***}$ & $0.010^{***}$ & $-0.004^{***}$ & $-0.011^{***}$ & 0.020 \\
$\mathcal{M}_{20, c}$ &$0.005^{\ \ \ }$ & $-0.026^{***}$ & $0.010^{***}$ & $0.003^{\ \ \ }$ & 0.000 \\
\hline
\end{tabular}

    \hspace{5in}
    \tablecomments{Same as \autoref{tab:S24}, but for metallicities calculated using strong line calibrations from \citet{Nakajima2022} (N=2632). }
\end{table*}

\section{Binned linear trends}\label{app:bin}

The linear models of Section 5 assume that the morphological parameters depend linearly on $\log M_\star/M_\odot$, log sSFR, and metallicity, which is a clear oversimplification. To verify that the trends we report are not artifacts of this parameterization, we examine the data directly in bins of the predictor variables.

\begin{figure*}
    \centering
    \includegraphics[width=1\linewidth]{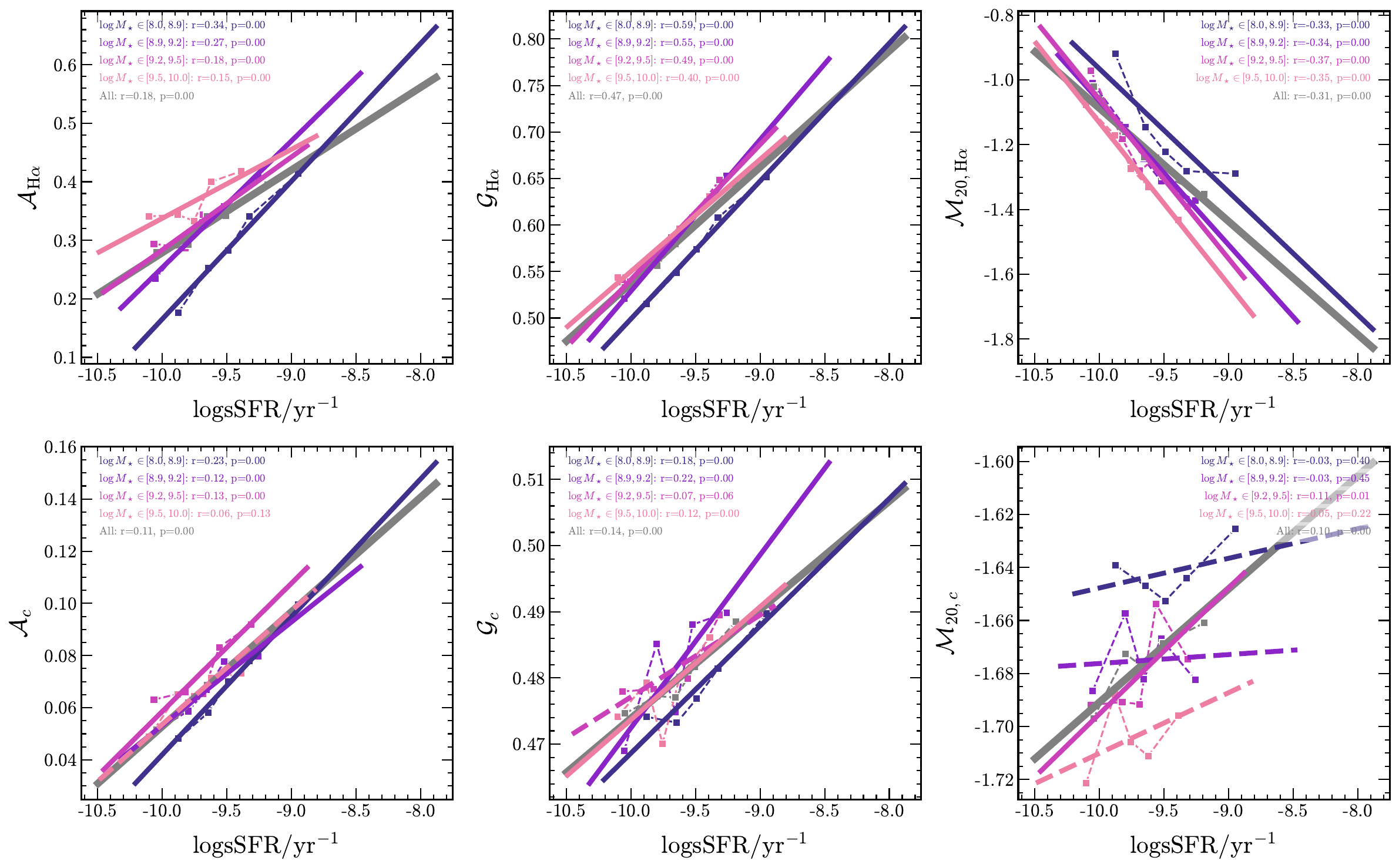}
    \caption{Binned trends of the morphological parameters with sSFR (derived from Merian photometry as described in Section~\ref{subsec:ha_measure}) in bins of stellar mass. Each panel shows one morphological statistic (for \ha in the top row and continuum maps in the bottom row) as a function of sSFR. Colored points and dashed lines show the binned medians within each stellar mass bin, and thick lines show linear fits to the individual galaxies in each bin, with the correlation coefficient and p-value for each bin indicated in the legend. Gray lines show the fit to the full sample. Fits with $p > 0.05$ are shown as dashed. The trends identified by the linear models of Section~\ref{subsec:morphtrends} are recovered clearly for the \ha morphologies in the individual mass bins and in most bins for the continuum.}
    \label{fig:ssfrtrends}
\end{figure*}

\begin{figure*}
    \centering
    \includegraphics[width=1\linewidth]{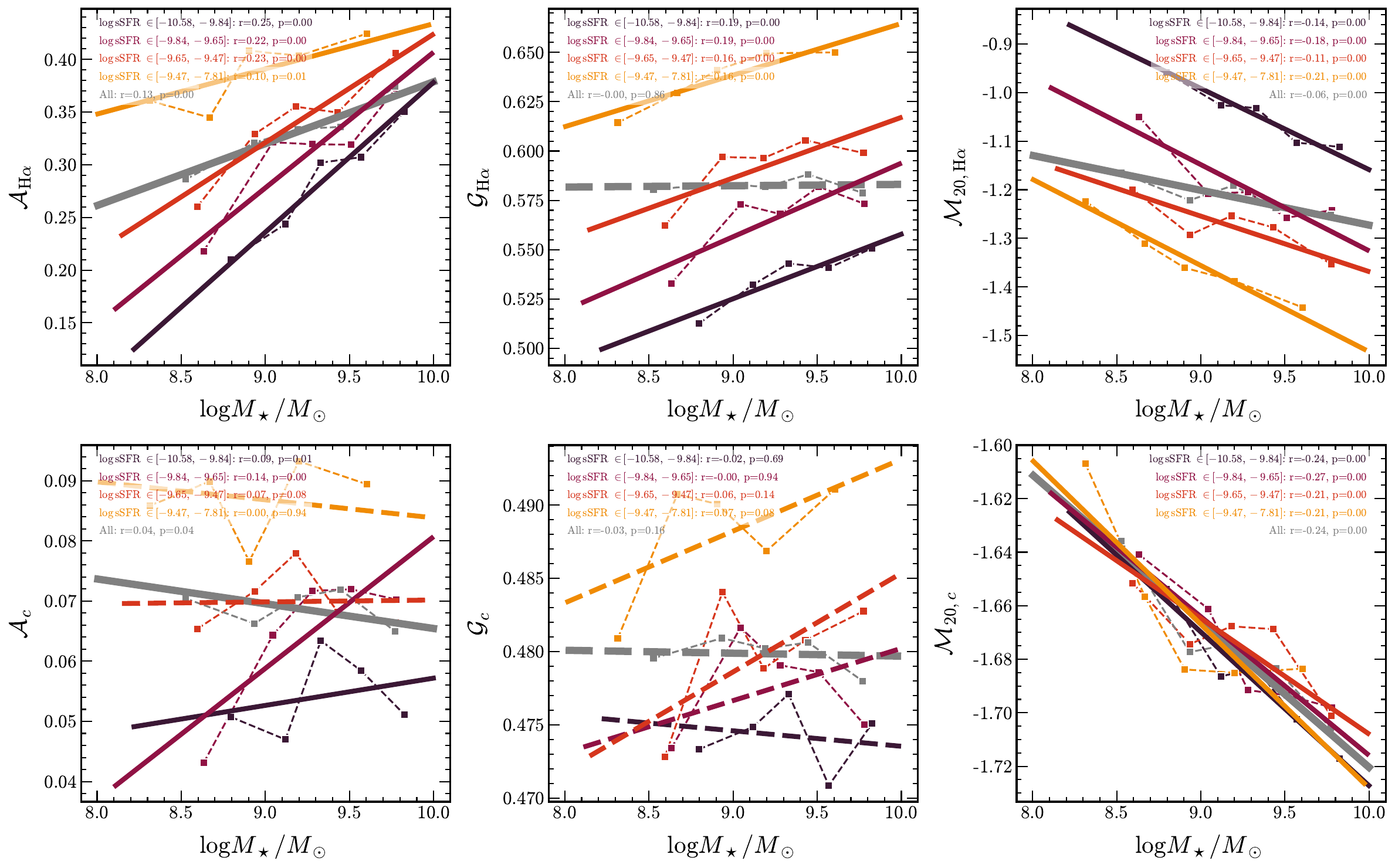}
    \caption{Same as \autoref{fig:ssfrtrends}, but showing the morphological parameters as a function of stellar mass in bins of sSFR.}
    \label{fig:masstrends}
\end{figure*}

\autoref{fig:ssfrtrends} shows each morphological parameter as a function of sSFR in four bins of stellar mass, and \autoref{fig:masstrends} shows the same parameters as a function of stellar mass in four bins of sSFR. In each panel we plot the binned medians together with a linear fit to the individual galaxies within each bin, and report the Pearson correlation coefficient and $p$-value for each bin in the legend. The gray line shows the fit to the full sample.

The trends identified in Section 5.1 are recovered clearly for the \ha morphologies. The continuum trends are weaker, as expected from their smaller fitted coefficients and lower $R^2$ values, but are recovered in most bins. 

\end{document}